\pdfoutput=1

\documentclass[12pt]{iopart}
\usepackage{cite}
\usepackage{graphicx}
\usepackage{subfigure}
\usepackage{color}
\definecolor{Blue}{rgb}{0.3,0.3,0.9}
\definecolor{Red}{rgb}{0.9,0.3,0.3}

\newcommand{\revision}[1]{#1}

\begin{document}
\title{Rapid simulation of protein motion: merging flexibility, rigidity and normal mode analyses}
\author{J. E. Jimenez-Roldan$^{1,2}$, R. B. Freedman$^2$, R. A. R\"{o}mer$^1$, S. A. Wells$^1$}
\address{1 Department of Physics and Centre for Scientific Computing, University of Warwick,
Coventry, CV4 7AL, UK}
\address{2 School of Life Sciences, University of Warwick, Coventry, CV4 7AL, UK}
\ead{j.e.jimenez@warwick.ac.uk}

\begin{abstract}
Protein function frequently involves conformational changes with large amplitude on time scales which are difficult and computationally expensive to access using molecular dynamics. In this paper we report on the combination of three computationally inexpensive simulation methods --- normal mode analysis using the elastic network model, rigidity analysis using the pebble game algorithm, and geometric simulation of protein motion --- to explore conformational change along normal mode eigenvectors. Using a combination of {\sc ElNemo} and {\sc First}/{\sc Froda} software, large-amplitude motions in proteins with hundreds or thousands of residues can be rapidly explored within minutes using  desktop computing resources. We apply the method to a representative set of $6$ proteins covering a range of sizes and structural characteristics and show that the method identifies specific types of motion in each case and determines their amplitude limits.

\indent{$Revision: 1.77 $, compiled \today}

\end{abstract}
\noindent{\it Protein Rigidity, protein flexibility, normal mode analysis, conformational change,
domain motion, rapid simulation, coarse-grained methods, multi-scale methods.\/}
\section{Introduction}

Protein conformational changes and dynamic behaviour are fundamental for processes such as
catalysis, regulation, and substrate recognition. The time scales of motions involved in enzyme
function span multiple orders of magnitude, from the picosecond timescale of local side groups
rotations to the milli-/microsecond timescale of the motion of entire domains \cite{HenK07}.
Empirical-potential molecular dynamics (MD) has proved to be a valuable tool for investigating
molecular motions, but specialized expertise, large-scale computing resources and weeks or months of
compute time are required to explore protein motion on simulation time scales greater than tens of
nanoseconds \cite{ShaMLS10}. There is clearly a need for methods that
that permit exploration of possible large conformational motions
of proteins in a rational, albeit somewhat simplified, fashion with minimal computational resources.
Such explorations allow for the generation of hypotheses about large conformational motion
and protein function that can then be investigated with detailed MD simulations and by experimental
methods such as FRET \cite{GieAET06}.

The computational cost of simulations can be reduced by coarse-graining (CG) --- averaging out
atomic degrees of freedom so as to represent groups of atoms by a single site \cite{RusLPS09,Cle08}
--- and/or by simplifying the intersite interactions.
Different levels of simplification can then be combined in multi-scale methods \cite{YalB07,SheBS08}. Here, we shall consider three such methods in particular. Pebble-game rigidity analysis, implemented in {\sc First} \cite{JacRKT01}, provides valuable information on the distribution of rigid and flexible regions in a structure \cite{WelJR09}. Geometric simulation in the {\sc Froda} algorithm \cite{WelMHT05} uses rigidity information and explores flexible motion \cite{JolWHT06,JolWFT08}. Normal mode analysis of a coarse-grained elastic network model (ENM), implemented in {\sc ElNemo} \cite{SuhS04,SuhSB04}, generates eigenvectors for low-frequency motion which are potential sources of functional motion and conformational change \cite{BahLBS10,Ma05,RueCO07,NakMKH10,DykS10}.

Our approach in this study is to bias the generation of new
conformations in {\sc First/Froda} along an eigenvector predicted by {\sc ElNemo} as a low-frequency mode.
The bias directs the motion of the atoms in the direction of the eigenvector while the geometric
constraint system maintains rational bonding and sterics and prevents the build-up of distortions
that occur in a linear projection. 
The method is outlined schematically in Figure \ref{fig:method}.
We apply our method to a set of six proteins of various sizes from $58$ to $1605$ residues.
We find that flexible motion in an all-atom model can be explored to large amplitudes in a few
CPU-minutes, until further motion is limited by bonding or steric constraints, representing the calculated limit of motion along that vector.

\section{Methods}

\subsection{Protein selection}

\begin{figure}[p]
\centering
(a)\includegraphics[width=0.2\textwidth]{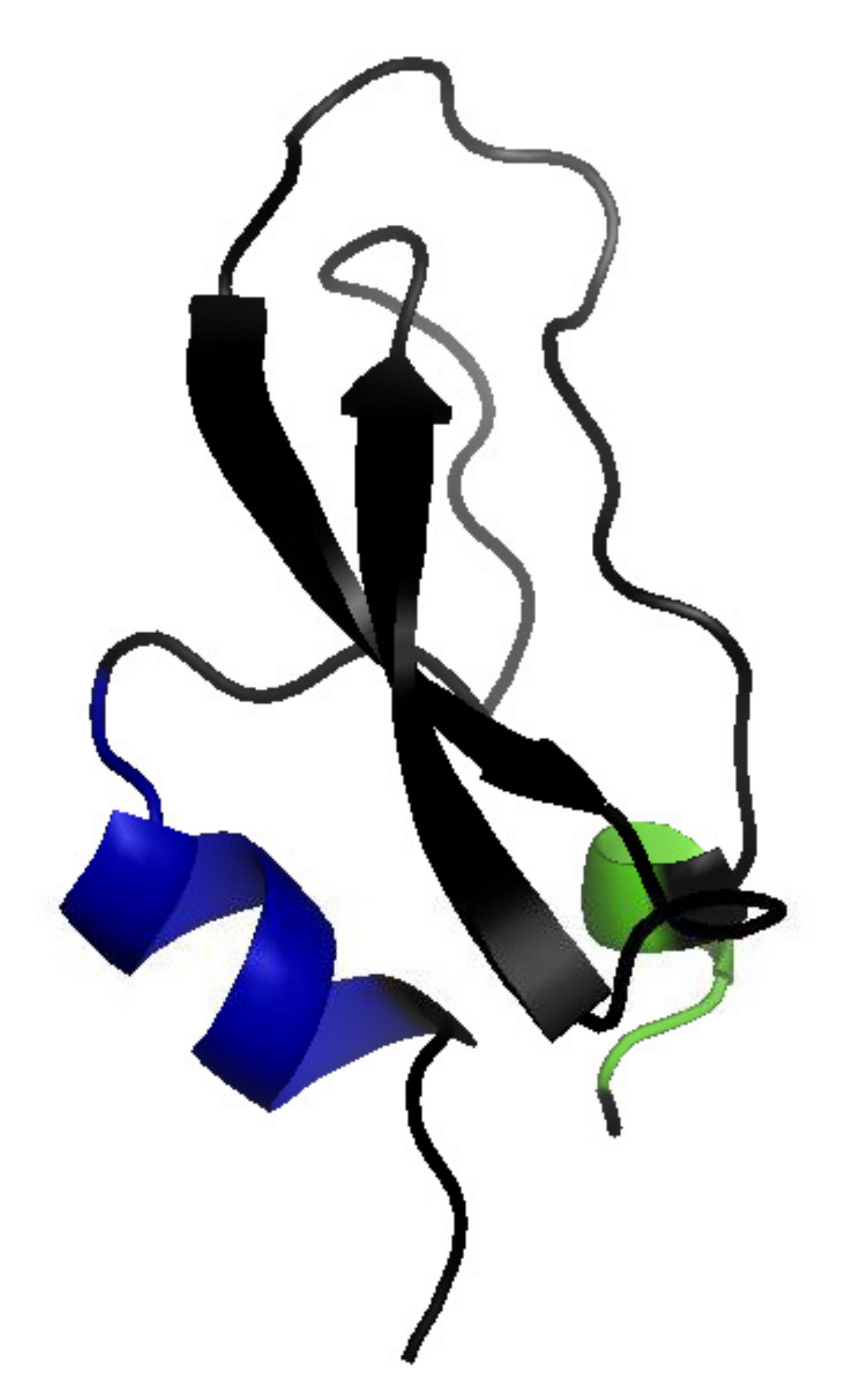}
\hfil 
(b)\includegraphics[width=0.36\textwidth]{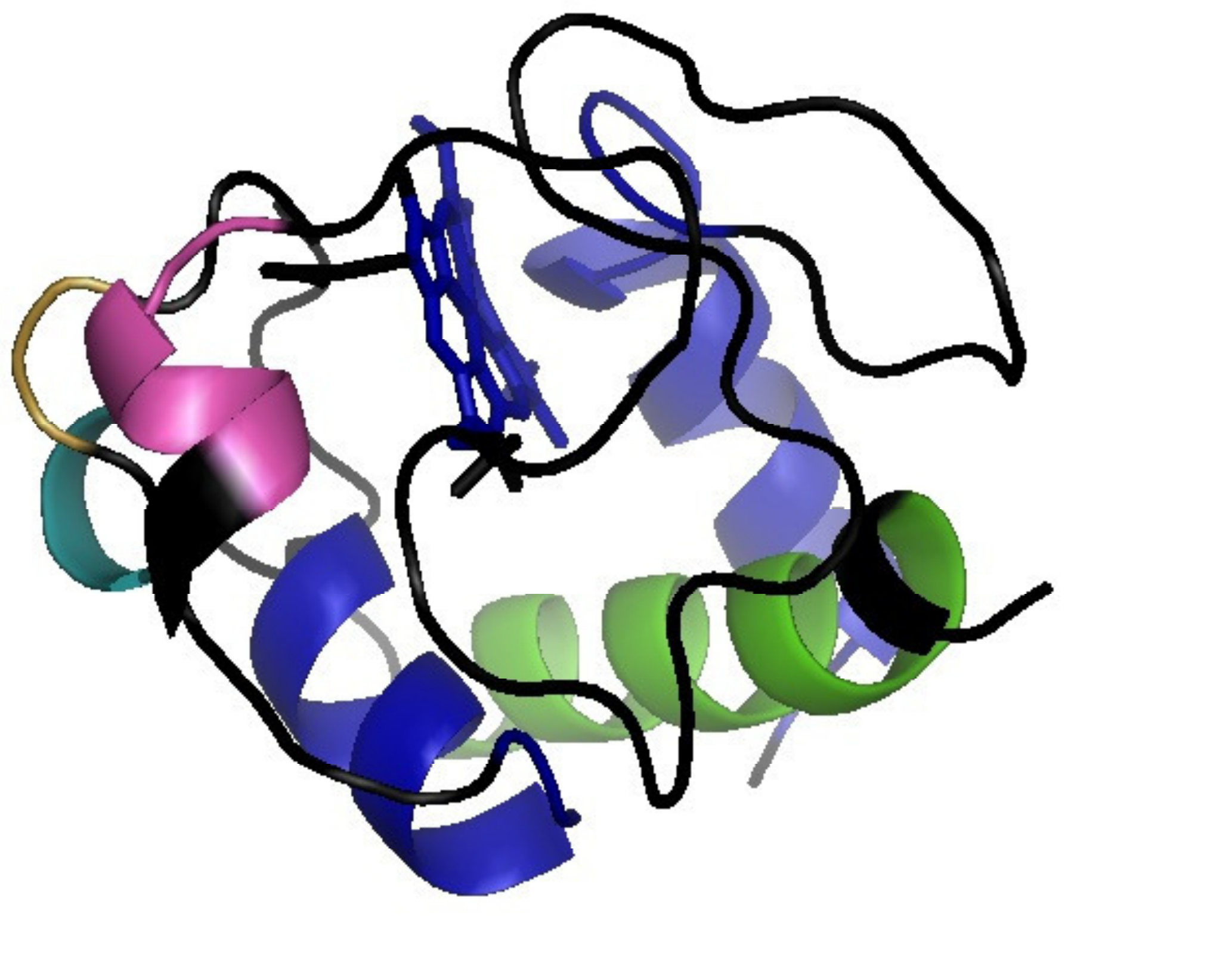}
(c)\includegraphics[width=0.36\textwidth]{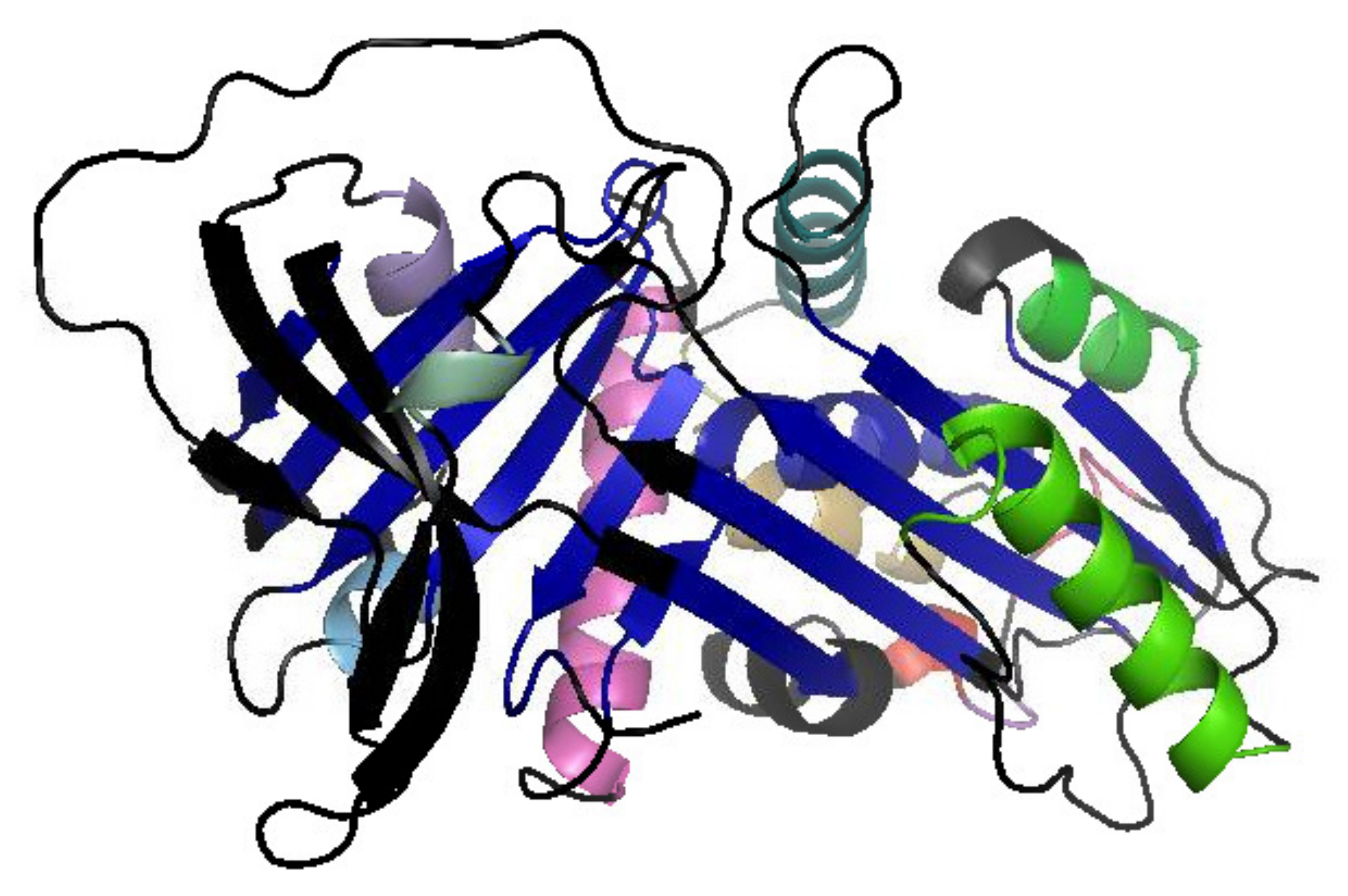}
(d)\includegraphics[width=0.42\textwidth]{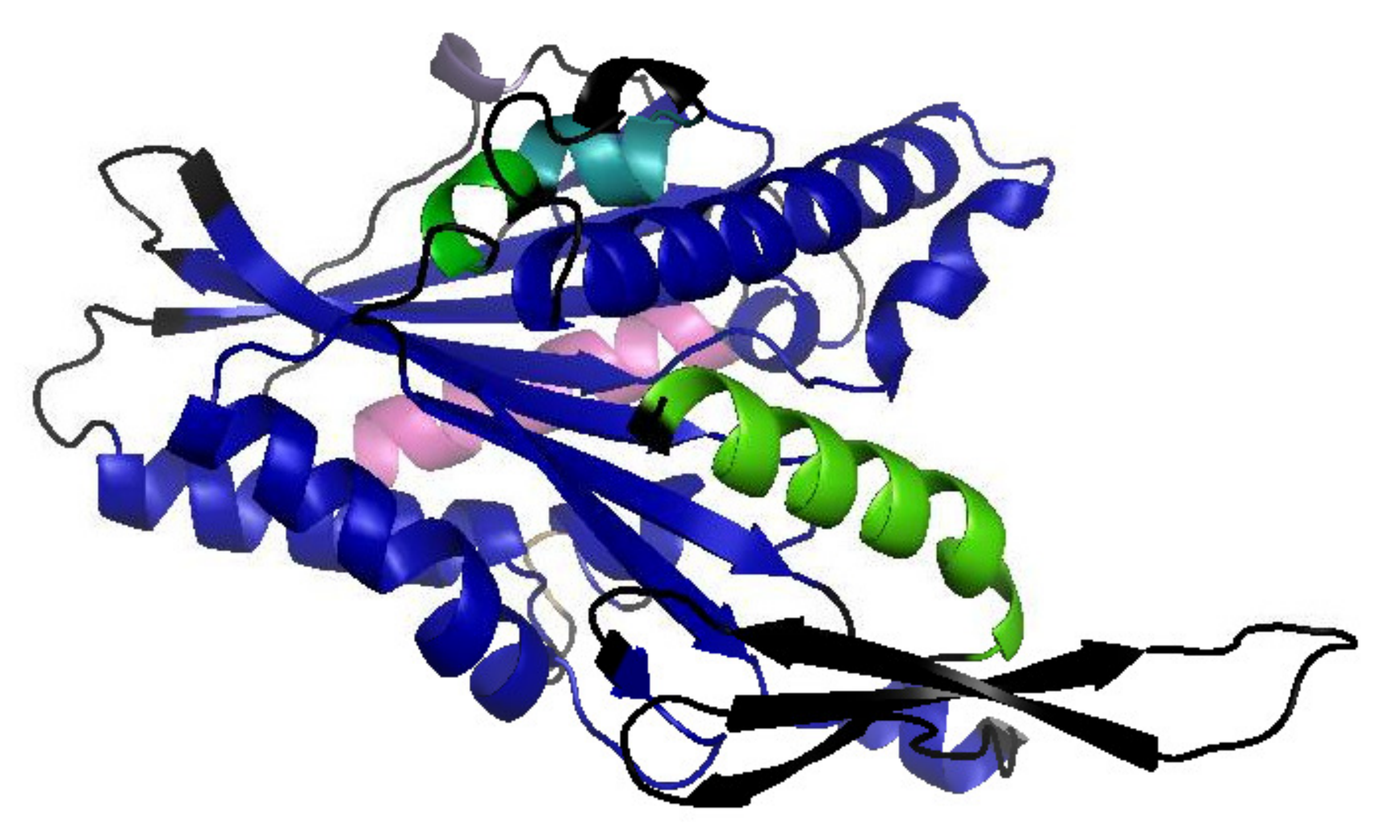}
(e)\includegraphics[width=0.42\textwidth]{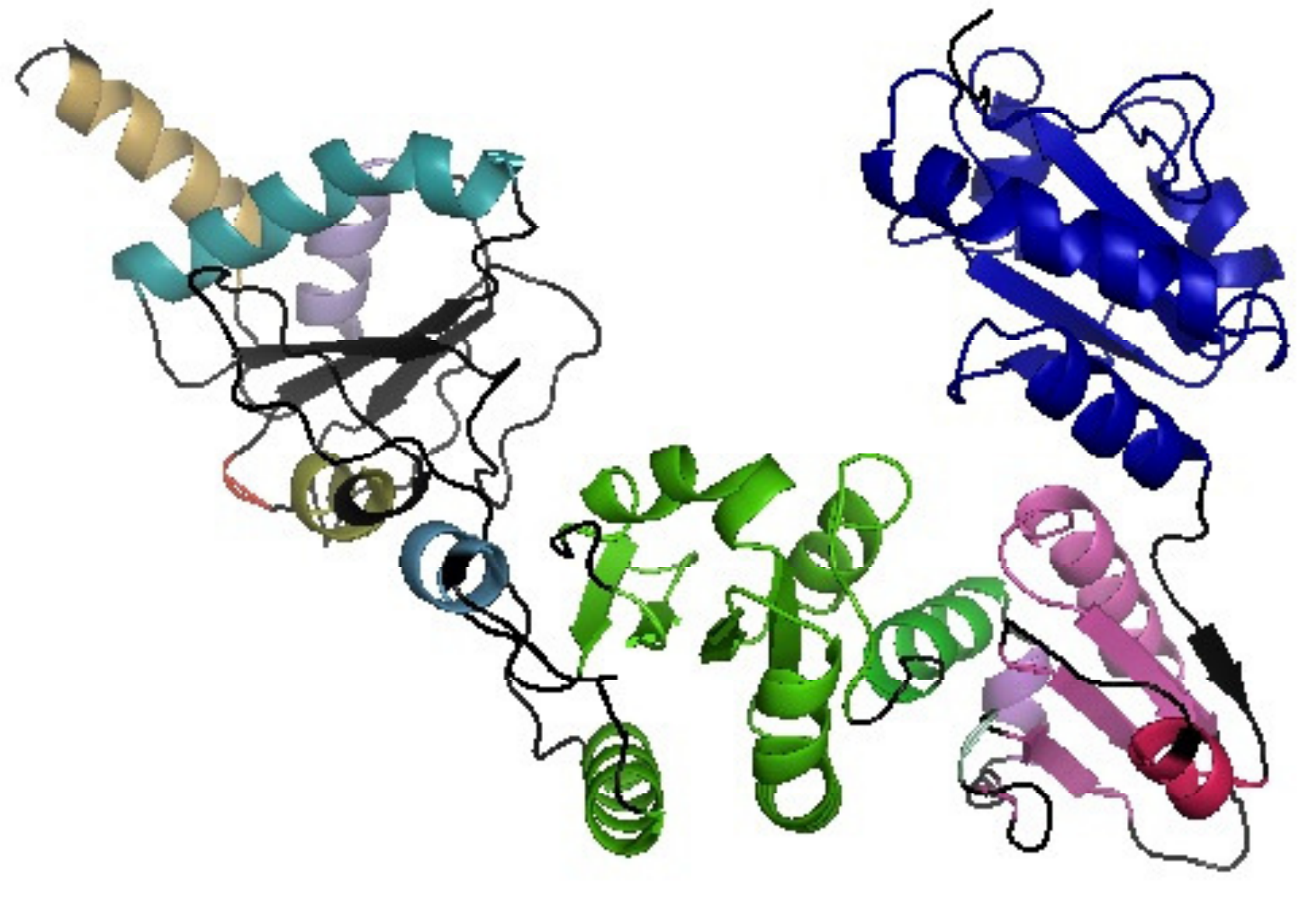}
(f)\includegraphics[width=0.40\textwidth]{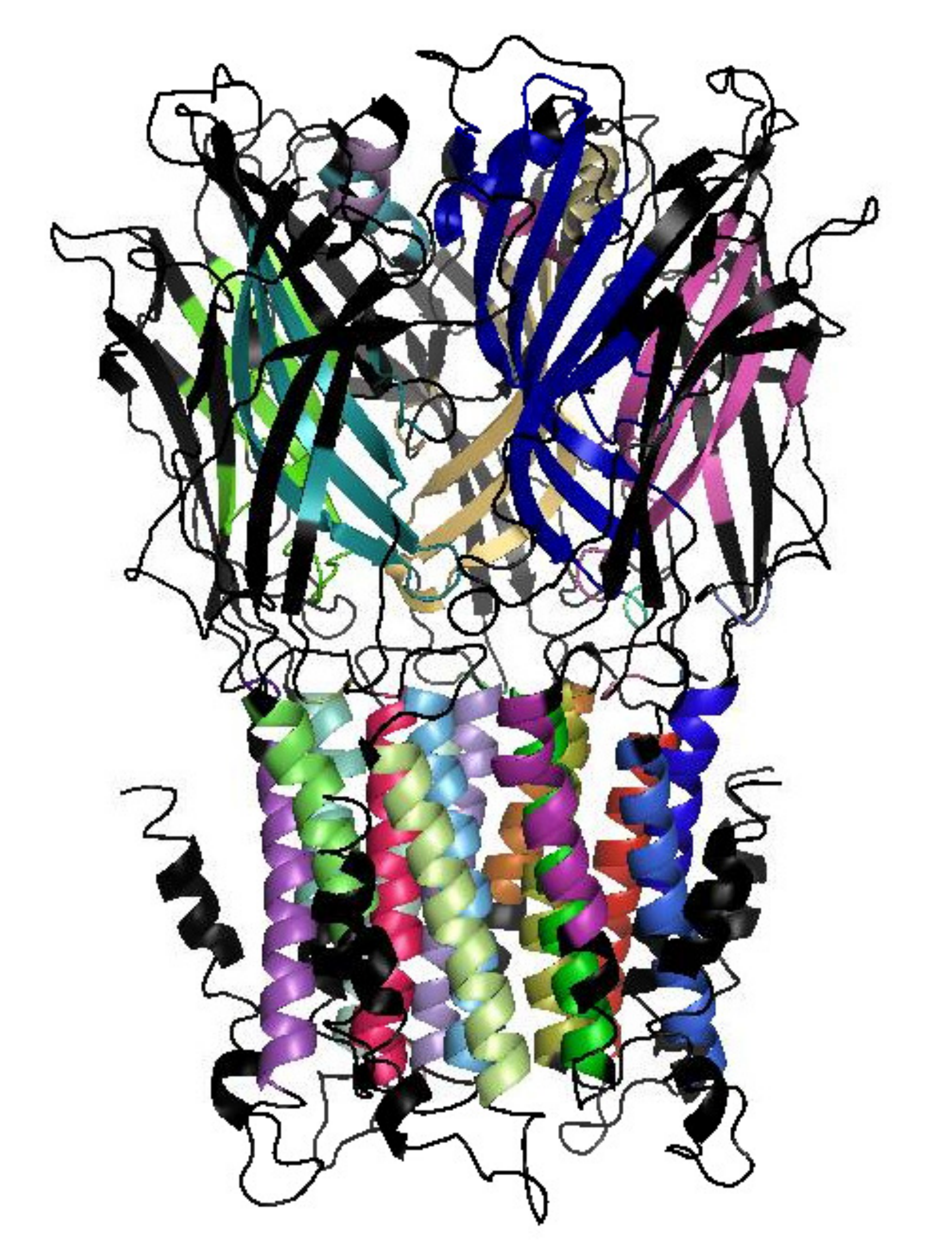}
\caption{\label{fig:proteins} 
Tertiary structure of all six protein structures (a) BPTI (1BPI), (b) cytochrome-c (1HRC), (c) $\alpha$1-antitrypsin (1QLP), (d) kinesin (1RY6,) (e) yeast PDI (2B5E) and (f) pLGIC (2VL0). The structures are given in standard {\sc Pymol} \cite{Del02} format but broken into rigid clusters according to the rigidity analysis (cp.\ Fig.~\ref{Fig:RCD_plot}) at the specific values of $E_{\rm cut}$
shown in Table \ref{Tab:cutoffs}. 
Each rigid cluster is represented in a different colour with the largest rigid cluster indicated
in blue and flexible regions shown in black.}
\end{figure}

We deliberately selected six proteins for analysis that are diverse in function, structural characteristics and size, ranging from $58$ to $1605$ residues. For each protein, we selected a representative high-resolution structure from the Protein Data Bank (PDB) \cite{BerWFG00}. The proteins and their PDB codes are listed in Table \ref{Tab:cutoffs} and their structures are shown in Figure \ref{fig:proteins} with colour coding according to the results of rigidity analysis.

Bovine pancreatic trypsin inhibitor (BPTI) is a small well-studied protease inhibitor of $58$ amino
acids, comprising mainly random-coil structure plus two antiparallel $\beta$-strands and two short
$\alpha$-helices; the protein has only a small hydrophobic core, but is additionally stabilized by
$3$
intra-chain disulphide bonds \cite{WloWHS84,AmiH88}. Mammalian mitochondrial cytochrome-c is a
classic electron-transfer protein containing a redox-active haem group bound within a primarily
$\alpha$-helical protein fold. These two were selected as contrasting small proteins.

As medium size proteins we selected $\alpha$1-antitrypsin and the core catalytic domain of
the motor protein kinesin \cite{ShiHTM04}. The former is a protease inhibitor of the serpin family
\cite{EllPDL00} which operates via a `bait' mechanism comparable to that of a mouse-trap, involving
a very significant conformational change, whereas the latter is a mechanochemical device that
transduces the chemical energy of ATP hydrolysis into mechanical work, specifically the
depolymerisation of microtubules in the case of this kinI kinesin. Both these proteins comprise an
extensive $\beta$-sheet core flanked by several $\alpha$-helices.

Protein disulphide-isomerase (PDI) is a large protein (with more than $500$ residues) comprising $4$ distinct
domains each with a thioredoxin-like fold, connected by two short and one longer linker
\cite{FreKR02}; the protein has both redox and molecular chaperone activity and intramolecular
flexibility is essential for its action in facilitating oxidative folding of secretory proteins
\cite{TiaKLS08,TiaXNLS06}. The largest protein selected is an integral membrane protein (a bacterial
protein
of $1605$ residues) that operates as a pentameric ligand-gated ion channel (pLGIC); it comprises an
extracellular --- mainly $\beta$-sheet --- domain and a membrane-embedded domain, mainly comprising
$\alpha$-helices which form the lining of the ion-channel; the mechanisms of ion permeation and
channel gating are not yet completely understood but it is clear that a conformational change is
required for function \cite{HilD08}.

\subsection{Rigidity analysis and energy cutoff selection}
\label{rigidityanalysis}

We add the hydrogen atoms absent from the PDB X-ray crystal structures using the software {\sc
Reduce} \cite{WorLRR99} and remove alternate conformations and renumber the hydrogen atoms in {\sc
Pymol} \cite{Del02}. This produces usable files for {\sc First} rigidity analysis. For each protein
we produce a ``rigidity dilution'' or rigid cluster decomposition (RCD) \cite{JacRKT01} plot
(displayed in the supplementary Figure \ref{Fig:RCD_plot}). The plots show the dependence of the
protein rigidity on an energy cutoff parameter, $E_{\rm cut}$, which determines the set of hydrogen
bonds to be included in the rigidity analysis. The tertiary structures with the residues coloured by
the rigid clusters they belong to are shown in Figures \ref{3D_1BPI_Ec}--\ref{3D_2VL0_Ec} for each
of the selected energy cutoffs. 

Previous studies \cite{JacRKT01} suggested that $E_{\rm cut}$ should be at least $-0.1$ kcal/mol in
order to eliminate a large number of very weak hydrogen bonds, and that a natural choice is near the
`room temperature' energy of $-0.6$ kcal/mol. \revision{We have recently discussed the criteria
for a robust selection of $E_{\rm cut}$ \cite{WelJR09}.}
For each protein we have selected several energy cutoffs at which to explore flexible motion, as
listed in Table \ref{Tab:cutoffs}. A higher cutoff energy increases the number of constraints
included in the simulation, and this is expected to restrict protein motion. We have used in each
case at least one cutoff at which the protein is largely rigid \revision{(in the range $-0.1$
kcal/mol to $-0.7$ kcal/mol)} and at least one lower cutoff at which the protein is largely flexible
\revision{(in the range $-0.5$ kcal/mol to $-2.2$ kcal/mol).}

\subsection{Normal modes of motion}

We obtain the normal modes of motion using the ENM \cite{Tir96} implemented in {\sc ElNemo}
software \cite{SuhS04,SuhSB04}. This generates, for each protein, a set of eigenvectors and
associated eigenvalues. Other implementations of elastic network models are also
available, for example the AD-ENM of Zheng et al.\ \cite{ZheD03}.

The low-frequency modes are expected to have the largest amplitudes and thus be most significant for large conformational changes. However, the six lowest-frequency modes (modes $1$ to $6$) are trivial combinations of rigid-body translations and rotations of the entire protein. For illustration, here we consider the \emph{five} lowest-frequency non-trivial modes, that is modes $7$ to $11$ for each protein. We will denote these modes as $m_7$, $m_8$, \ldots, $m_{11}$.

The mode eigenvectors are predicted on the basis of a single protein
conformation. The amplitude to which a mode can be projected may be limited by bonding and/or
steric constraints that are not evident in the input structure or fully captured by the ENM. A
linear projection of all the residues in the protein along a mode eigenvector introduces
unphysical distortions of the interatomic bonding. Typically, to avoid this and project a mode to finite amplitude requires one or more cycles of a combined method; the mode is projected linearly until distortions become evident and the resulting structure is relaxed using constrained MD/molecular mechanics \cite{CheLGL06,XuTB03,MiyOW03}. We explore an alternative method for projection of modes to large amplitudes, using rigidity analysis and
geometric simulation.

\subsection{{\sc Froda} mobility simulation}
\label{mobility}

Geometric simulation, implemented in the {\sc Froda} module within {\sc First} \cite{WelMHT05,FarST10},
explores the flexible motion available to a protein with a given pattern of rigidity and
flexibility. New conformations are generated by applying a small random perturbation to all atomic positions;
{\sc Froda} then reapplies bonding and steric constraints to produce an acceptable new conformation.
Motion can be biased by including a directed component to the perturbation. The capability to use a
mode eigenvector as a bias was implemented in {\sc First/Froda} by one of us (SAW) and has been
briefly reported previously \cite{JimWFR11}. The combination of {\sc ElNemo} and {\sc First/Froda},
illustrated schematically in Figure \ref{fig:method}, is described in detail in supplementary
material (section \ref{Supp:settings}).

Since the displacement from one conformation to the next is small, we record only every 100th conformation and
continue the run for typically several thousand conformations. The run is considered complete when
no further projection along the mode eigenvector is possible (due to steric clashes or bonding
constraints) which manifests itself in slow generation of new conformations and poor reproducibility
in the results of independent runs. We have performed {\sc Froda} mobility simulation for each
protein at several selected values of $E_{\rm cut}$, see section \ref{rigidityanalysis}.

During conformation generation we track the fitted RMSD between $\alpha$ carbons of the initial and current conformation. This measure is discussed in more detail in supplementary material (section \ref{Supp:RMSD}). To project a mode to an amplitude of several {\AA} in fitted RMSD typically takes a few CPU-minutes on a single
processor. We carry out five parallel simulations for each structure, mode and direction of
motion and monitor the evolution of fitted RMSD during each run, as illustrated in Figures
\ref{Fig:RMSD:SL}-\ref{Fig:RMSD:2VL0}.
 
\subsection{Raw and fitted RMSD}
\label{Supp:RMSD}

\revision{
The RMSDs reported in \revision{Table \ref{RMSD_table}} are $\alpha$ carbon RMSDs from the input
structure to a generated conformation, obtained after least-squares fitting using the {\sc PyMOL} {\tt
intra\_fit} command. These values differ somewhat from the raw RMSD values reported by {\sc Froda}
in its output files, which are calculated without any fitting being carried out. In particular, the
fitted RMSD saturates once further motion along the mode direction is no longer possible, due to
steric clashes or limits imposed by covalent or noncovalent bonding constraints. The raw RMSD is
greater than the fitted RMSD and tends not to saturate, but rather to continue to increase slowly,
once the motion is effectively jammed. The reason for this different behaviour is a small difference
in the statistical weighting given to each residue by {\sc ElNemo} and by {\sc First/Froda}. In the
elastic network modelling, every residue is given equal statistical weight. The non-trivial mode
eigenvectors thus generated have no significant component of rigid-body motion for the whole
structure. In {\sc Froda}, however, the bias is applied to an all-atom representation of the
structure; and thus the bias applied to a residue with many atoms affects the whole-body motion of
the structure more than the bias applied to a residue with few atoms. The motion in {\sc Froda}
therefore acquires a small component of rigid-body translation and rotation, which increases the raw
RMSD. Least-squares fitting to the input structure removes the rigid-body components, so the fitted
RMSD detects actual conformational change.}

\revision{The effects of fitting are
illustrated in Figure \ref{RawvsfittedRMSD}a for mode $m_7$ of structure 1BPI. The raw RMSD values increase
almost linearly during the generation of $10000$ conformers, whereas the fitted RMSD values
saturate for conformers from $\approx$ $5000$ up to $10000$. Conformers $5000$ and $10000$ differ by $\approx3 \AA$ in raw RMSD but by only  $\approx0.8 \AA$ in fitted RMSD.
Superpositions of conformers $0$, $5000$ and $10000$ with and without fitting, shown in Figures
\ref{RawvsfittedRMSD}b,c, show that conformers $5000$ and $10000$ are indeed very similar to each other. 
Hence, fitting structures before calculating RMSD values allows us to identify real conformation change and remove the component of rigid-body motion introduced by {\sc Froda}. }

\section{Results}

\subsection{Conformer generation with mode bias}

The output of the mobility simulations for BPTI (1BP1) is summarized in Figures
\ref{Fig:RMSD:SL}a--c. The evolution of RMSD for each of  $m_7$, \ldots,$m_{11}$, during
runs at two
selected values of $E_{\rm cut}$, is represented in Figures \ref{Fig:RMSD:SL}b and
\ref{Fig:RMSD:SL}c
respectively. In all cases, we observe an initial phase in which the RMSD increases almost linearly,
as the protein explores the mode direction without encountering significant steric or bonding
constraints on the motion. During this phase, generation of new conformations in {\sc Froda} is very
rapid and the RMSDs from different runs are very similar to each other.
The RMSD then displays an inflection, ceasing to rise linearly, and approaching an asymptote; this
indicates that steric clashes and bonding constraints (such as hydrophobic tethers) are preventing
further exploration along the mode direction. The asymptote is thus an amplitude limit on the mode.
In this phase the generation of new conformations in {\sc Froda} becomes slower as the fitting
algorithm has increasing difficulty finding a valid conformation, and the RMSDs achieved by different runs differ.
In the regime of slow conformation generation, the mode bias is forcing the structure
into a regime of steric clashes and/or of bonding constraint limits, for example when residues
connected by a hydrophobic tether are being pushed apart. This regime cannot correspond to the
low-frequency flexible motion which we wish to explore. Our presentation of RMSD data for larger
proteins is therefore truncated once this ``jamming`` starts. For most of our proteins,
$2500$ conformations is sufficient to cover the regime of rapid conformation
generation. For our largest protein, pLGIC (2VL0), we find that $1000$ conformations was
sufficient.

\subsection{Small loop motion}

In Figure \ref{Fig:RMSD:SL}a we show an ensemble of structures for BPTI generated by exploring the
lowest-frequency nontrivial mode, $m_7$, and in Figures \ref{Fig:RMSD:SL}b,c we show the RMSDs
achieved for the five lowest-frequency nontrivial modes.
The amplitudes of flexible motion for BPTI at the cutoffs $-0.2$ kcal/mol and $-2.2$
kcal/mol reach an asymptote at RMSD values around $2$ to $4${\AA}.
Considerable asymmetry, a factor of $2$
difference in the achievable RMSD, is observable in some modes between the two possible directions
of motion. Let us emphasize here that in general the amplitudes of flexible motion are not
available from either the elastic network model or the rigidity analysis without simulation of
flexible motion.

BPTI (1BPI) is a small protein with some relatively flexible loop regions. For contrast,
we now examine a compact globular protein, cytochrome-c (1HRC). RMSD results are shown
in Figure \ref{Fig:RMSD:SL}e,f truncated once jamming had begun. Although this protein is twice as
large as BPTI in terms of residues, its capacity for flexible motion is visibly more limited, with
amplitudes below $2${\AA} in all modes. This result is important in validating our
method of projecting modes to large amplitude; if geometric simulation were capable of reaching
unphysically large amplitudes, the method would lose its value.

\subsection{Large loop motion}

RMSDs for low-frequency modes of internal kinesin motor domain protein (1RY6) are shown in
Figure \ref{Fig:RMSD:LL}b, c for two different energy cutoffs. The amplitudes achievable for these
modes differ little between the different energy cutoffs; motion occurs principally in a flexible
loop region around residues $37$--$46$. An exploration of $m_7$ is presented in Figure \ref{Fig:RMSD:LL}a
for an energy cutoff of $-1.1$ kcal/mol, clearly showing the loop motion.
We find that the combination of the mode bias and the bonding constraints naturally causes the large flexible loop to follow a curved trajectory.

As shown in Figures \ref{Fig:RMSD:LL}e, f, $\alpha$1-antitrypsin (1QLP) displays several low-frequency modes which easily explore amplitudes of up to $2$--$2.5${\AA} RMSD depending on the rigidity cutoff. The motion shown in Figure \ref{Fig:RMSD:LL}d again involves the easy motion of large flexible loops with respect to the relatively rigid $\beta$-sheet core of the protein.

\subsection{Domain motion}

Protein disulphide isomerase (2B5E) is an interesting case for protein mobility.
The protein consists of four domains (a--b--b'--a') connected by flexible linkers. Biological
evidence indicates that conformational flexibility is vital to the function of the enzyme
\cite{FreKR02}. Rigidity analysis immediately brings out the flexibility of the molecule (see
Figure \ref{Fig:RCD_plot}e). Even at very high $E_{\rm cut}$ values ($E_{\rm cut}= -0.015$
kcal/mol), the rigidity analysis reveals the domain organisation of the protein with each domain
corresponding to a distinct rigid cluster flanked by flexible linkers. The RMSD achievable by
low-frequency flexible motion therefore does not depend significantly on the energy cutoff; motion
is slightly limited at the weakest cutoff (Figure \ref{Fig:RMSD:2B5E}b), but at other cutoffs the
achievable amplitudes are essentially the same (Figure \ref{Fig:RMSD:2B5E}c and d). Close
examination of the conformation generation in {\sc First}/{\sc Froda} indicates that the amplitudes
are limited eventually as further motion along the mode would over-extend covalent and
hydrophobic-tether constraints.

The inter-domain nature of the flexible motion is detailed in Figure \ref{Fig:RMSD:2B5E}a which
shows
structures at the amplitude limits of $m_7$ in the positive and negative directions. The
structures
are aligned on the b--b' domains, bringing out the motion of the a domain and particularly of the
a' domain. The CPU time required to project this protein with more than $500$ residues along $m_7$ to an amplitude of more than $20${\AA} is less than $15$ minutes.

The largest protein we have investigated is pLGIC (2VL0). Its pentameric structure includes a
transmembrane domain composed of $\alpha$-helices and an extracellular domain consisting largely of
$\beta$-sheets. The major rigidity transition in the protein identified by {\sc First} occurs
between a cutoff of $-0.4$ kcal/mol, when almost the entire
structure forms a single rigid cluster, and $-0.5$ kcal/mol, when the five backbone sections linking
the two major domains have become flexible and the many $\alpha$-helices in the transmembrane are
mutually flexible. At the lower energy cutoff it is possible for the two domains to move relative to
each other, and the transmembrane helices are also capable of relative motion. The increased RMSD
possible at the lower $E_{\rm cut}$ is visible in Figure \ref{Fig:RMSD:2VL0}.

The flexible motion at the higher cutoff, with the protein largely rigid, involves only the motion
of a few flexible loops. The motion at the lower cutoff is far more biologically interesting. The
lowest-frequency non-trivial mode, $m_7$, involves a counter-rotation of the transmembrane
and
extracellular domains, including a change in the relative tilt of transmembrane helices lining the ion
channel. This flexible motion is shown
in Figure \ref{Fig:RMSD:2VL0}a, b. The CPU time to project this large protein with more than $1600$
residues along $m_7$ to its amplitude limit is less than twenty minutes.

\section{Discussion}

\subsection{Extensive RMSD as a \revision{characterisation} of total flexible motion}

\begin{figure}[p]
\centering
BPTI \hfil  \mbox{   } \hspace*{16ex} Cytochrome-c\\
(a)\includegraphics[width=0.45\textwidth]{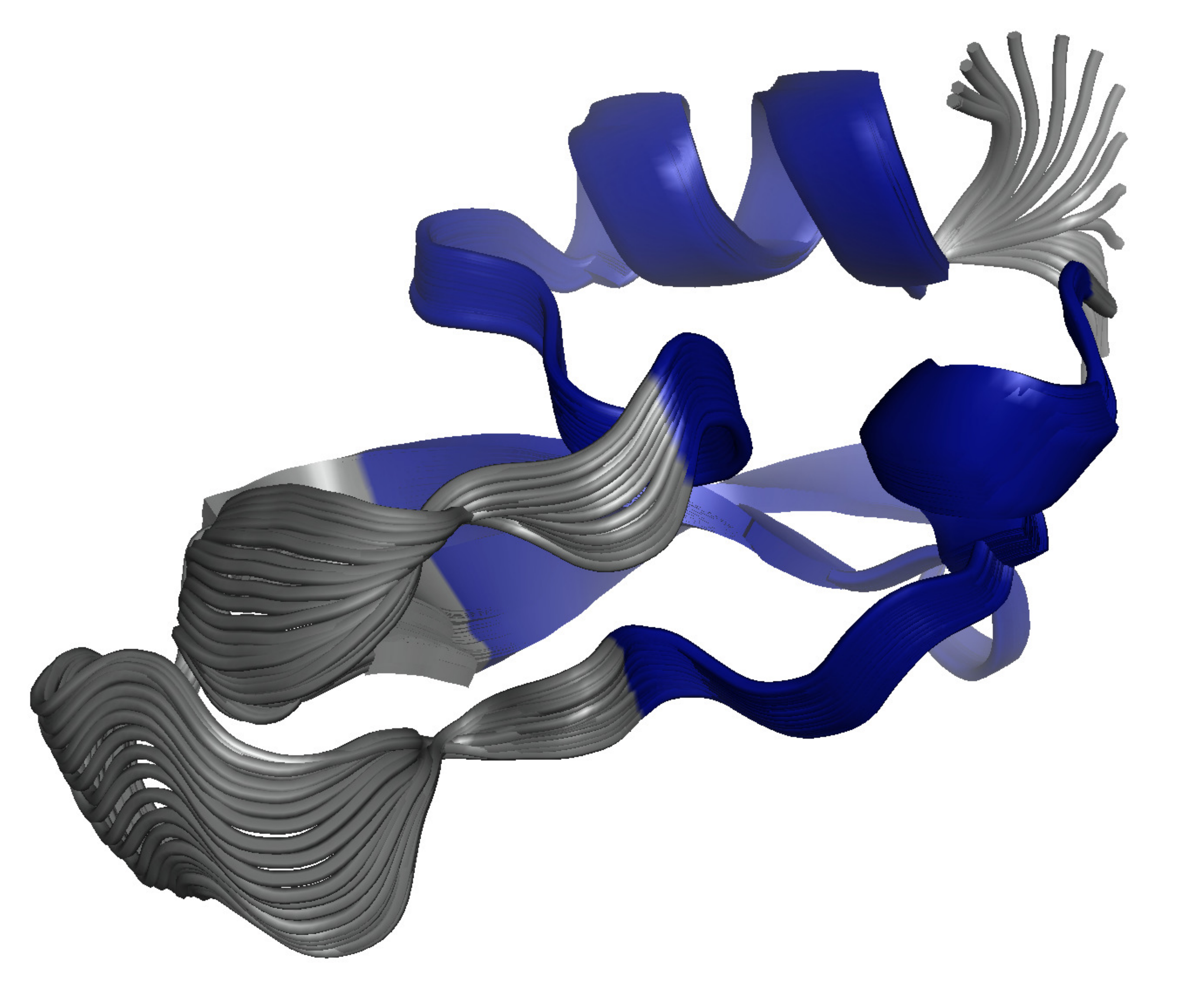}
(d)\includegraphics[width=0.45\textwidth]{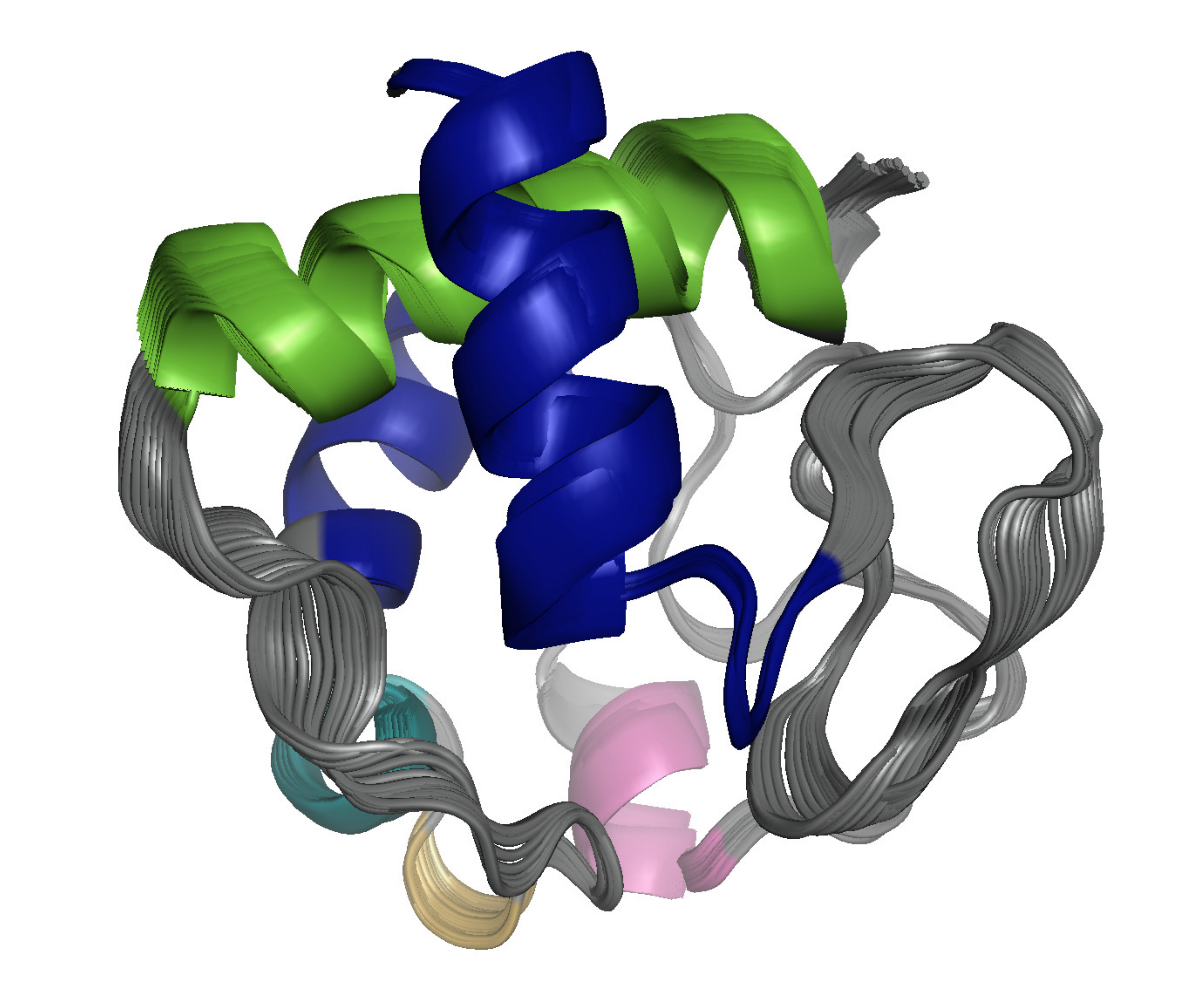}
\\[2ex]
(b)\includegraphics[width=0.45\textwidth]{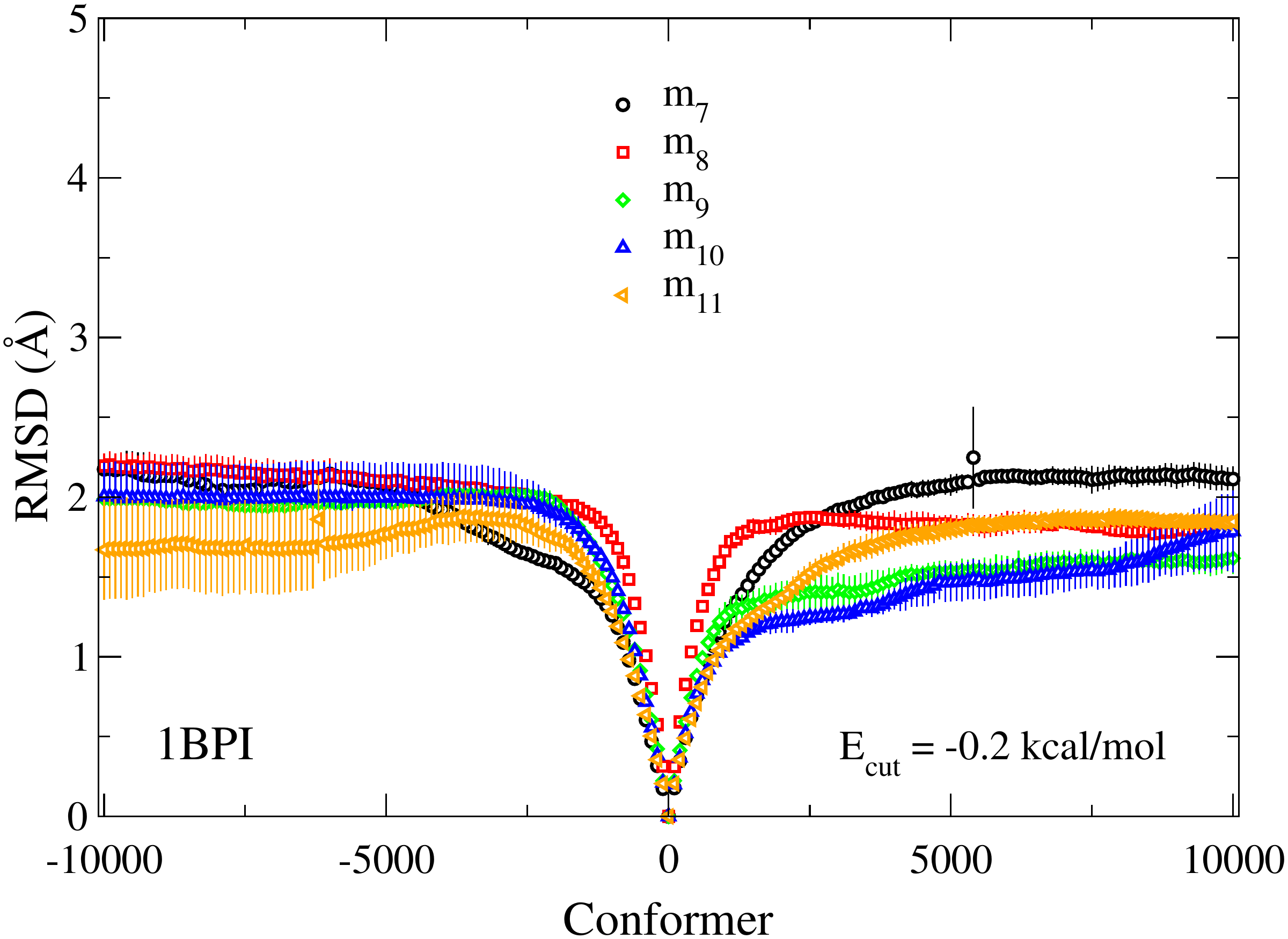}
(e)\includegraphics[width=0.45\textwidth]{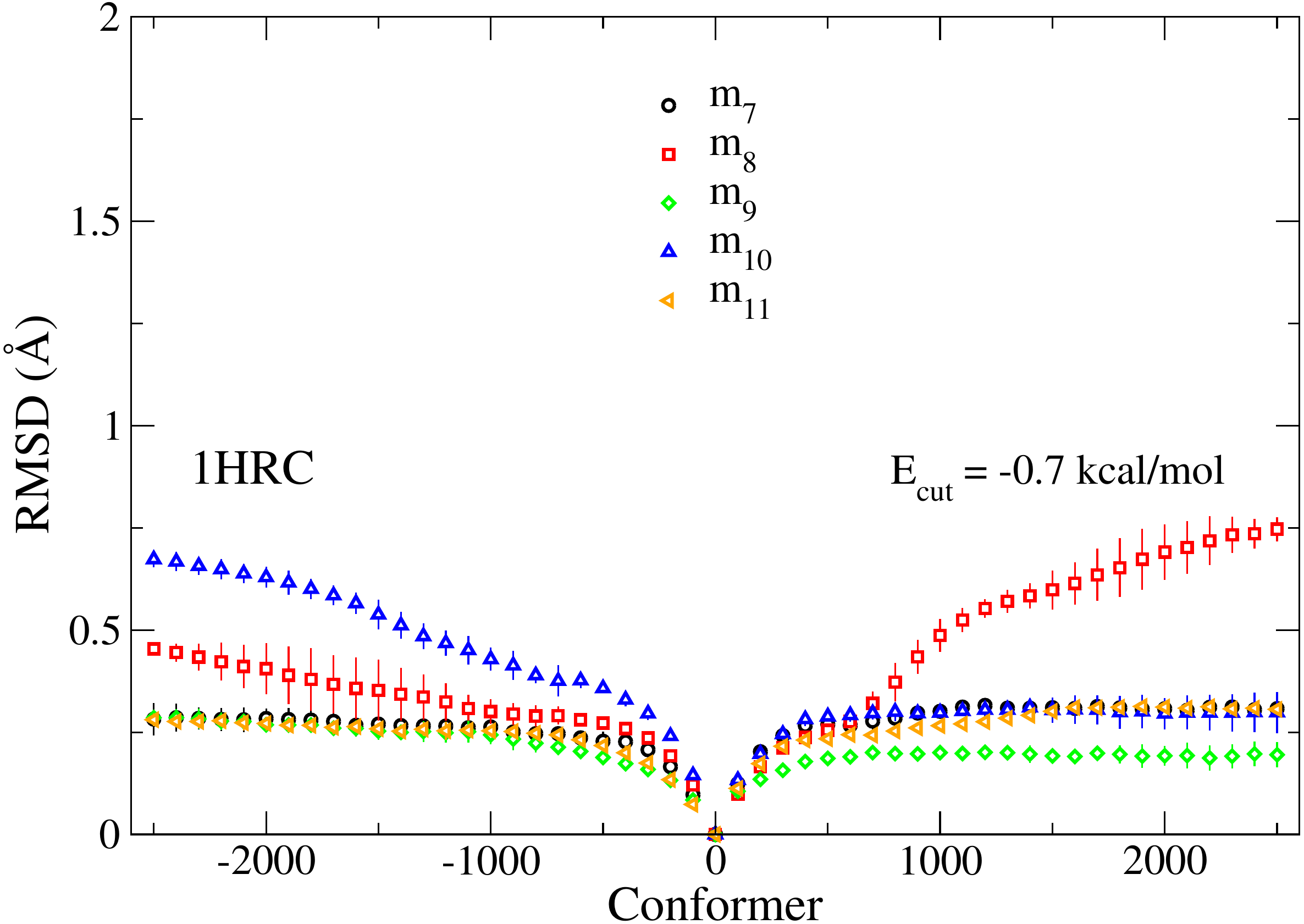}

(c)\includegraphics[width=0.45\textwidth]{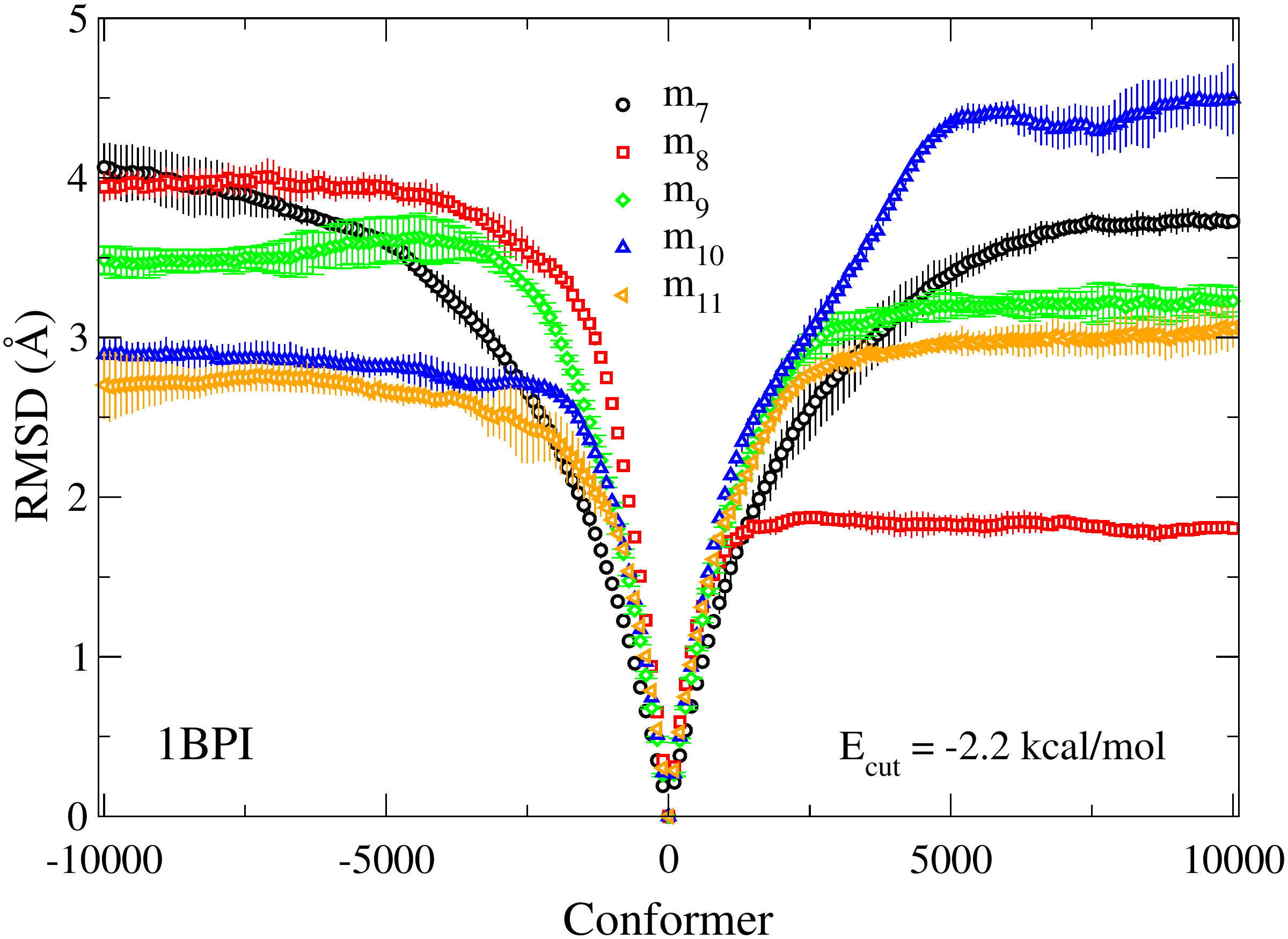}
(f)\includegraphics[width=0.45\textwidth]{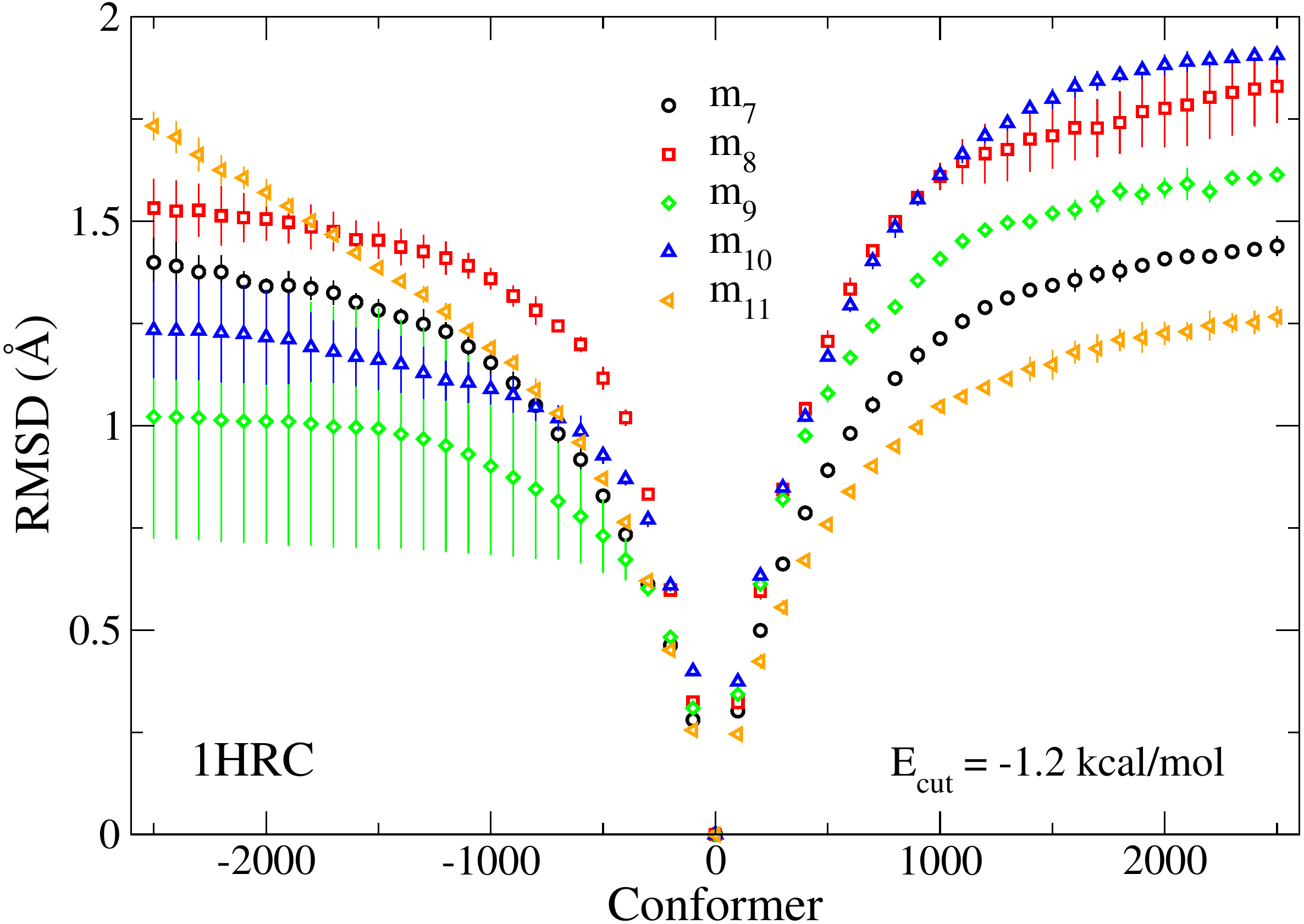}

\caption{\label{Fig:RMSD:SL} 
Superimposed structural variations and fitted RMSD for small loop motion as found in BPTI and cytochrome-c.
Panels (a) and (d) indicate the range of projected tertiary structure for motion along mode $m_7$ at $E_{\rm cut}=-2.2$ kcal/mol for BPTI and at $E_{\rm cut}=-1.2$ kcal/mol for cytochrome-c, respectively.
Panels (b,c) and (e,f) show the fitted RMSD as a function of {\sc Froda} conformations for BPTI (1BPI) and cytochrome-c (1HRC), respectively, for the non-trivial modes $m_7$, \ldots, $m_{11}$ at two values of $E_{\rm cut}$ as shown. Positive conformation values indicate motion along the direction of the corresponding {\sc ElNemo} mode, whereas negative conformation values indicate motion in the opposite direction.
Points and error bars indicate mean and standard deviation obtained from five runs of the conformation generation for each mode.}

\end{figure}

\begin{figure}[p]
\centering
Kinesin \hfil \mbox{   } \hspace*{16ex} Antitrypsin\\
(a)\includegraphics[width=0.45\textwidth]{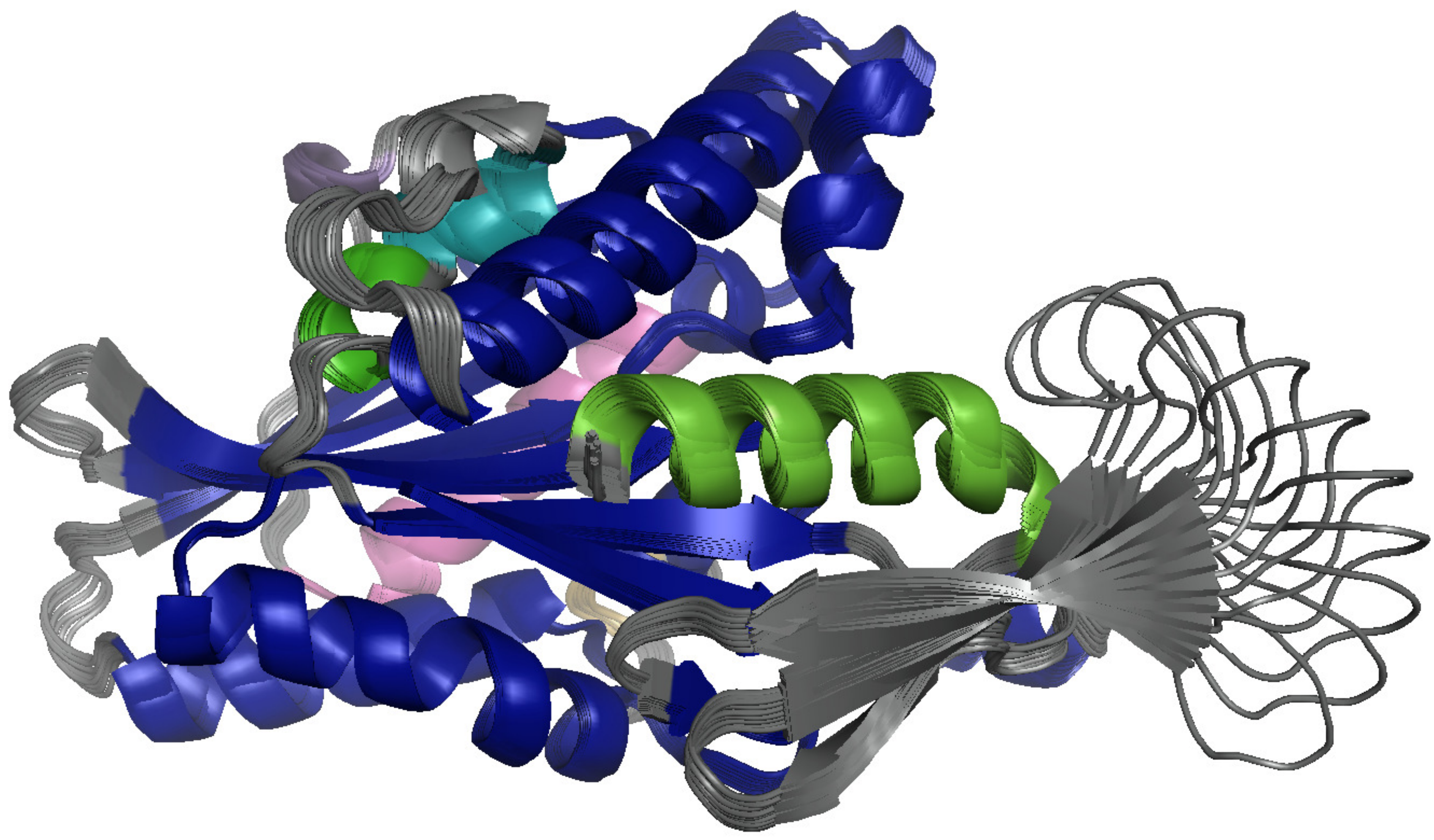}
(d) \includegraphics[width=0.45\textwidth]{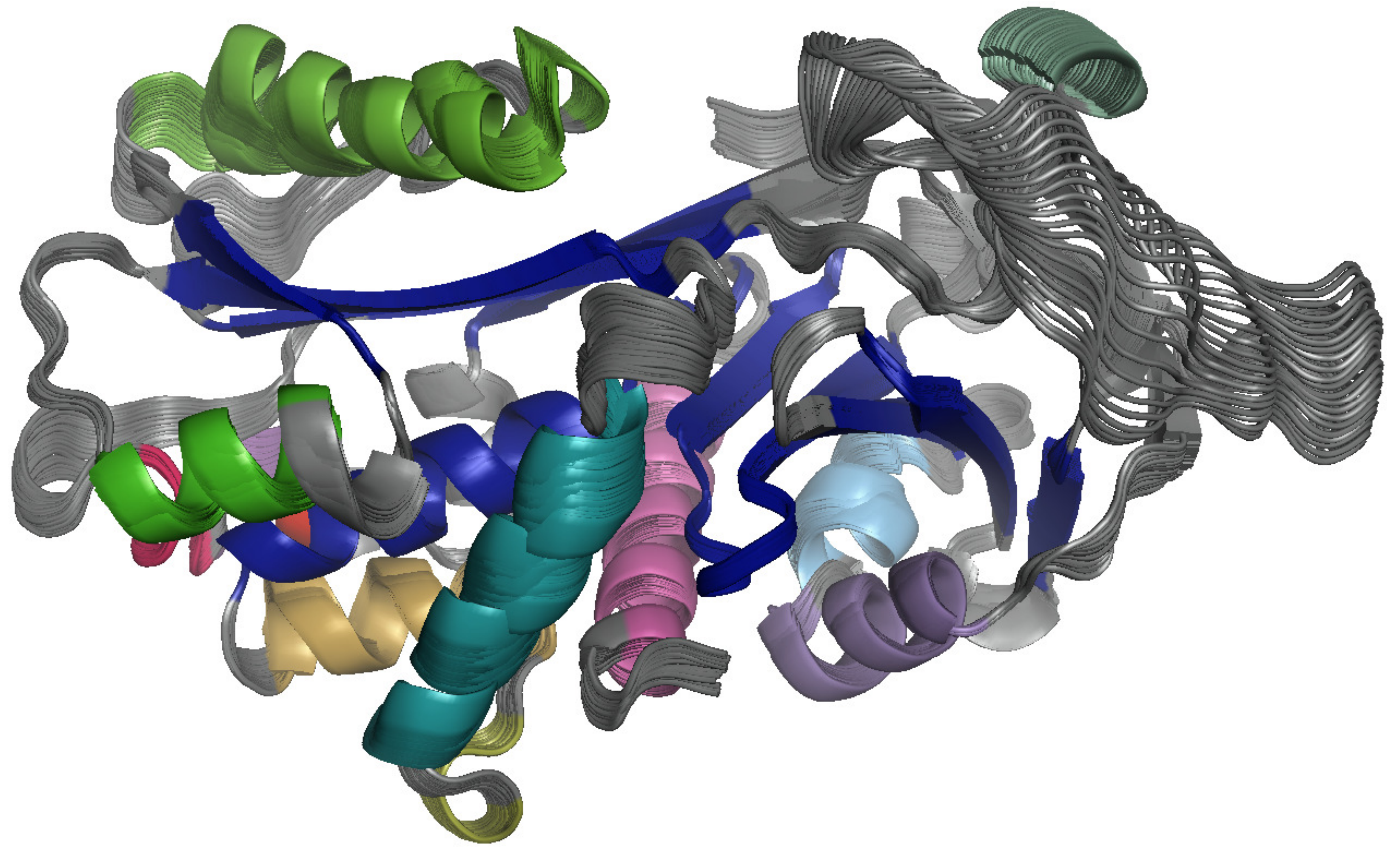}
\\[2ex]
(b)\includegraphics[width=0.45\textwidth]{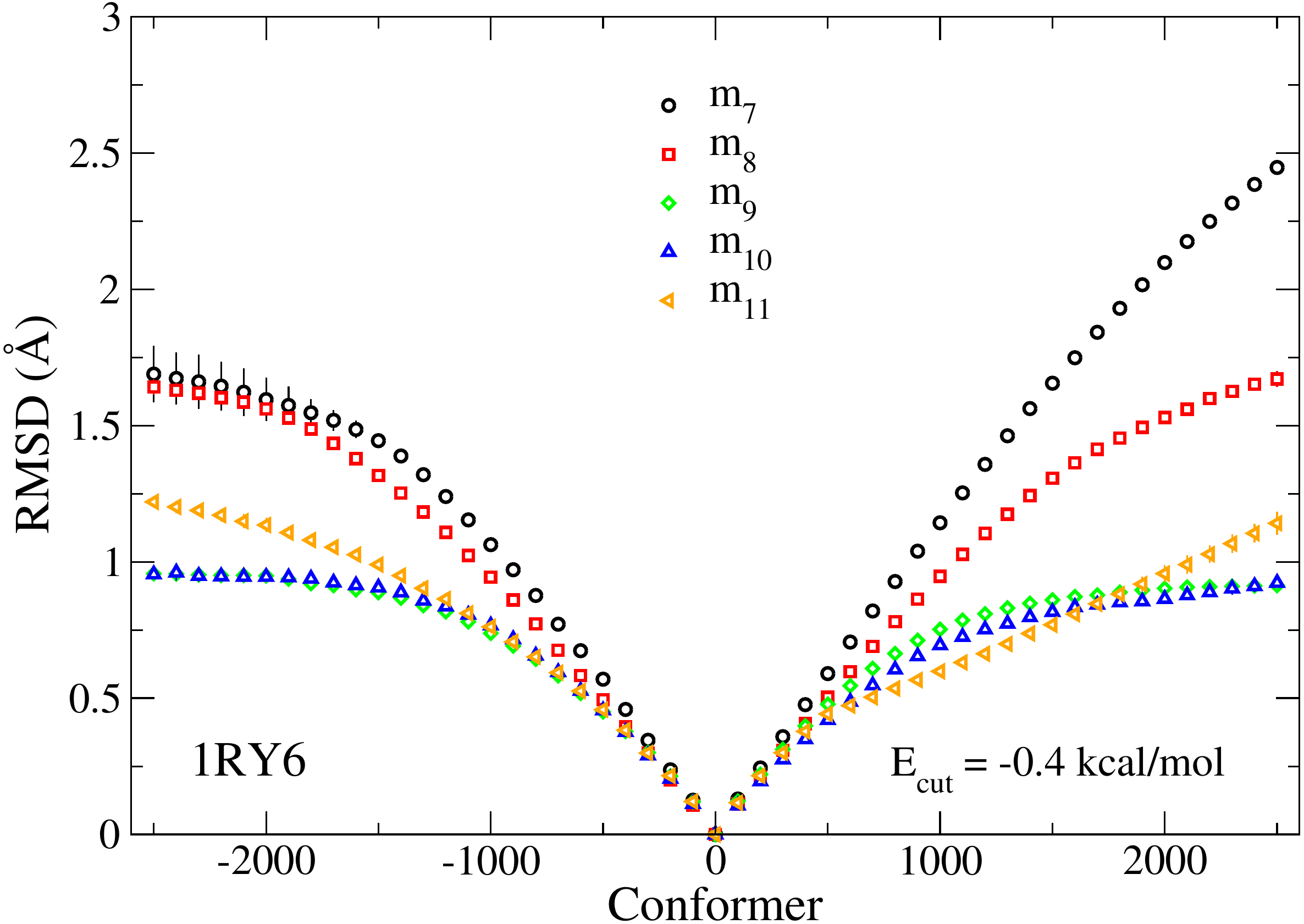}
(e)\includegraphics[width=0.45\textwidth]{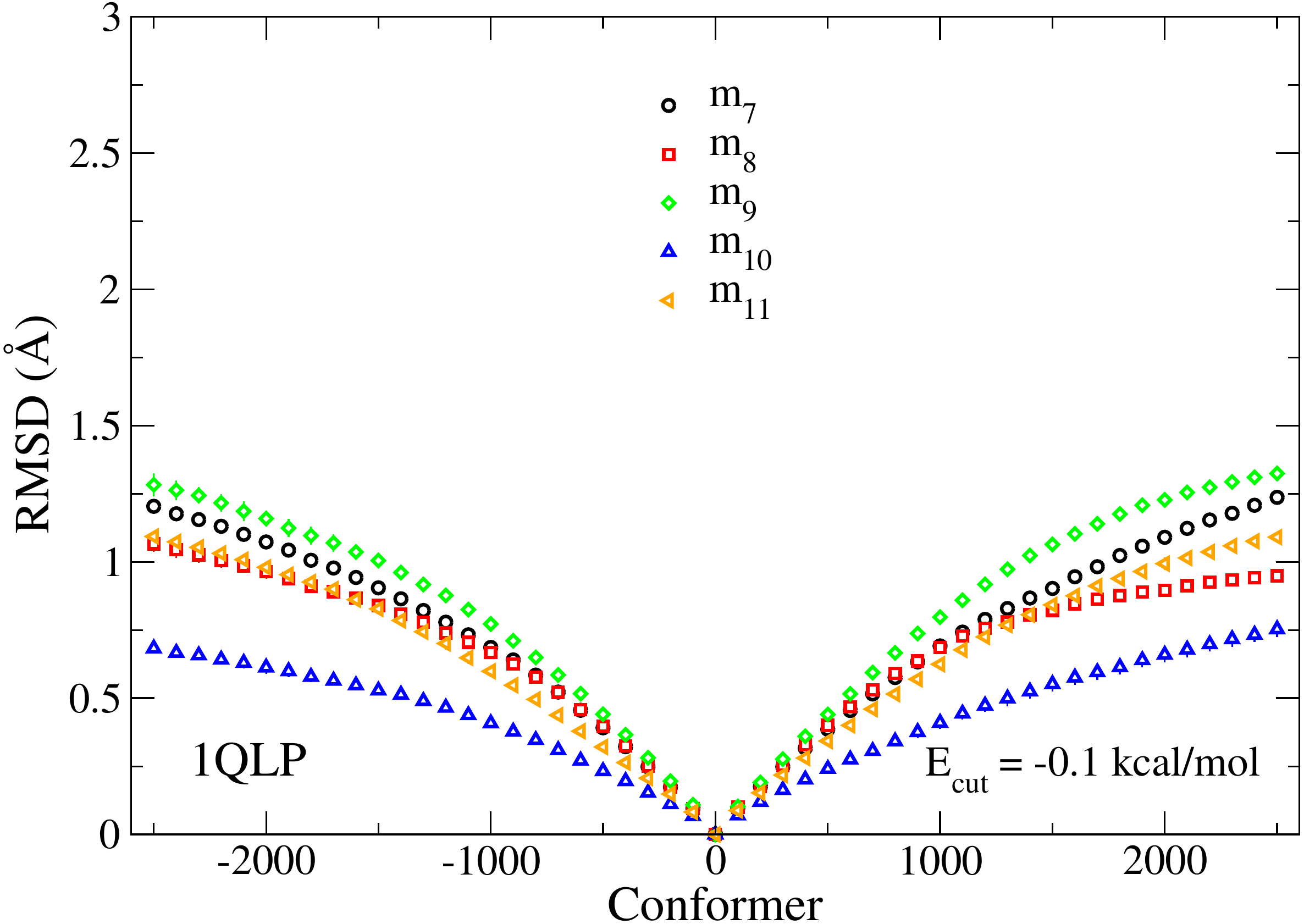}
(c)\includegraphics[width=0.45\textwidth]{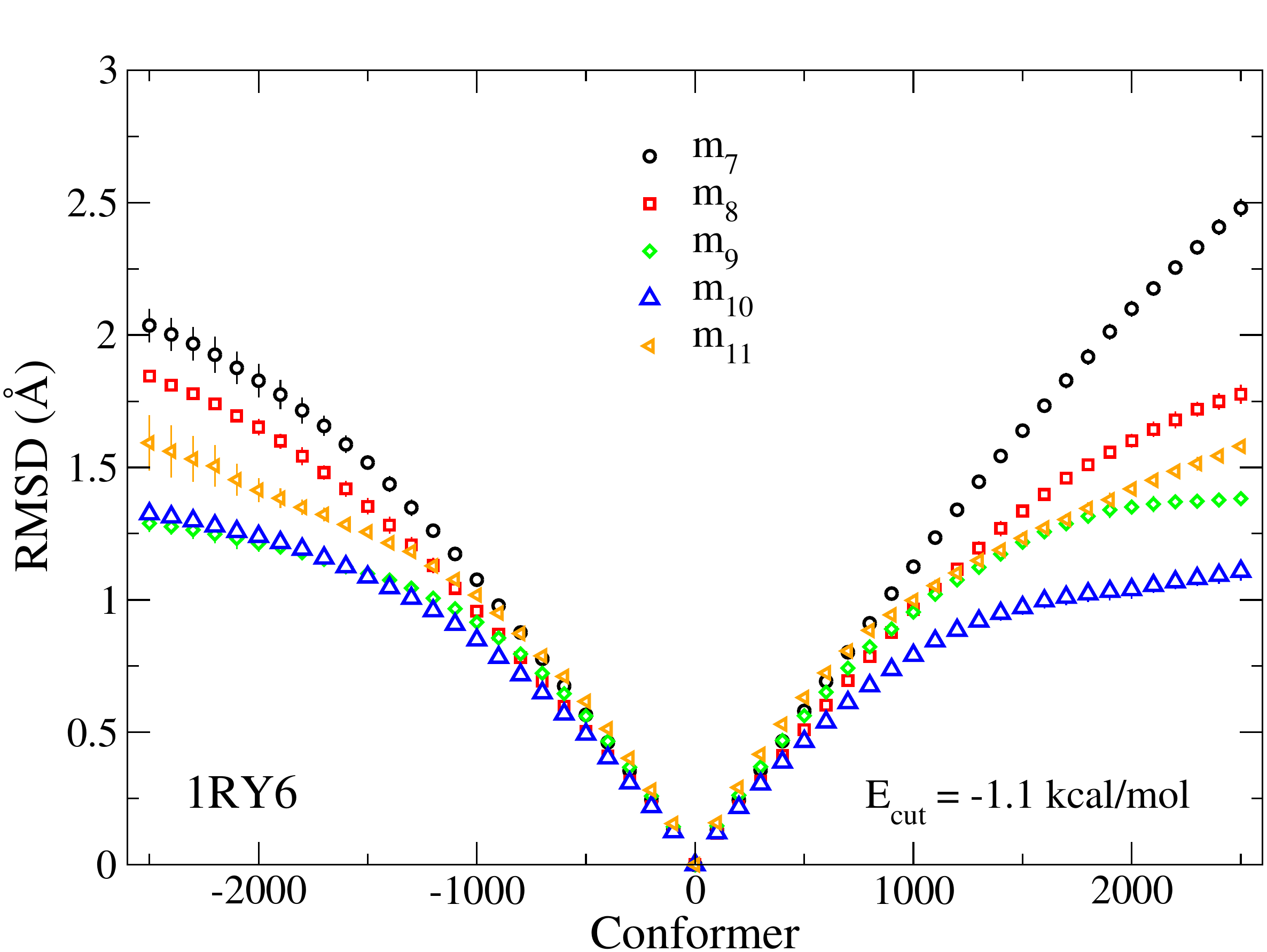}
(f)\includegraphics[width=0.45\textwidth]{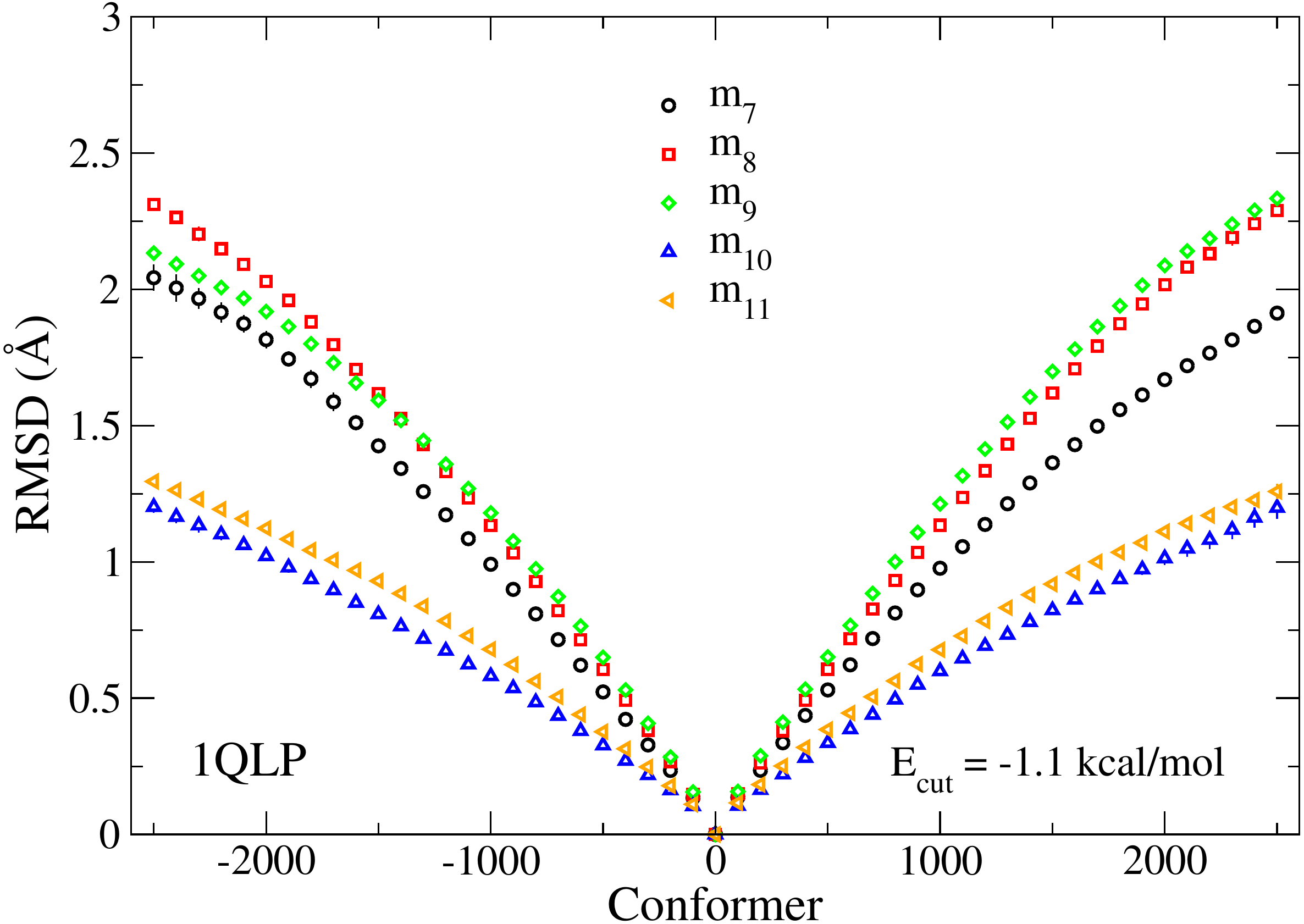}

\caption{\label{Fig:RMSD:LL} 
Superimposed structural variation and fitted RMSD for large loop motion as in (a) kinesin (1RY6) and (d) antitrypsin (1QLP) for $E_{\rm cut}=-1.1$ kcal/mol. Panels (b,c) and (e,f) represent --- as in Figure \ref{Fig:RMSD:SL} --- the fitted RMSD at two values of $E_{\rm cut}$ for kinesin and antitrypsin, respectively. Points and error bars have been determined as in Figure \ref{Fig:RMSD:SL}.
}
\end{figure}

\begin{figure}[p]
\hspace*{-10ex}Yeast PDI\\
\centering
(a)\includegraphics[width=0.47\textwidth]{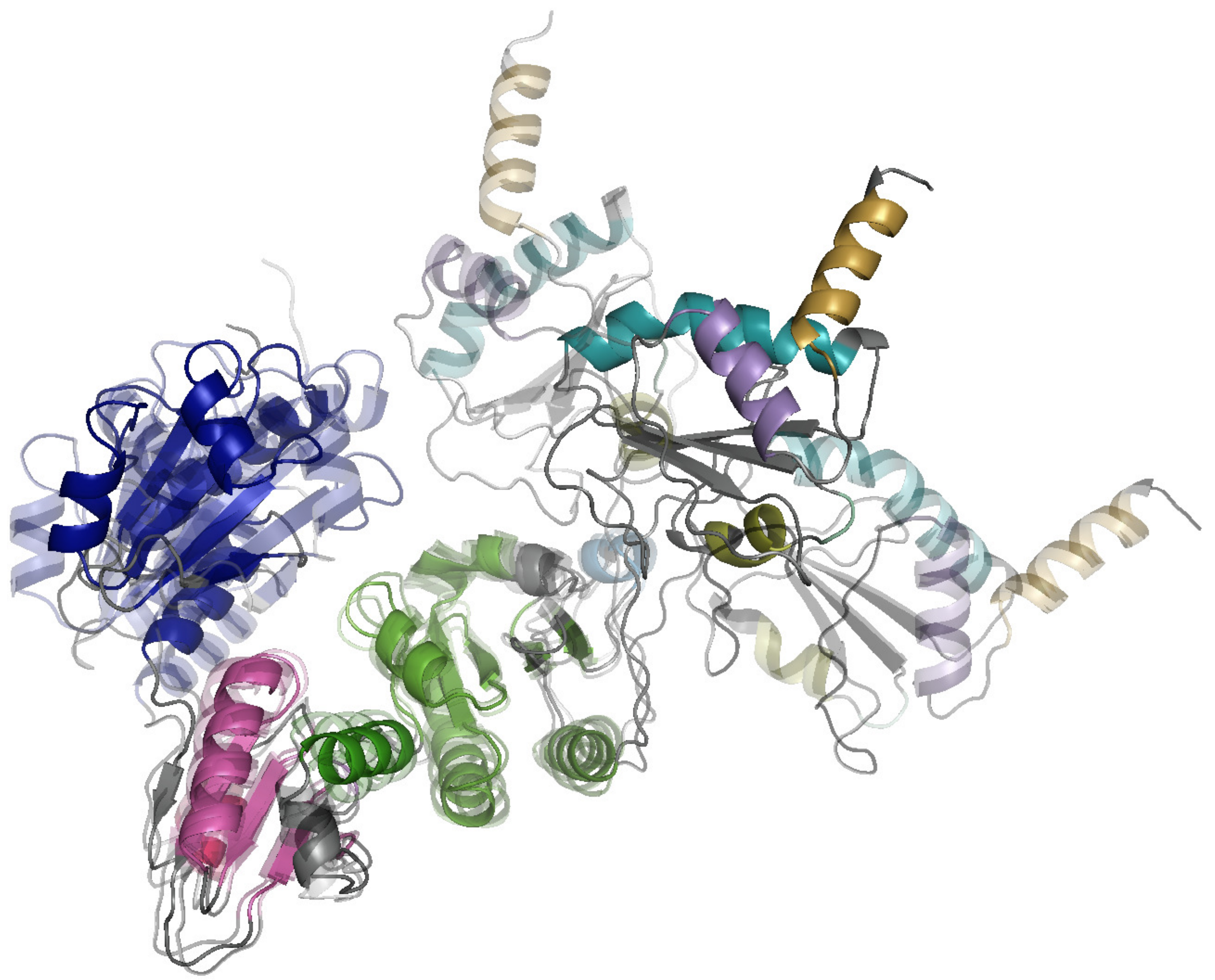}
(b)\includegraphics[width=0.45\textwidth]{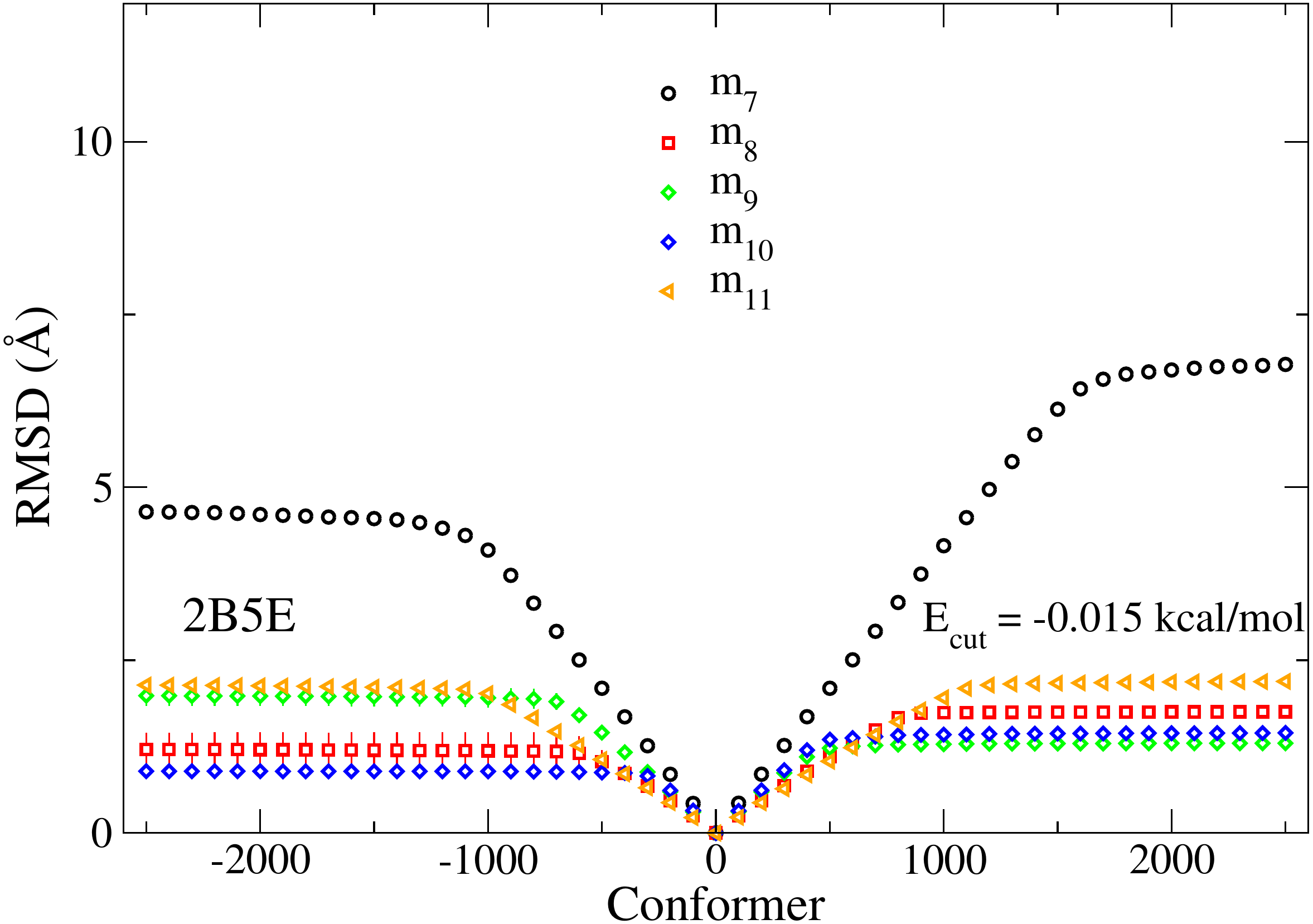}
(c)\includegraphics[width=0.45\textwidth]{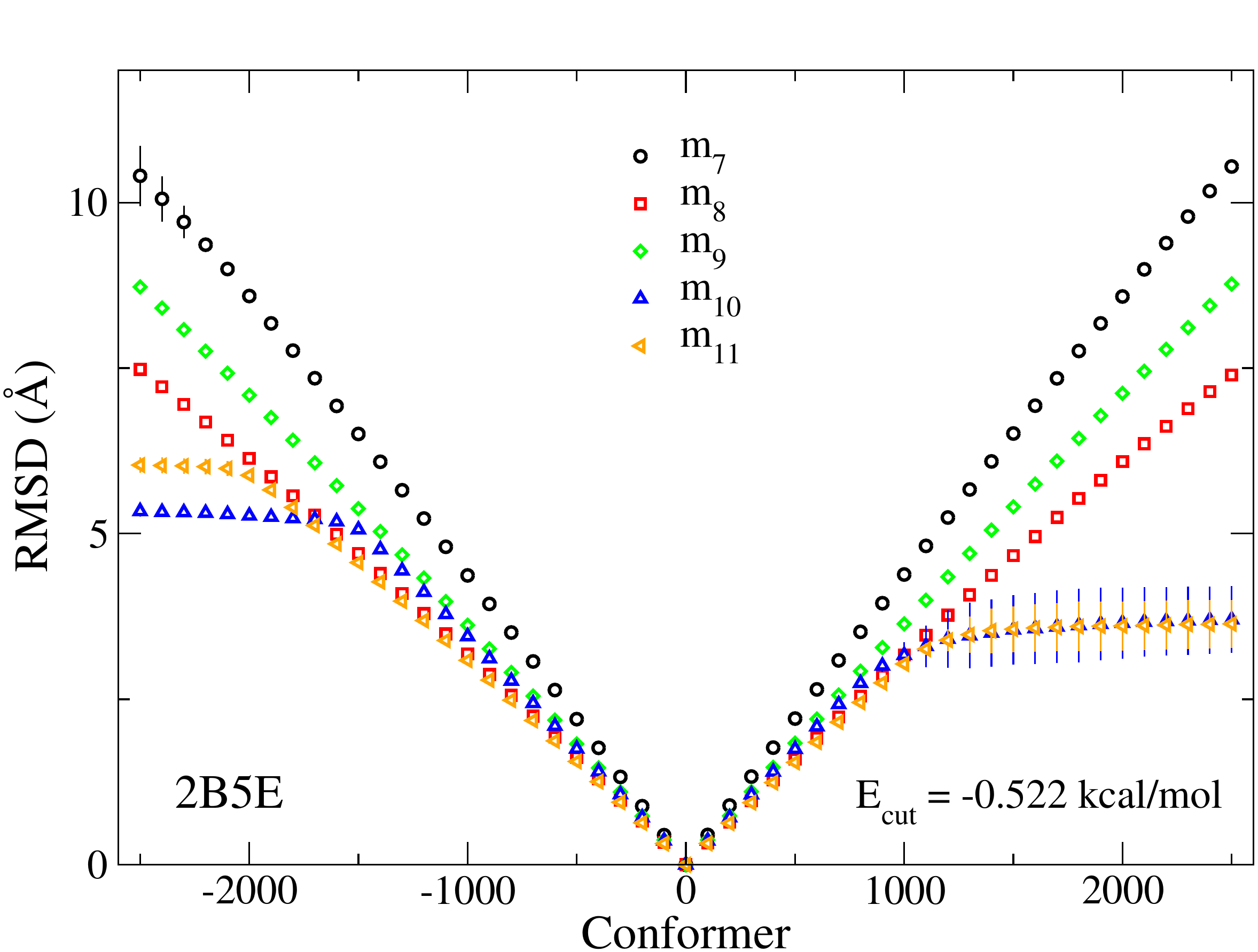}
(d)\includegraphics[width=0.45\textwidth]{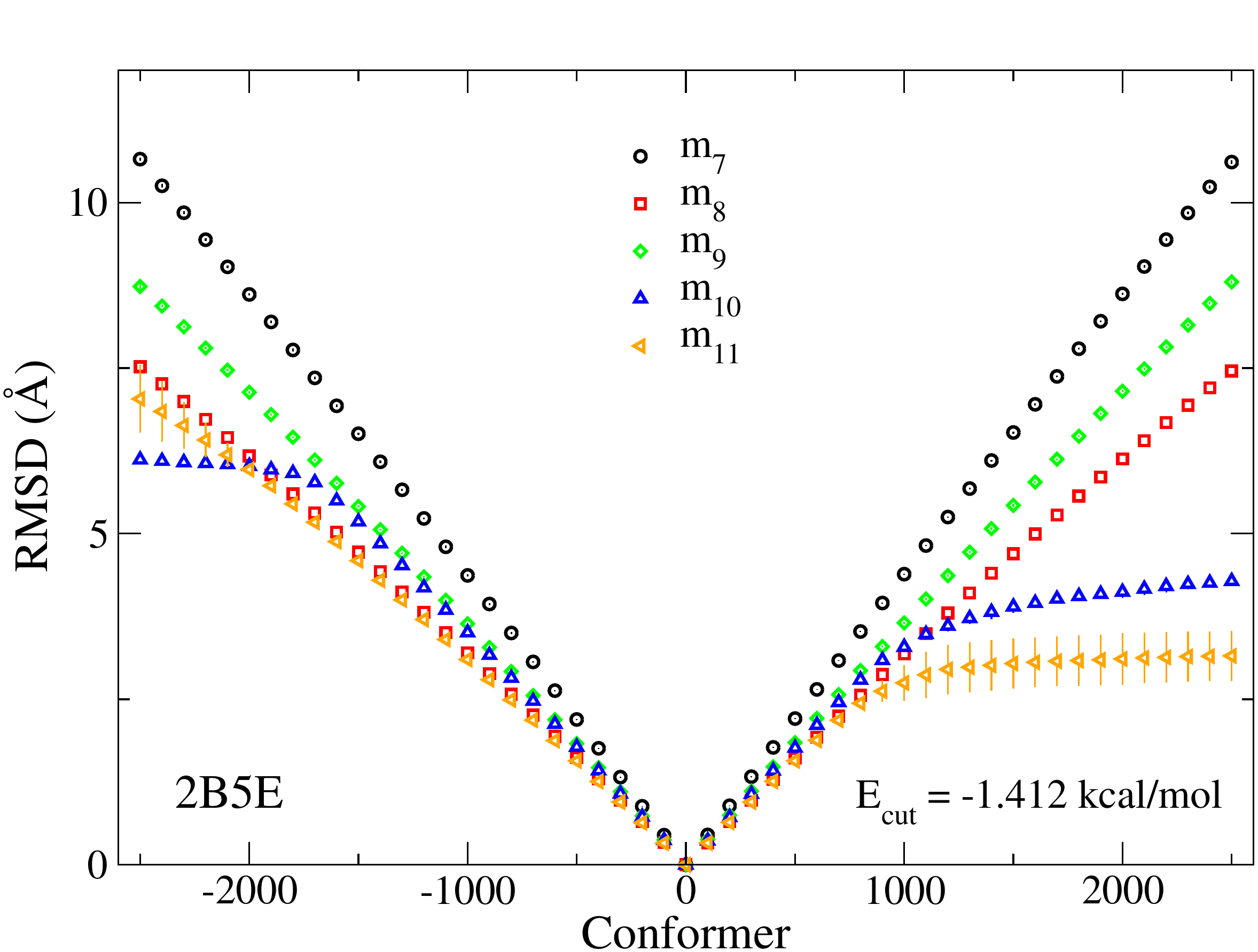}
\caption{\label{Fig:RMSD:2B5E} 
Superimposed structural variation of large domain motion and fitted RMSD for yeast PDI (2B5E). 
(a) We show the initial tertiary structure as opaque and the projected structures as partially transparent. All structures are aligned on the central two
domains b--b' to highlight the motion of the a and a' domains. Motion represents large conformational change along $m_7$.  
Panels (b), (c) and (d) show the fitted RMSDs relative to the initial conformation for three values of $E_{\rm cut}$. Points and error bars as in Figure \ref{Fig:RMSD:SL}.}
\end{figure}

\begin{figure}[p]
\centering
pLGIC (topview) \hfil \hspace*{3em} pLGIC (sideview)\\
(a) \includegraphics[width=0.45\textwidth]{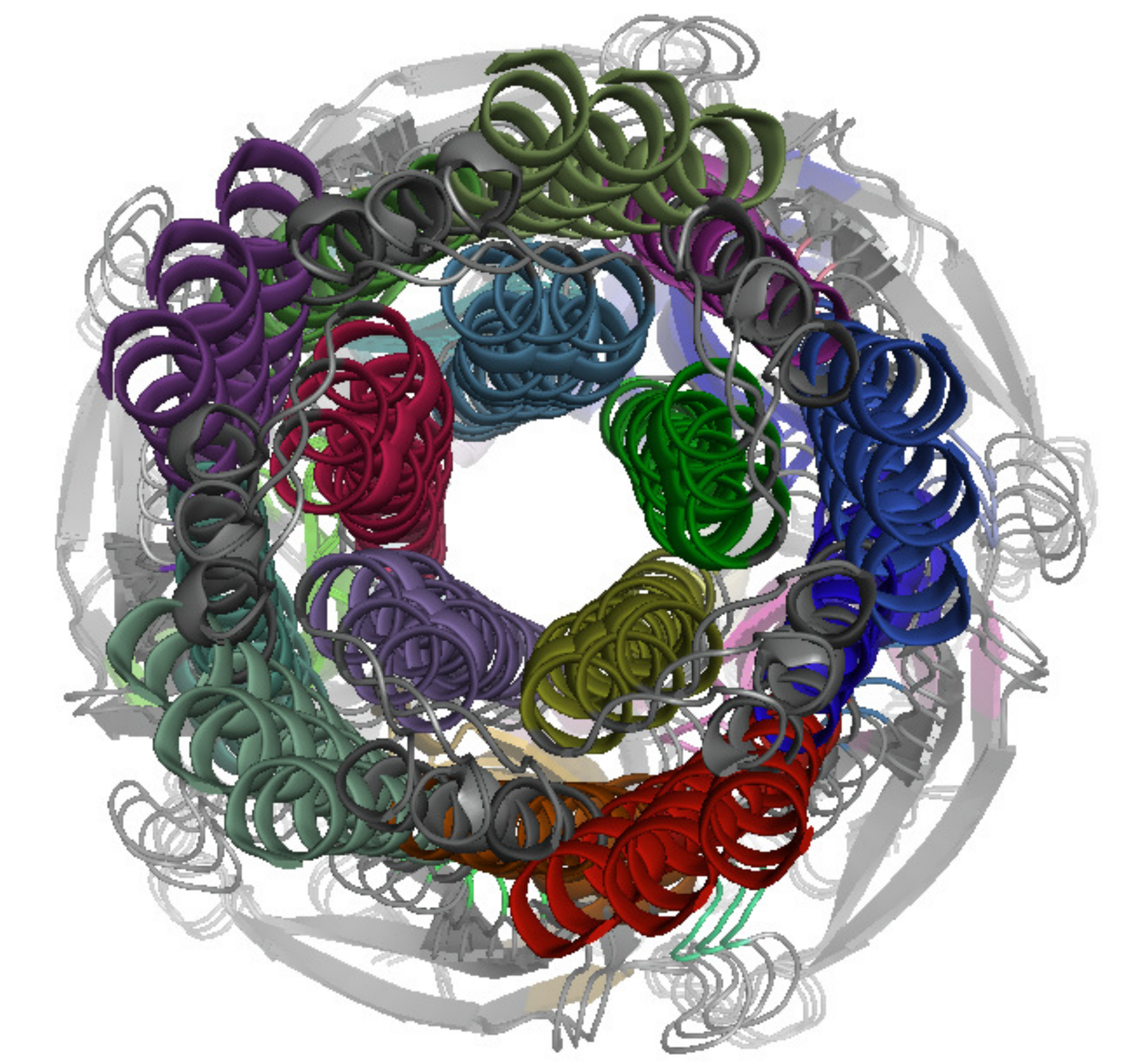}
(b) \includegraphics[width=0.4\textwidth]{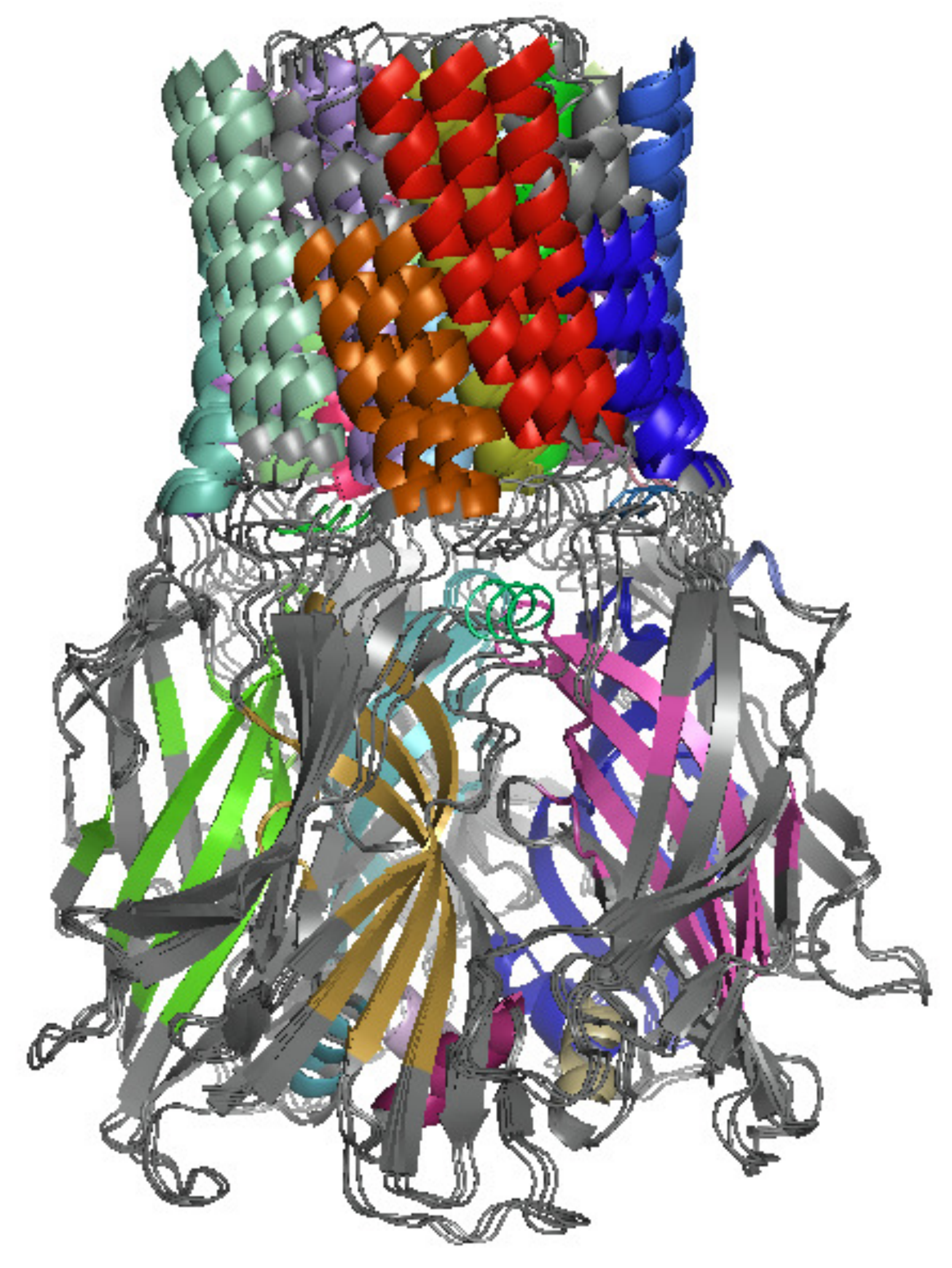}\\

(c)\includegraphics[width=0.45\textwidth]{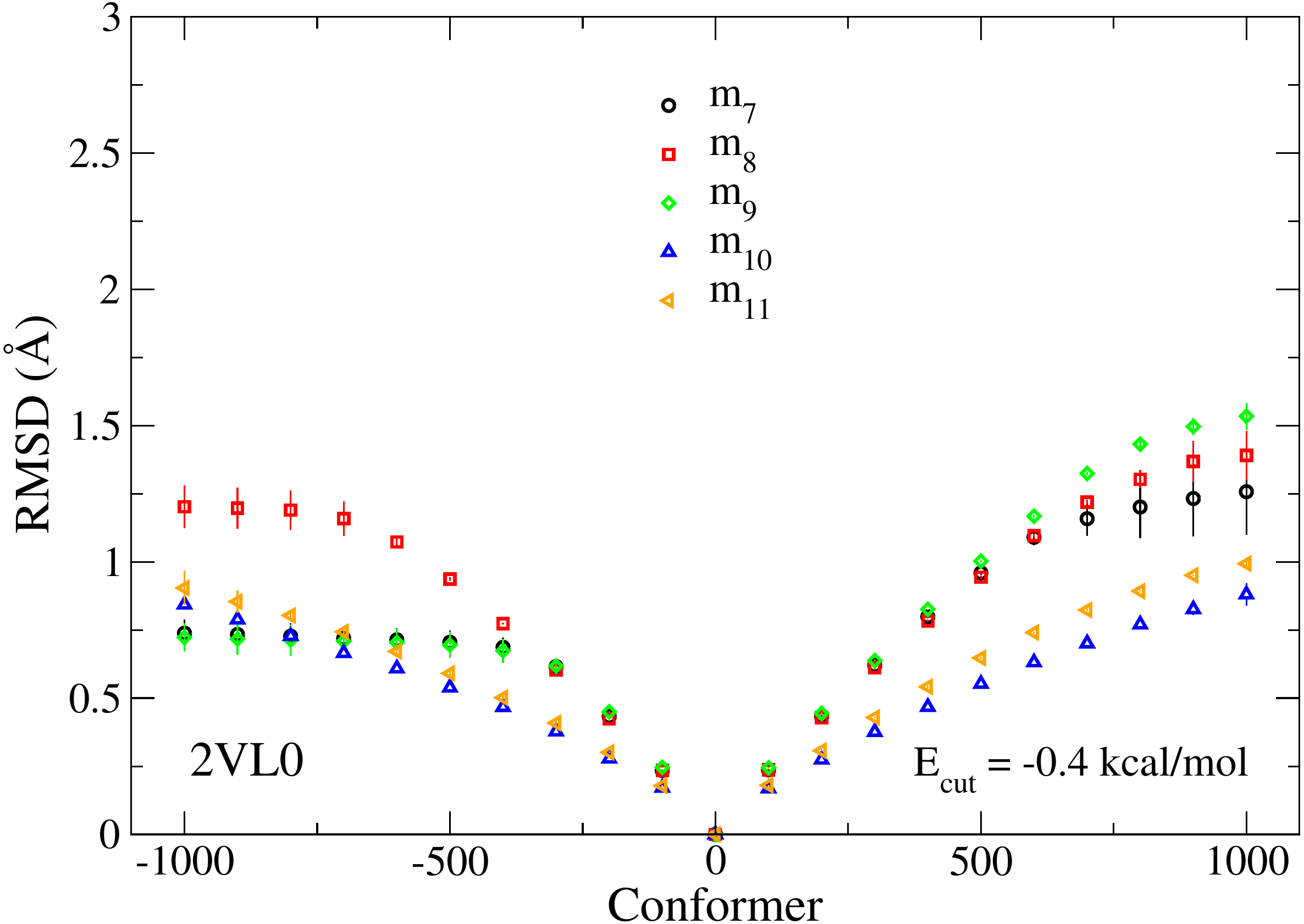}
(d)\includegraphics[width=0.45\textwidth]{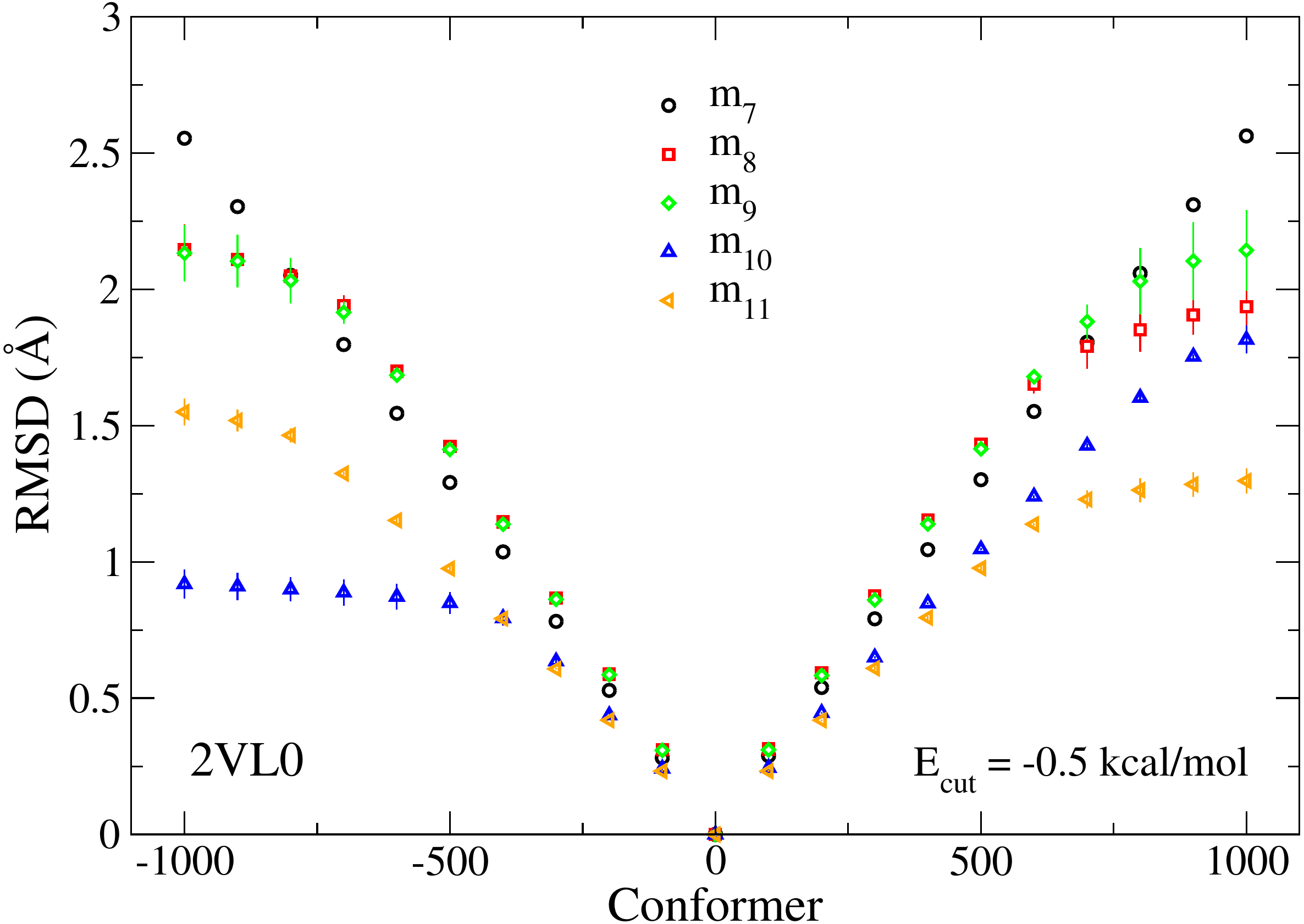}
\caption{\label{Fig:RMSD:2VL0} 
Large scale twist motion in a ligand gated ion channel (2VLO). 
(a) View down the transmembrane channel in its initial state and projected
along $m_7$ in two directions. 
(b) Side view showing tilting of the helices during the motion. 
In
both images the structures have been aligned on the extracellular $\beta$-sheet portion so as to
highlight the relative motion of the domains, and residues from number $283$ upwards in each chain
are not shown to make the major helices visible.
(c and d) Fitted RMSDs relative to initial conformation for low-frequency modes $m_7$, \ldots, $m_{11}$ and two cutoff energies. Points and error bars as in Figure \ref{Fig:RMSD:SL}.}
\end{figure}

The  \revision{evolution of} RMSD values \revision{is} displayed in Figures
\ref{Fig:RMSD:SL}--\ref{Fig:RMSD:2VL0}  \revision{and the maximum values achieved by these
motions, which range from $1.5${\AA} to $10${\AA}, are shown in Table \ref{RMSD_table}}.
However, the character of the flexible motion does not seem well reflected by the RMSD
values. \revision{For example, a} small protein \revision{of 58 residues} without a large
conformational change \revision{such as} BPTI shows RMSD values of up to
$3.5${\AA} in its small loop motion (Figure \ref{Fig:RMSD:SL});  \revision{whereas} the channel
protein pLGIC, a thirty times larger protein by residue count (\revision{1605 residues}), shows a
substantial domain motion, \revision{in which a large proportion of the atoms undergo relative
motion as the transmembrane protein and extra cellular domains counter-rotate (see Figure
\ref{Fig:RMSD:2VL0}a, b).} Yet \revision{ the channel protein pLGIC} shows maximum RMSD values of
around $2.5${\AA} \revision{only}. So, although RMSD is a good measure for comparing two similar
structures, it does not necessarily capture the scale of motion in different structures.

\begin{figure}[p]
\centering
\includegraphics[width=0.95\textwidth]{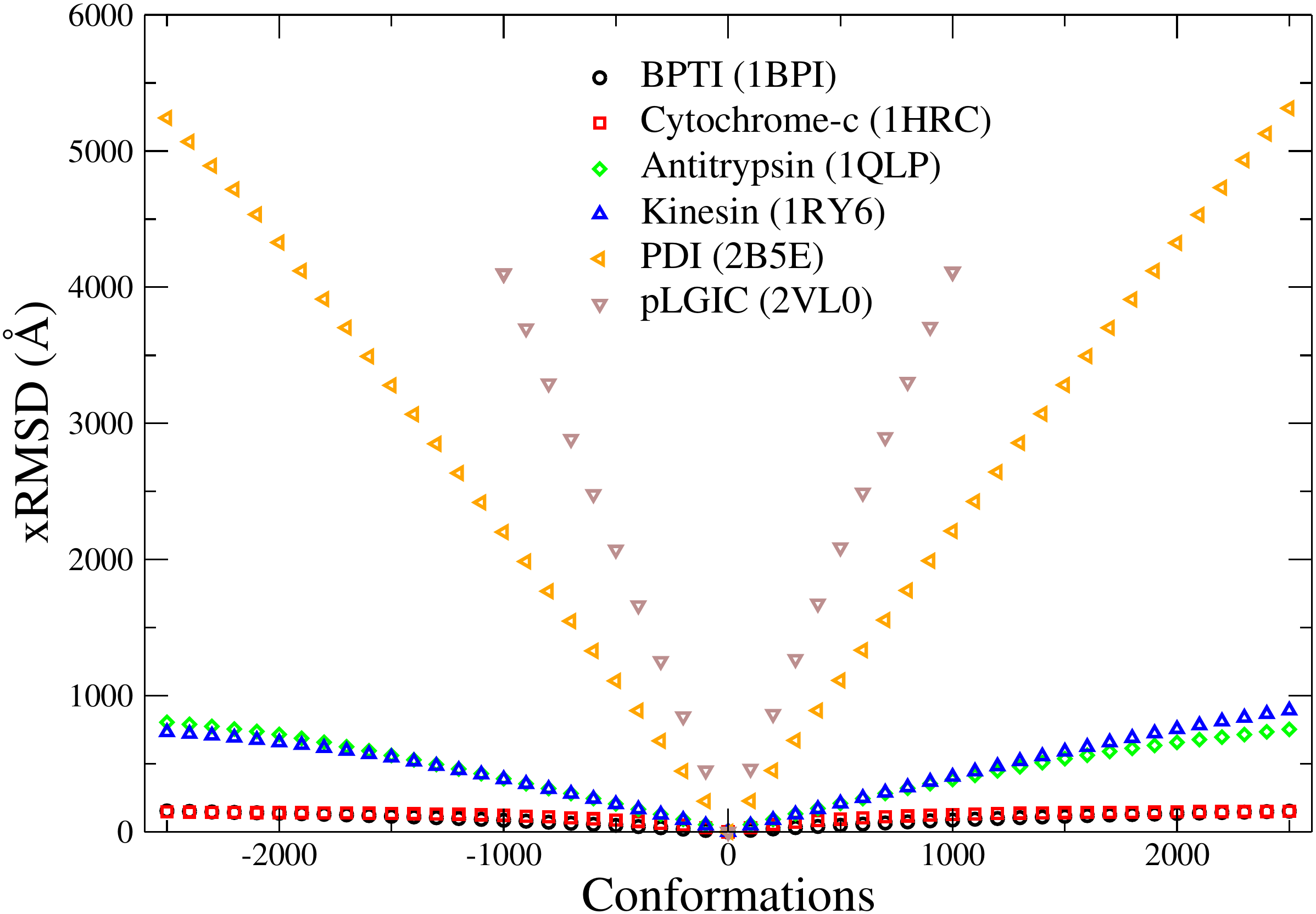}
\caption{Extensive RMSD as a function of {\sc Froda} conformations for all six proteins moving along
mode $m_7$. The maximum xRMSD values range from $150${\AA} for BPTI to $5243${\AA} for PDI. There
are three clear categories of protein motion: large conformational changes achieved by domain motion
(PDI and pLGIC), large loop motions (antitrypsin and kinesin) and small loop motions (BPTI and
cytochrome-c). Theselected $E_{\rm cut}$ for each protein are $E_{\rm cut}^{\rm 1BPI} = -2.2
$kcal/mol, $E_{\rm cut}^{\rm 1HRC} = -1.2$ kcal/mol, $E_{\rm cut}^{\rm 1RY6} = -1.1$ kcal/mol,
$E_{\rm cut}^{\rm 1QLP} = -1.1$ kcal/mol, $E_{\rm cut}^{\rm 2B5E} = -0.522$ kcal/mol and $E_{\rm
cut}^{\rm 2VL0} = -0.5$ kcal/mol. The XRMSD values obtained for $m_8$, \ldots, $m_11$ for the
selected proteins are consistent with $m_7$ xRMSD.
} 
\label{Fig:xRMSD}
\end{figure}

We introduce an \emph{extensive} RMSD measure by multiplying the RMSD (which describes the average displacement of atoms) by the number of
residues in the protein. Figure \ref{Fig:xRMSD} shows these xRMSD values for all the selected
proteins moving along $m_7$%
\footnote{For the proteins with domain motion, we have chosen values of $E_{\rm cut}$ which correspond to lower flexibility. The observed large variation in xRMSD is therefore taking place despite a restrictive bond network. On the other hand, for the proteins with loop motion we selected energy cutoffs which allow more structural flexibility. In this situation, although larger regions of the protein could become mobile, we still only observe localized loop motion.}.
The three categories of motion which we have discussed --- small loop motion, large loop motion and
domain motion --- become clearly visible in xRMSD. The xRMSD results for BPTI and cytochrome-c
closely resemble each other even though cytochrome-c is almost double the size of BPTI. Similarly,
the kinesin protein and the $\alpha$1-antitrypsin display similar xRMSD behaviour to each other in
their large loop motion. PDI and the pLGIC likewise have similar xRMSD behaviour reflecting their
domain motion. \revision{ Thus the character and extent of the flexible motions in proteins of various sizes, as
shown in Figures \ref{Fig:RMSD:SL}a,d,  \ref{Fig:RMSD:LL}a,d, \ref{Fig:RMSD:2B5E}a and
\ref{Fig:RMSD:2VL0}a,b, is better reflected by the xRMSD than by the RMSD alone. }

\subsection{Monitoring the evolution of normal modes}

It is implicit in normal-mode analysis of protein conformational change that a mode eigenvector should be a valid direction for motion over some non-zero amplitude. For example, Krebs {\it et al.} \cite{KreAWE02}
have surveyed a large number of known conformational changes, using paired crystal structures, comparing the vector describing the observed conformational change to the low-frequency elastic network mode eigenvectors using a dot product. In many cases the observed change had a large dot product ($>0.5$) with only one or two normal modes.

In each of our simulations we use an \emph{initial} normal mode, $m_{j}^{(i)}$, as a bias throughout the simulation. We calculate a new set of \emph{current} normal modes, $m_{j}^{(c)}$, for each newly generated conformation.
We compute the dot product of the bias vector, $m_{j}^{(i)}$, with $m_{j}^{(c)}$, that is, the current normal mode with the same mode number $j$, as in $m_{j}^{(i)} \cdot m_{j}^{(c)}$. Graphs of these dot products are shown for all protein structures investigated here in the supplementary Figures \ref{Fig:dot1BPI}--\ref{Fig:dot2VL0}.

\begin{figure}[p]
\centering
(a)\includegraphics[width=0.45\textwidth]{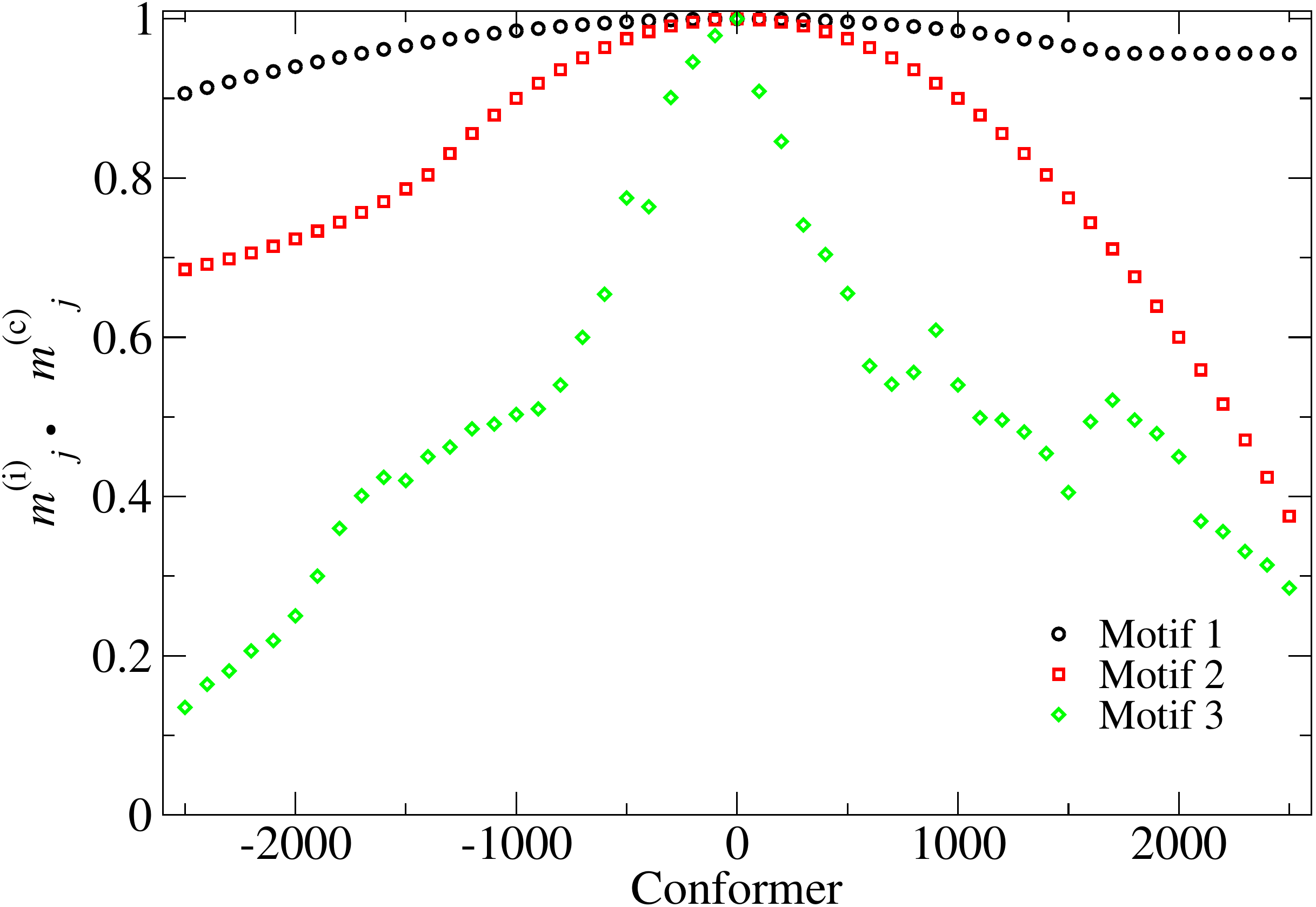}
(b)\includegraphics[width=0.45\textwidth]{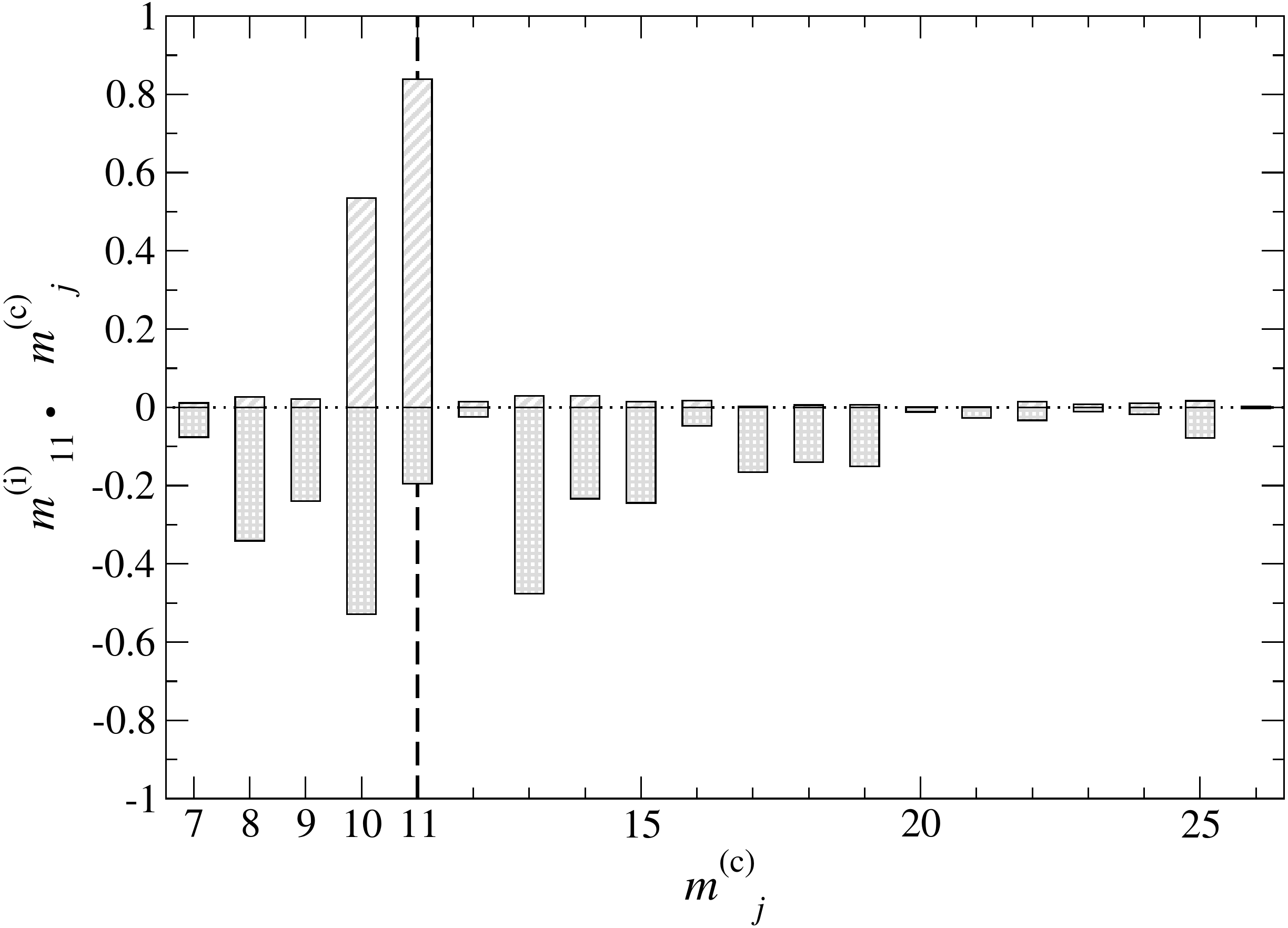}
\caption{Dot product motifs and illustration of mode mixing. 
(a) Schematic representation of the typical behaviours of the dot product
 of an initial mode with a current mode during projection along the initial mode eigenvector.
 Motif $1$: gradual, nearly quadratic reduction in the dot product due to a progressive rotation of
the current mode compared to the initial one. Eventual constant behaviour indicates that the
motions has reached its amplitude limit. Motif $2$: more rapid roughly quadratic reduction. Motif
$3$: sudden collapse of the dot product and the initial mode no longer resembles the current mode
with the same mode number.
(b) Dot products computed for initial normal mode $m_{11}^{(i)}$ of cytochrome-c and current normal modes from $m_7^{(c)}$ to $m_{26}^{(c)}$. Columns in the positive overlap direction denote projection
in the positive direction. Columns in the negative overlap direction indicate projection into the opposite direction. 
The vertical dashed line marks mode $m^{(c)}_{11}$ and the horizontal dotted line indicates a $0$ dot product value.
}
\label{Fig:paraboles}
\label{Fig:dots}
\end{figure}

We find that three main classes of dot product behaviour emerge, shown schematically in
Figure \ref{Fig:paraboles}a. In motif $1$, the dot product remains close to $1$ throughout the
simulation, indicating the initial and current modes remain very similar. In this case
the motion of the protein is not introducing significant changes to the elastic network model and
the mode eigenvector remains almost unchanged. In motif $2$, there is a
gradual decline in the dot product, of quadratic or cosine character; this suggests a gradual
rotation of $m_{j}^{(c)}$ relative to $m_{j}^{(i)}$. Perhaps the most interesting case is motif $3$,
in which the dot product $m_{j}^{(i)} \cdot m_{j}^{(c)}$ collapses rapidly; this can occur at any
point in the simulation, even if the RMSD between the initial and current conformations is small.
Examples of these motifs can be observed for all our proteins. We find, e.g.\ a motif $1$ behavior
in mode $m_7$ for PDI (cp.\ Figure \ref{Fig:dot2B5E}b, c, d), a motif $2$ behaviour in $m_7$ for
kinesin (cp.\ Figure \ref{Fig:dot1RY6}) and motif $3$ character in $m_7$ for antitrypsin (cp.\
Figure \ref{Fig:dot1QLP}) as well as in BPTI (cp.\ Figure \ref{Fig:dot1BPI}). Similar agreement with
all motifs can be found for higher modes, although, as a general tendency, the smooth motifs $1$ and
$2$ become gradually less visible and the more rapid changes exemplified by motif $3$ more
pronounced.

These sudden collapses do not indicate that the initial normal mode eigenvector has ceased to be a
valid direction along which flexible motion is possible. Rather, the eigenvector has ceased to
represent a single pure mode; $m_{j}^{(i)}$ now has significant overlap with multiple other modes
$m_{k}^{(c)}$. This \emph{mode mixing} is illustrated, e.g., in  Figure \ref{Fig:dots}b for
projection of cytochrome-c (1HRC) along $m_{11}^{(i)}$ at a cutoff energy $E_{\rm cut} = -1.2$
kcal/mol. After projection in the positive direction over about $0.25$\AA\ RMSD, $m_{11}^{(i)}$ has
a large overlap with $m_{11}^{(c)}$ and also with $m_{10}^{(c)}$. After a similar projection in the
negative direction, $m_{11}^{(i)}$ has only a small overlap with $m_{11}^{(c)}$, and instead has
some overlap with many low-frequency modes $m_{k}^{(c)}$ such as $k= 8$, $9$, $10$, $12$, $13$,
$14$.
Thus, a pure mode calculated on one structure may be a mixture of multiple modes when calculated on a very similar conformation. 

These results clarify that while the dot products provide useful additional information about the stability of the initial modes during the simulated motion, they are not simply correlated with loop or domain motion in contradistinction to the xRMSD.

\subsection{Significance of rigidity-analysis energy cutoff}

In the case of small loop motion, it is clear that lowering the rigidity-analysis energy cutoff --- thus making the structure more flexible --- increases the amplitude of flexible motion, as one might expect. The simulation of flexible motion can thus add value to rigidity analysis of protein structures by identifying the constraints that must be eliminated in order for two residues to become independently mobile. In the case of large domain motion, however, the most important criterion appears to be whether the domains are mutually rigid or not. 

We can see in the case of PDI (Figure \ref{Fig:RMSD:2B5E}) that the amplitude of flexible motion for
the lowest-frequency modes is almost unaffected by the choice of the energy cutoff ($E_{\rm cut}$)
provided it is set at a reasonable value of energy cutoff which represents each domain as a number
of separate small rigid clusters. This conclusion can also be drawn in the case of the ligand-gated
ion channel protein. 

\section{Conclusions}

We have reported a hybrid method to explore protein motion by integrating both rigidity constraints
from {\sc First} and directional information, in the form of low-frequency elastic network mode
eigenvectors obtained using {\sc Elnemo}, into the
geometric simulation method {\sc Froda}. The exploration brings out features of the motion that could not be inferred using {\sc First} or {\sc ElNemo} alone.
In order to illustrate the method, we have applied it here to a diverse selection of proteins whose flexible motion ranges from small loop motion (BPTI, cytochrome-c) and large loop motion (a kinesin and an antitrypsin) to large motions of entire domains (protein disulphide isomerase and a transmembrane pore protein).
Detailed studies of dynamics in relation to function of particular proteins are currently in progress \cite{JimBVF11,BelCM11}. The combined method can rapidly explore motion to large amplitudes in an all-atom model of the protein structure, maintaining steric exclusion
and retaining the covalent and noncovalent bonding interactions present in the original structure. Significant amplitudes of motion are achieved with only CPU-minutes of computational effort
even in a pentameric pore protein with more than $1600$ residues.
The amplitude of motion that can be achieved by flexible loops increases as the rigidity-analysis energy cutoff is lowered. For large-scale motion of domains, the most important criterion is that the energy cutoff should be low enough that different domains do not form a single rigid body.
We note that RMSD, a measure of structural similarity, does not properly reflect the scale of
flexible motion between different proteins; this is better captured by an extensive measure, xRMSD,
which reflects both the size of the protein and the amplitude of its motion.
Examination of the behaviour of the elastic network eigenvectors during the motion shows many examples of mode mixing, so that a given vector of motion can change from being a pure mode to a mixed one
after quite small displacements, without losing its character as an ``easy'' direction for flexible motion.

We believe that the ability to explore large amplitudes of flexible motion in an all-atom model with minimal computational resources will be of great use in biochemistry, structural biology, and biophysics, as a generator of new hypotheses and intuitions about protein structure and function, and as an ally to other simulation methods such as molecular dynamics, Monte Carlo folding simulations \cite{BurVWW11} and {\it ab-initio} simulations. Such a study is currently in preparation for PDI \cite{JimBVF11}.

\section*{Acknowledgements} 

We gratefully acknowledge support from the Royal Society through the India-UK International
Scientific Seminar scheme. SAW thanks the Leverhulme Trust for support through an Early Career
Fellowship. JEJ thankfully acknowledges EPSRC and BBSRC funding support during his PhD.
\revision{The authors would like to thank two anonymous reviewers for their very helpful
comments.}

\section*{References}

\clearpage

\begin{table}[p]
\caption{\label{Tab:cutoffs}The proteins and specific structures selected for this study. For each
protein, various $E_{\rm cut}$ values (kcal/mol) were chosen on the basis of rigidity analysis
(Figure \ref{Fig:RCD_plot}); the Table presents those values used in the simulations of motion that
are presented in the main text. Bold values for $E_{\rm cut}$ have been used to compute the xRMSD.
Visualization of structures for additional cutoffs (see Table \ref{Tab:cutoffs:complete}) are shown
in the supplementary material, see Figures \ref{3D_1BPI_Ec} -- \ref{3D_2VL0_Ec}.}
\begin{center}
\begin{tabular}{lllll}
\br
Protein&PDB &Resolution&Residues& $E_{\rm cut}$ (kcal/mol)\\
\mr
BPTI&1BPI&$1.1$\AA & $58$& $-0.2$, $\mathbf{-2.2}$\\
Cytochrome-c&1HRC&$1.9$\AA &$105$&  $-0.7$, $\mathbf{-1.2}$\\
Kinesin &1RY6&$1.6$\AA & $360$& $-0.4$, $\mathbf{-1.1}$ \\
$\alpha$1-antitrypsin & 1QLP & $2.0$\AA & 394& $-0.1$, $\mathbf{-1.1}$ \\
PDI&2B5E&$2.4$\AA &$504$& $-0.015$, $\mathbf{-0.522}$, $-1.412$ \\
pLGIC&2VL0&$3.3$\AA &$1605$& $-0.4$, $\mathbf{-0.5}$ \\
\br
\end{tabular}
\end{center}
\end{table}

\clearpage

\begin{figure}[p]
 \centering
\includegraphics[width=\textwidth]{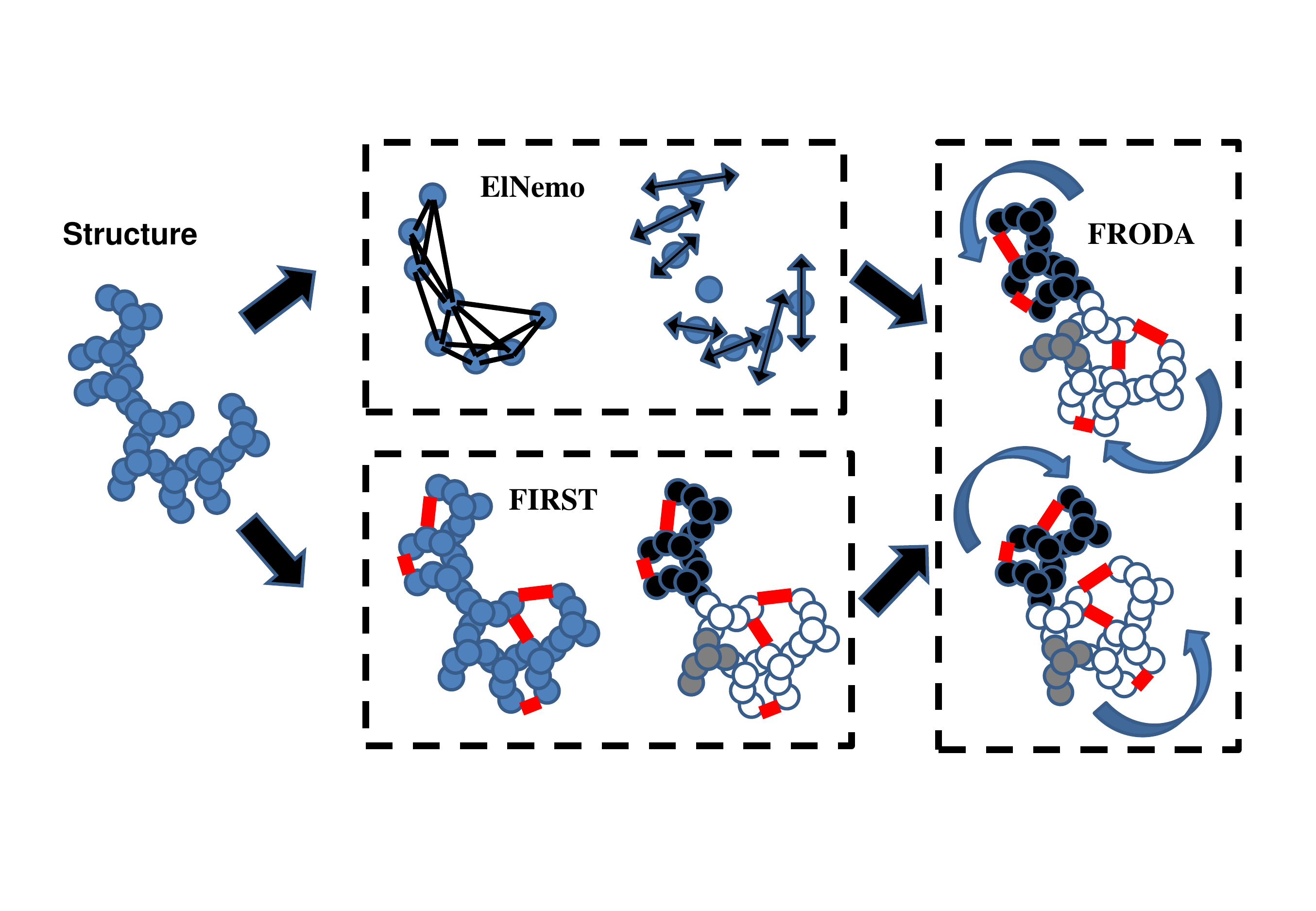}
\caption{\label{fig:method} Schematic of the method. The input (at left) is an all-atom protein
structure. Normal mode analysis (above) models the protein with a one-site-per-residue coarse
graining and a simple spring model to produce an eigenvector for low-frequency motion. Rigidity
analysis (below) identifies noncovalent interactions in an all-atom model of the protein and
divides the protein into rigid clusters and flexible linkers. Geometric simulation (right) integrates normal-mode and
rigidity information to explore the flexible motion of the protein. }
\end{figure}

\pagebreak
\setcounter{page}{1}
\clearpage
\newpage
\setcounter{figure}{0}
\setcounter{section}{0}
\setcounter{table}{0}
\renewcommand{\figurename}{Supplementary figure}
\renewcommand{\tablename}{Supplementary table}
\def\thefigure{S\arabic{figure}}
\def\thetable{S\arabic{table}}
\def\thesection{S\arabic{section}}
\pagebreak
\setcounter{page}{1}

\section*{Supplementary information}

\section{Settings, input files and command line options}
\label{Supp:settings}

All simulations were carried out using the {\sc ElNemo} and {\sc First} software. {\sc Froda} is a routine available within {\sc First}. The file formats and command line options to apply a mode eigenvector as a bias in {\sc First/Froda} have not previously been published in the literature, and we therefore describe them here.

The required input for all calculations is a {\tt .pdb} format file containing an all-atom representation of the protein structure including hydrogen atoms, which we shall name {\tt protein.pdb}. Our usual procedure is to obtain a file containing the heavy atom positions from the Protein Data Bank; to remove alternate side chain conformations and nonbonded heteroatoms including water molecules; to add hydrogens using the {\sc Reduce} software, including flipping of side chains where necessary; and to renumber the atoms sequentially using {\sc PyMOL}.

Normal mode calculations were carried out using {\sc ElneMo } using the default setting in the { \tt
pdbmat.dat } input file of a $12$\AA\ cutoff in the spring network. The protein structure is given
as a {\tt pdbmat.structure } file, consisting of only the $\alpha$ carbon lines from the all-atom
structure. The output is a {\tt pdbmat.eigenfacs } file which, for a protein of $N$ residues,
contains $3N$ mode eigenvectors. Each eigenvector is described with a mode number, a frequency, and
$N$ lines each giving a Cartesian vector; the $i$th line is the displacement to be applied to the
$i$th residue. The vector is normalized so that the sum of the squares of all displacement vectors
is unity. The first few lines of a mode appear thus:
\begin{verbatim}
 VECTOR    7       VALUE  6.0869E-04
 -----------------------------------
  2.5267E-02  2.1069E-02  0.1020
  1.7347E-02  1.5141E-02  9.3303E-02
  3.5897E-02  2.5485E-02  9.2557E-02
 ...
\end{verbatim}

To pass this mode as a bias to {\sc First/Froda} we prepare a {\tt mode.in} file giving, for each residue, the identity of the $\alpha$ carbon atom for that residue in the {\tt pdb} file, followed by the displacement vector. The first few lines of a {\tt mode.in} file appear thus:
\begin{verbatim}
2 0.025267 0.021069 0.102
6 0.017347 0.015141 0.093303
20 0.035897 0.025485 0.092557
...
\end{verbatim}
In {\sc Froda}, the first displacement vector will be applied as a bias to the motion of all atoms in the first residue, and so forth.

The command line options for {\sc First} to carry out a {\sc Froda} simulation with a mode bias can be given as follows:
{\tt FIRST -non \$PROTEIN -E -\$CUT -FRODA -mobRC1  -freq \$FREQ -totconf \$TOTCONF -modei -step \$STEP -dstep \$DSTEP}

Taking terms in order, {\tt FIRST} is the {\sc First} executable; {\tt -non} is to run in noninteractive mode; {\tt \$PROTEIN} is the name of the all-atom {\tt .pdb} format input file; {\tt -E -\$CUT} runs the rigidity analysis with hydrogen-bond cutoff energy {\tt -\$CUT}; {\tt FRODA} invokes the {\sc Froda} algorithm to simulate flexible motion; {\tt -mobRC1} specifies that the largest rigid cluster is to be mobile, like the smaller rigid clusters, during the simulation; {\tt -freq \$FREQ} specifies the frequency with which newly generated conformations are to be written to file as new conformations; {\tt -totconf \$TOTCONF} specifies the total number of new conformations to generate; {\tt modei} specifies that {\tt mode.in} should be read and applied as a bias; {\tt -step \$STEP} specifies the magnitude of random perturbations of the atomic positions in the generation of each new conformation; and {\tt -dstep \$DSTEP} specifies the magnitude of the displacement of the structure along the mode eigenvector in the generation of each new conformation.

For all our calculations we have set {\tt \$FREQ} to 100 so as to save every $100$th conformation; we have used a {\tt \$STEP} of $0.1$\AA; and we have used {\tt \$DSTEP} values of $0.01$\AA\ and $-0.01$\AA\ so as to project each mode in both possible directions. Thus to explore the motion of {\tt protein.pdb} at an energy cutoff of $-1.0$ kcal/mol, we would use two command lines as follows
\begin{verbatim}
FIRST -non protein.pdb -E -1.000 -FRODA -mobRC1  -freq 100 -totconf 2500 
  -modei -step 0.1 -dstep 0.01

FIRST -non protein.pdb -E -1.000 -FRODA -mobRC1  -freq 100 -totconf 2500 
  -modei -step 0.1 -dstep -0.01
\end{verbatim}
\section{Raw vs fitted RMSD}
\label{Supp:Rawvsfitted}

\begin{figure}
\hspace{1pc}
\centering
(a)\includegraphics[width=0.95\columnwidth]{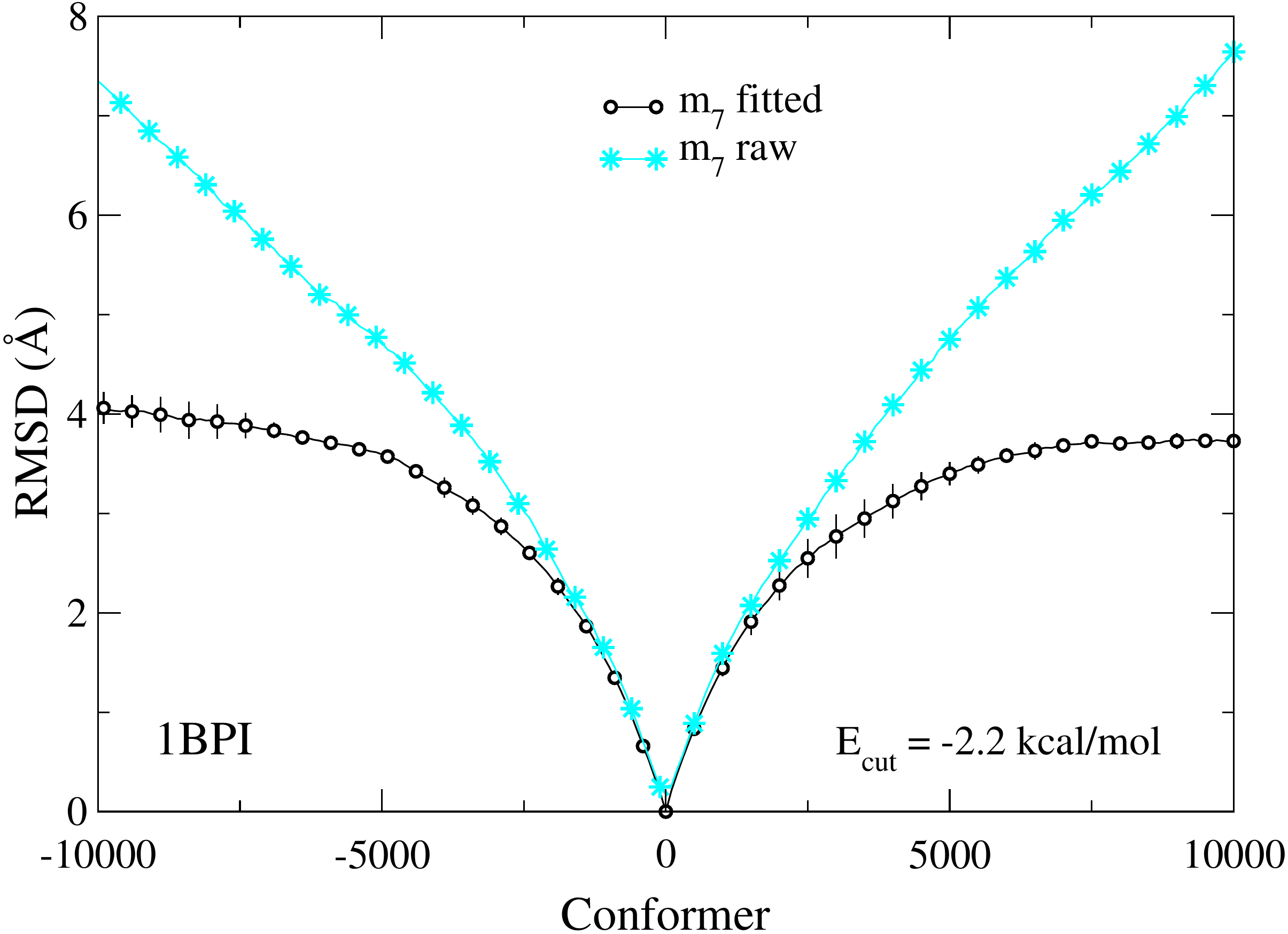}\\
(b)\includegraphics[width=0.45\columnwidth]{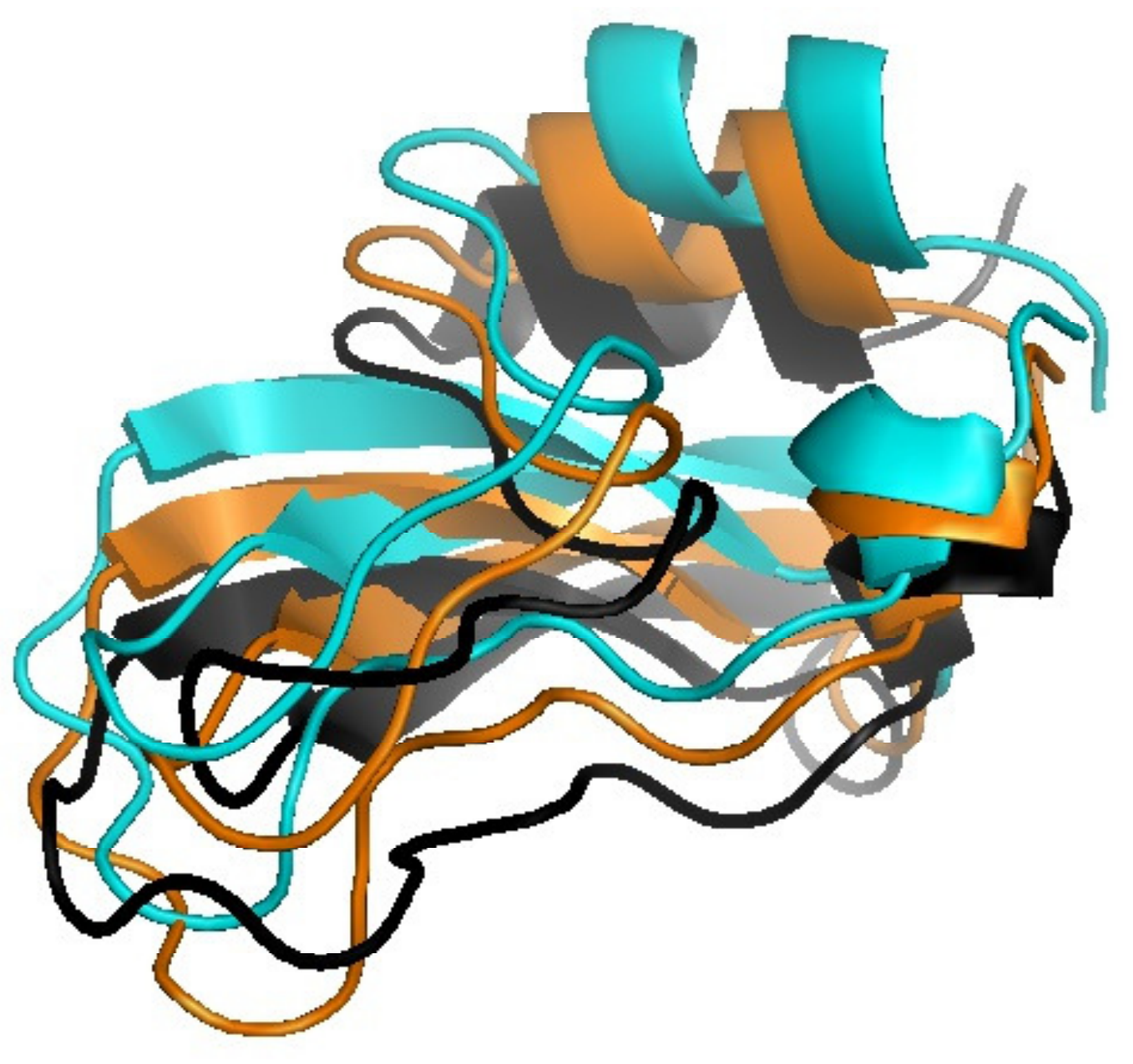}
(c)\includegraphics[width=0.45\columnwidth]{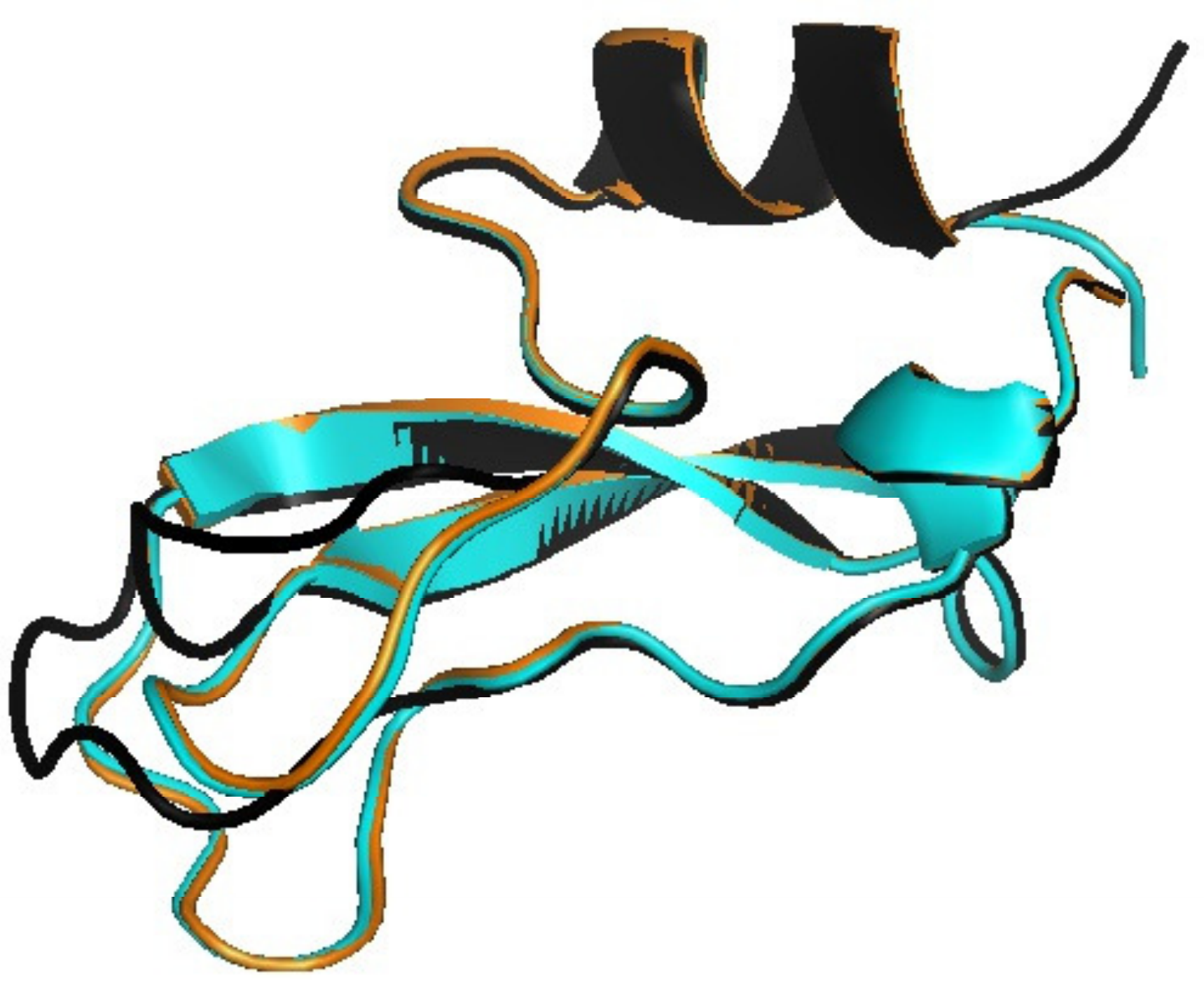}
\caption{Raw vs fitted RMSD for BPTI. (a) Comparison between the RMSD
  values obtained before (raw RMSD) and after fitting the new
  conformers to the initial structure. We report the fitted and raw
  RMSD values of mode ${m_7}$ for every 100th conformer (only every
  500th indicated by a symbol for clarity) and up to in total $10000$
  conformers at a cutoff energy of $E_{\rm cut}=-2.2$ kcal/mol. The
  error bars denote the standard deviation obtained from including $5$
  different initial random perturbations to the guided motion, see
  section \ref{mobility}. The fitted RMSD values, also shown in Figure
  \ref{Fig:RMSD:SL}e for modes $m_7$, \ldots, $m_{11}$, saturate
  whereas the raw RMSD values increase linearly, due to {\sc Froda}
  weighting each residue by the number of atoms that it contains, thus
  producing a component of rigid body motion. Panels (b) and (c) show
  the superimposed structures for conformers $0$ (black), $5000$
  (orange) and $10000$ (green) of (b) raw and (c) fitted structures.}
\label{RawvsfittedRMSD}
\end{figure}
\revision{Figure \ref{RawvsfittedRMSD} shows the importance of using a
  fitted RMSD to identify and subtract correctly the rigid-body motion during a
  simulation. The raw RMSD values (in cyan) do not saturate, whereas the fitted
  RMSD values (in black) do. As presented in the figure, the evolution of the
  fitted and raw RMSD values shows that raw RMSD
  values continue to increase linearly, hence accounting for network
  internal motion as well as spatial rigid-body translation (cp.\ figure
  \ref{RawvsfittedRMSD}b), whereas fitted RMSD values saturate as they
  account for network internal motion only (cp.\ figure
  \ref{RawvsfittedRMSD}c). The good overlap of the conformers $5000$
  (orange) and $10000$ (green) as shown in figure
  \ref{RawvsfittedRMSD}c correspond to the only minimal increase in
  fitted RMSD observed in figure \ref{RawvsfittedRMSD}a beyond
  conformer $5000$. The two structures overlap each other along the
  polypeptide chain, which indicates that there is little motion
  between the two conformers. However, conformer $5000$ and $10000$
  have substantially moved with respect to the initial structure,
  i.e.\ conformer $0$ (black). Hence, it is clear from this comparison
  that we account for the rigid-body motion effect by fitting all the
  conformers to the initial structure.}


\clearpage

\begin{table}[bt]

\caption{\label{RMSD_table}Extensive RMSD values, maximum RMSD values and the
cutoff energies choosen to calculate xRMSD for each protein based on the RCD graphs. For proteins
that are expected to be rigid we have choosen a higher $E_{\rm cut}$ and for proteins with an
expected conformational change we have choosen a more restrictive $E_{\rm cut}$.}

\begin{center}
\begin{tabular}{lllllll}
\br
Protein&Residues &$E_{\rm cut}$  &RMSD pos &RMSD neg& xRMSD pos & xRMSD neg \\
 & & (kcal/mol) &{\AA}&{\AA}&{\AA}&{\AA}
\\
\mr
BPTI& $58$&$-2.2$&$2.62$&$2.66$& $152$& $154$\\
Cytochrome-c& $105$&$-1.2$&$1.44$&$1.40$&$151$&$146$\\
Kinesin& $360$&$-1.1$&$2.48$&$2.04$& $892$&$733$\\
Antitrypsin & $394$&$-1.1$ &$1.91$&$2.04$& $753$&$804$\\
PDI&$504$&$-0.522$&$10.54$&$10.40$& $5314$&$5243$\\
pLGIC&$1605$&$-0.5$& $2.56$ & $2.55$ & $4113$&$4099$\\
\br
\end{tabular}
\end{center}
\end{table}
\begin{table}[tb]
\caption{\label{Tab:cutoffs:complete} Rigidity analysis cutoff energies. Summary of \emph{all}
cutoff energies
extracted from the rigidity analysis used in the geometric simulations. RMSD data for only a
subset of these are shown in the main text.}
\begin{center}
\begin{tabular}{lll}
\br
Protein&PDB code&Cutoffs (kcal/mol) \\
\mr
BPTI&1BPI&  $-0.200$, $-1.700$, $-2.200$\\
Cytochrome-c& 1HRC&$-0.700$, $-1.200$\\
Kinesin &1RY6&$-0.400$, $-0.600$, $-1.100$\\
Antitrypsin & 1QLP &$-0.100$, $-0.500$, $-1.100$\\
PDI&2B5E&$-0.015$, $-0.522$, $-0.885$, $-1.412$\\
pLGIC &2VL0&$-0.400$, $-0.500$\\
\br
\end{tabular}
\end{center}
\end{table}
\clearpage

\begin{figure}[tb]
\centering
(a)\includegraphics[width=0.45\columnwidth]{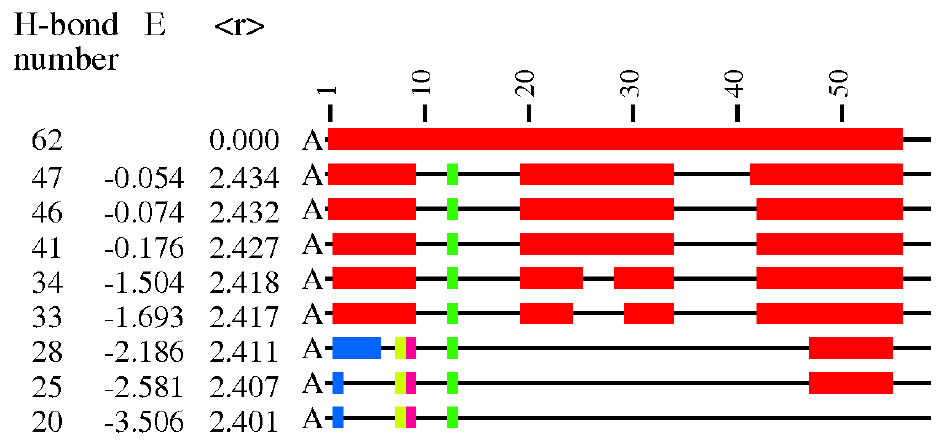}
(b)\includegraphics[width=0.45\columnwidth]{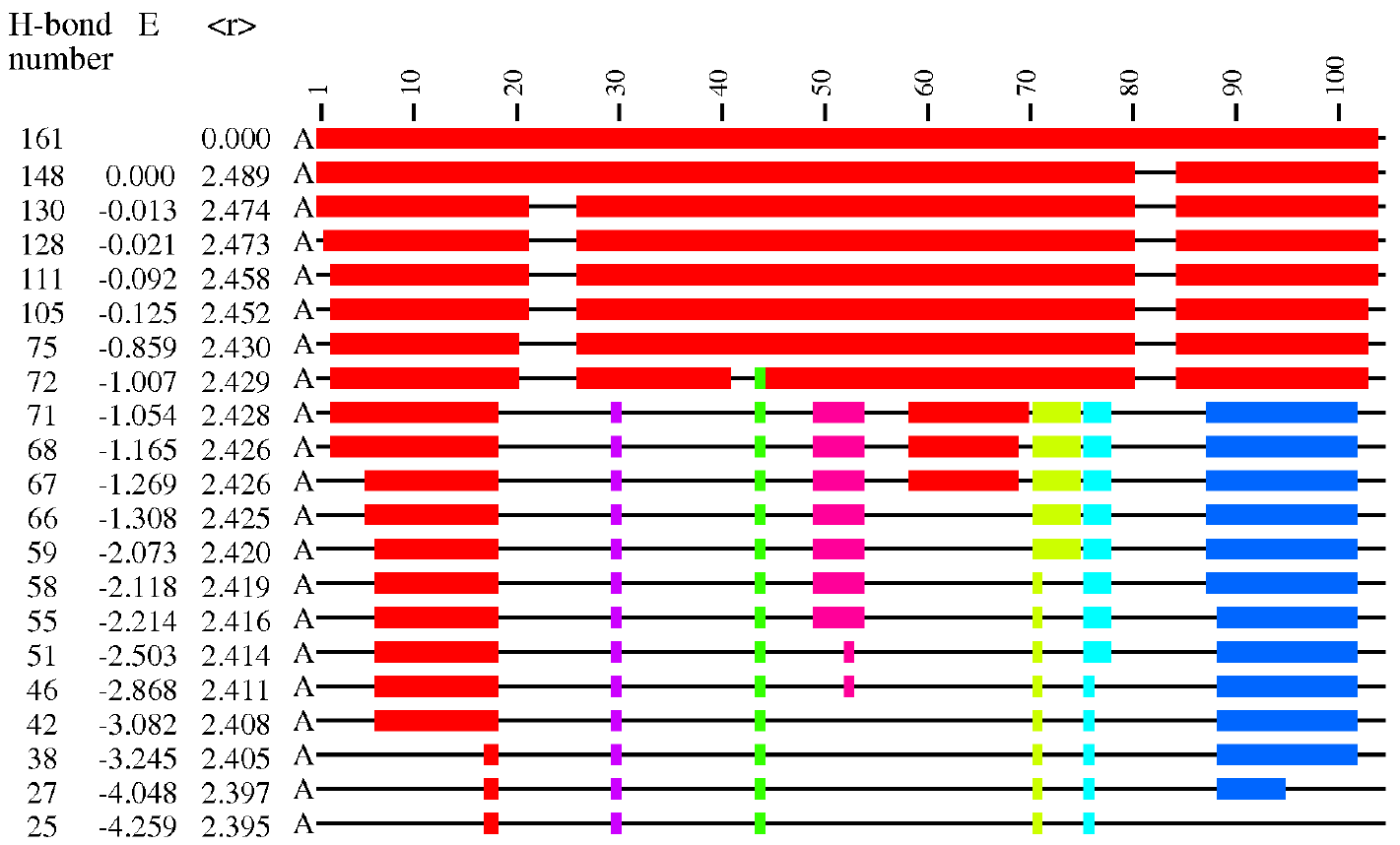}\\
(c)\includegraphics[width=0.45\columnwidth]{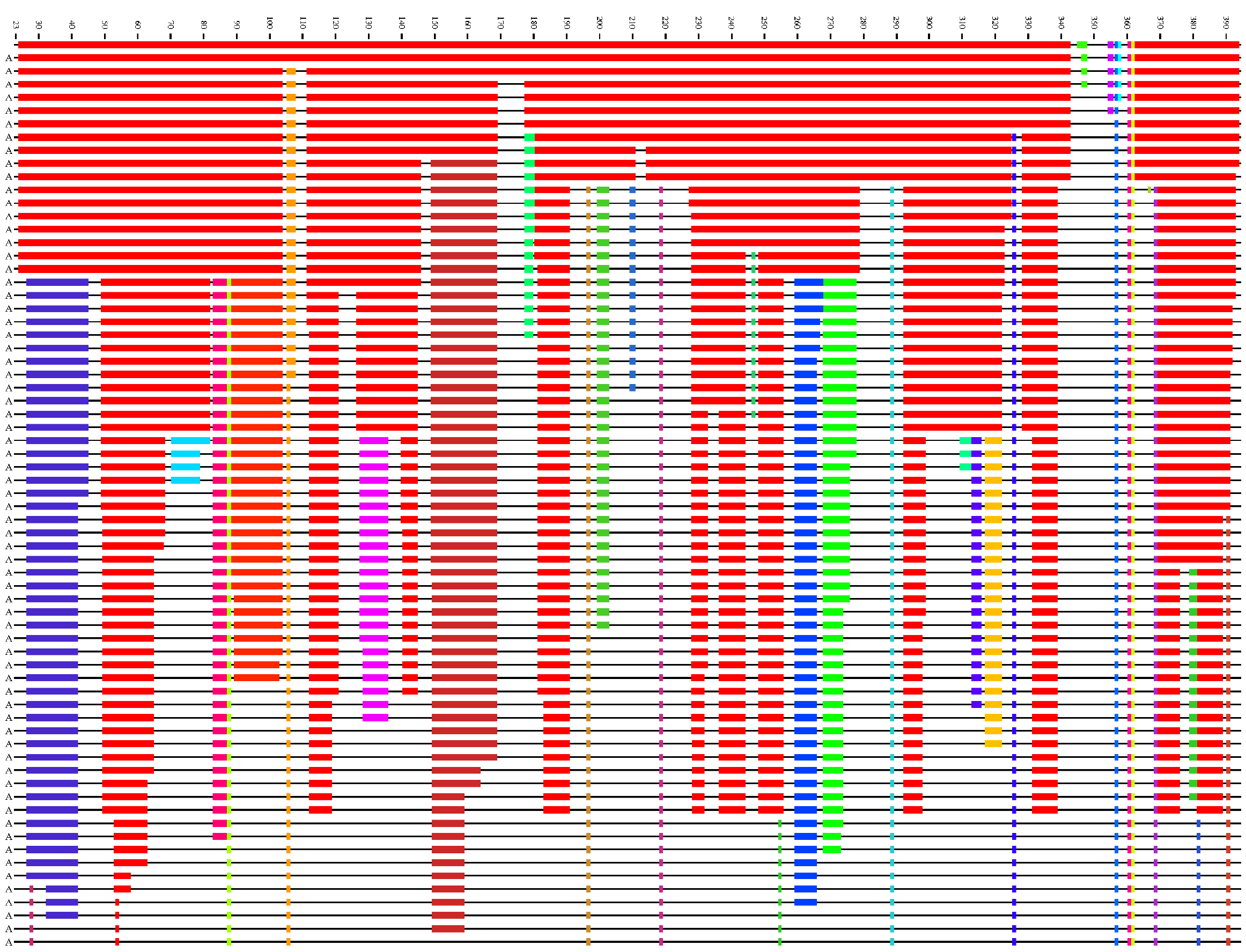}
(d)\includegraphics[width=0.45\columnwidth]{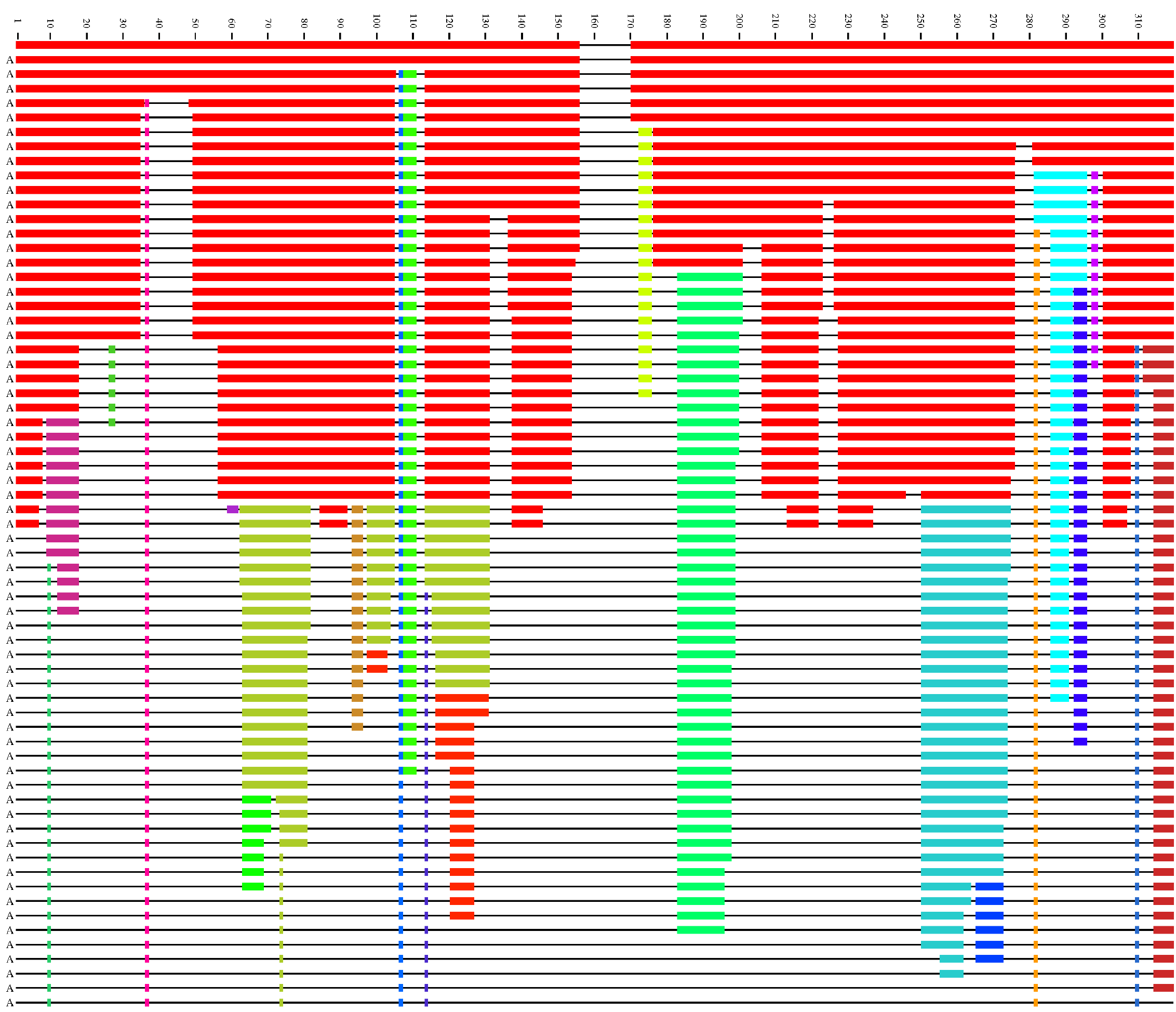}\\
(e)\includegraphics[width=0.45\columnwidth]{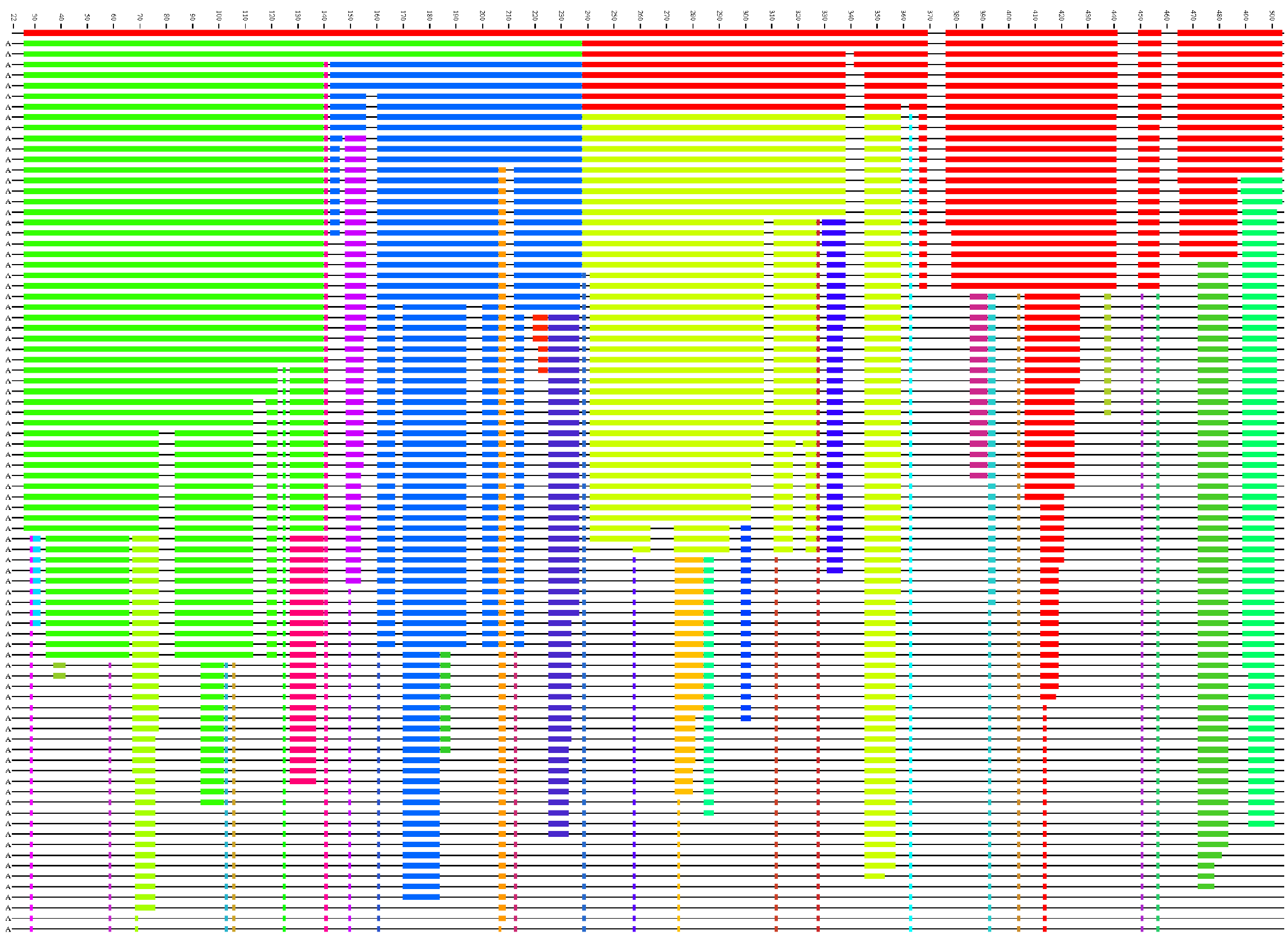}
(f)\includegraphics[width=0.45\columnwidth]{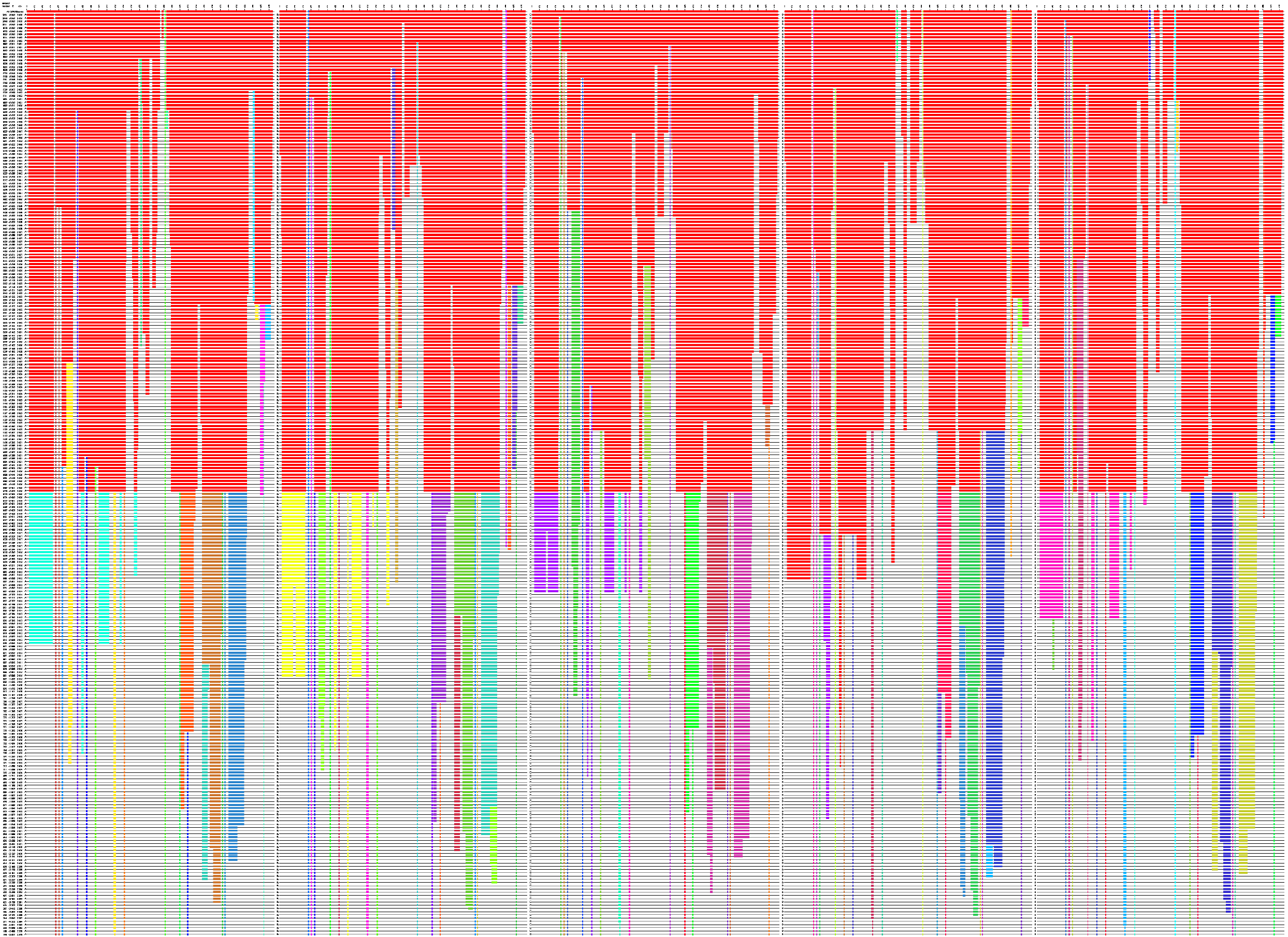}

\caption{\label{Fig:RCD_plot} Rigid cluster decomposition graphs for: (a) BPTI (1BPI) (b)
cytochrome-c (1HRC) (c) $\alpha$1-antitrypsin (1QLP) (d) internal kinesin motor
domain (1RY6) (e) yeast PDI (2B5E) and (f) pLGIC (2VL0). The \emph{x} axis represents the protein
backbone and the \emph{y} axis the energy, $E_{\rm cut}$, of the last hydrogen bond, which after
being removed provokes a change in the rigidity distribution. Each line represents the new rigidity
distribution of the polipeptide chain induced by removing a bond which alters the previous rigidity
configuration. The residues belonging to rigid
clusters are coloured --- with the biggest rigid cluster coloured in red, whereas the flexible
regions are shown as thin black lines. We choose the energy cutoffs defining the number of rigidity
constraints using the RCD plots. For a detailed description of RCD graphs see Refs.
\cite{JacRKT01,WelJR09}
}
\end{figure}


\begin{figure}
\centering
\hspace{0pc}
\centering
\subfigure{
(a)\includegraphics[width=0.3\columnwidth]{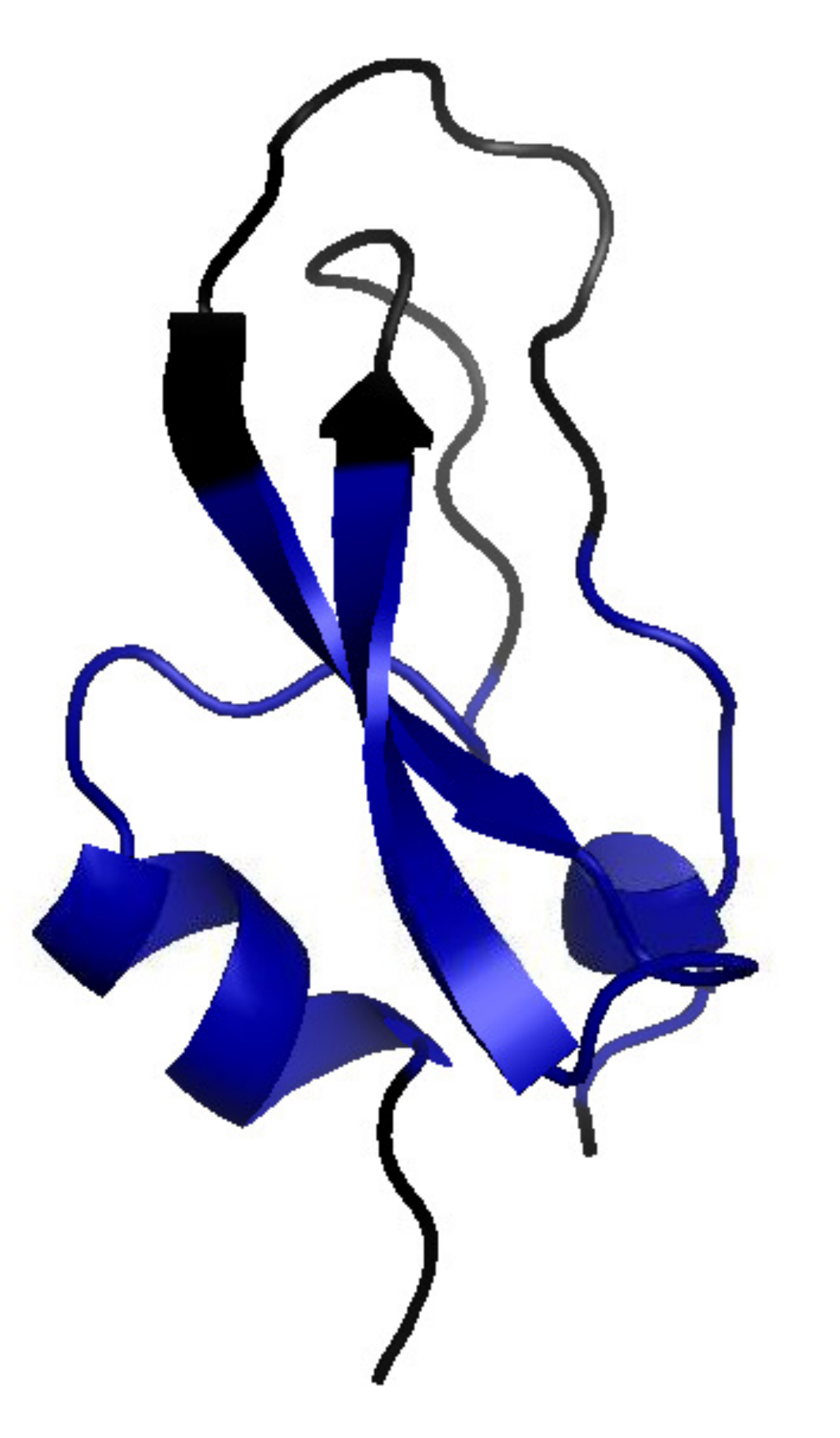}
(b)\includegraphics[width=0.3\columnwidth]{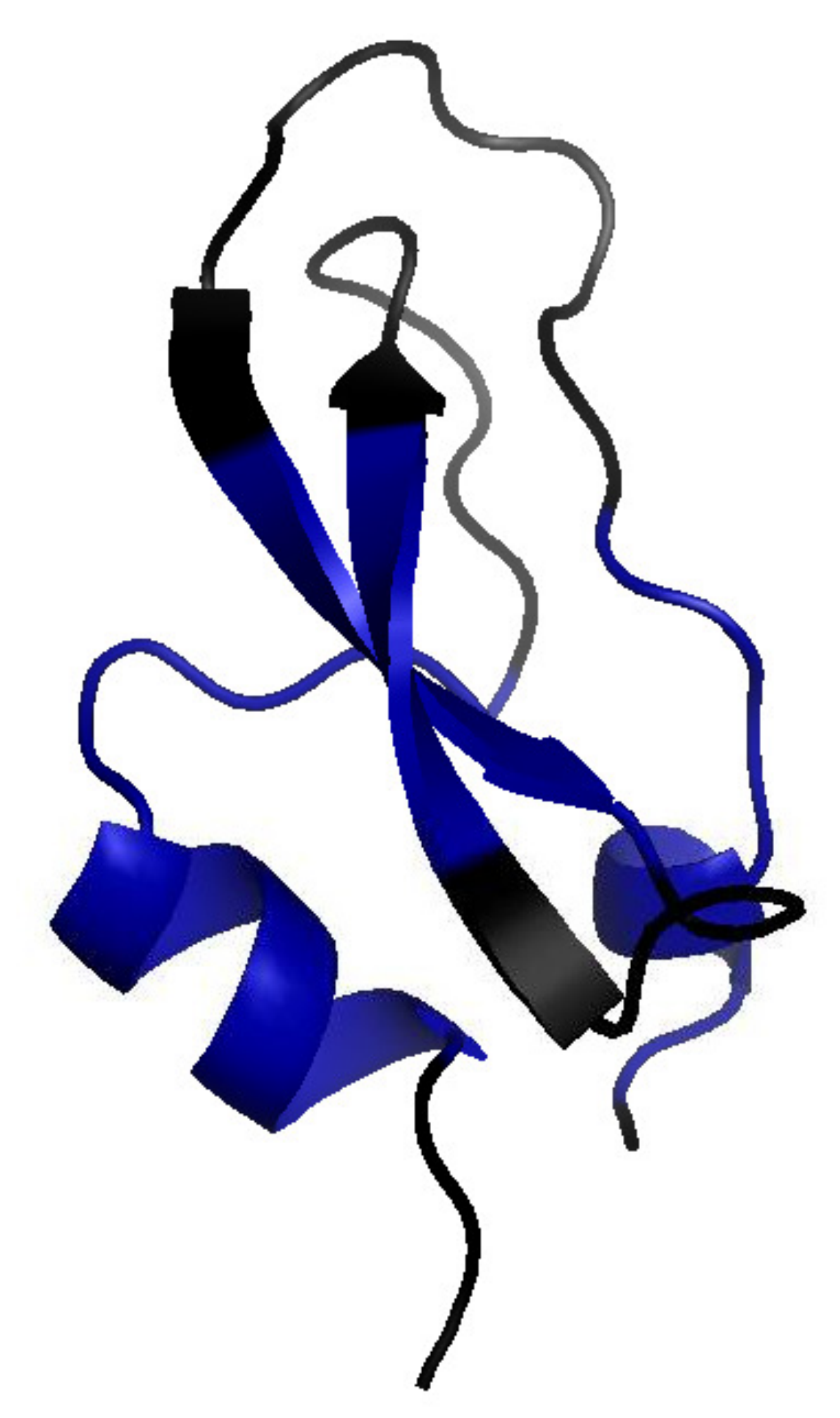}
(c)\includegraphics[width=0.3\columnwidth]{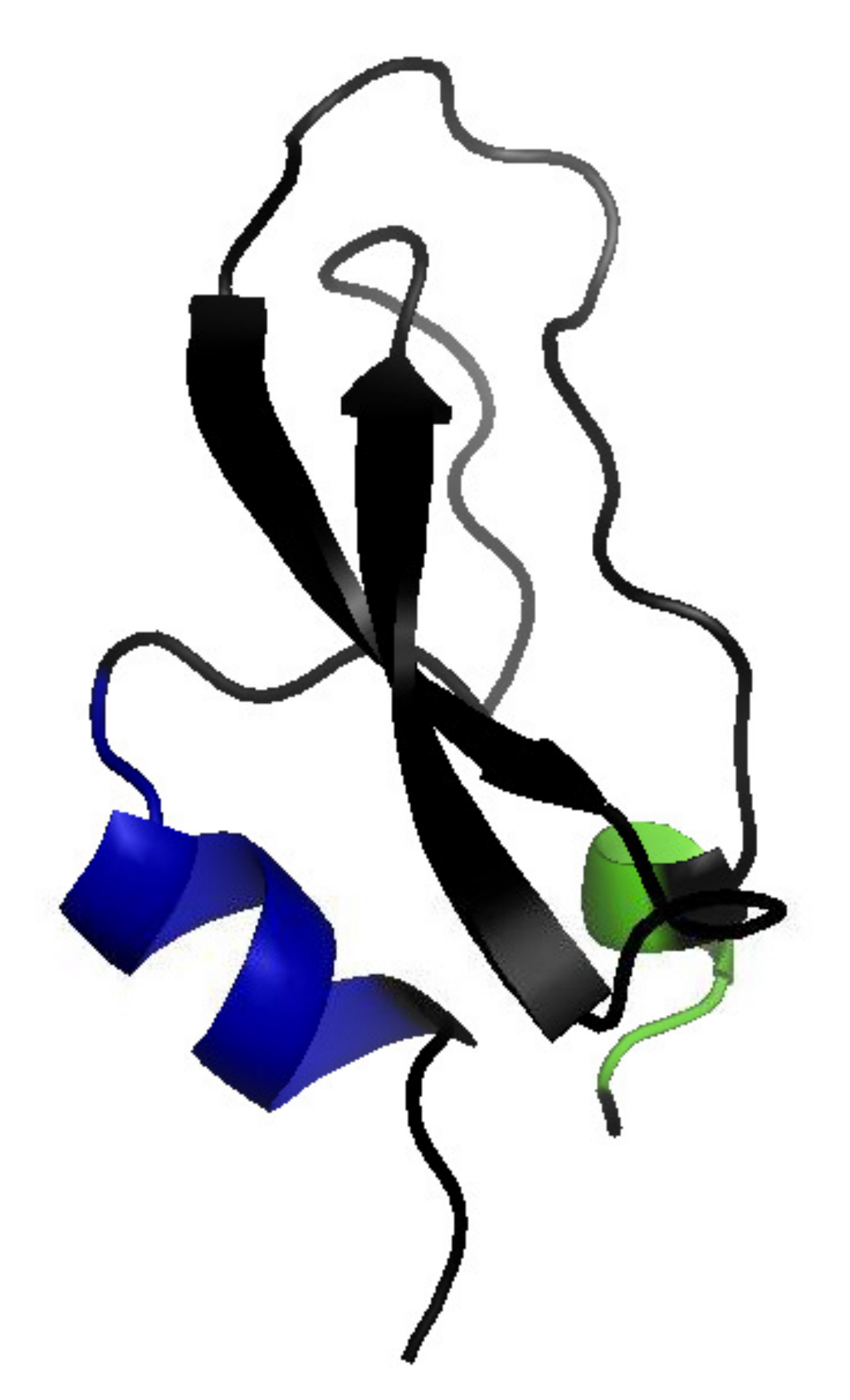}
}
\caption{\label{3D_1BPI_Ec}Tertiary structure of BPTI (1BPI). Colouring
is defined using the rigidigy analysis results shown in Figure \ref{Fig:RCD_plot}. Flexible
regions are illustrated in black whereas rigid residues are coloured as per the rigid cluster they
belong to. The biggest rigid cluster is coloured in blue. The number and size of the rigid clusters
vary depending on the chosen cutoff value, which for BPTI are (a) $E_{\rm cut}=-0.2$ kcal/mol, (b)
$E_{\rm cut}=-1.7$ kcal/mol and (c) $E_{\rm cut}=-2.2$ kcal/mol. Note that the colour code used to
represent residues within the same rigid cluster is not the same in the RCD and in the tertiary
structures.The biggest rigid cluster in the RCD graphs is noted in red and in the tertiary
structures is noted in blue.}
\end{figure}

\begin{figure}
\hspace{1pc}
\centering
\subfigure {
(a)\includegraphics[width=0.45\columnwidth]{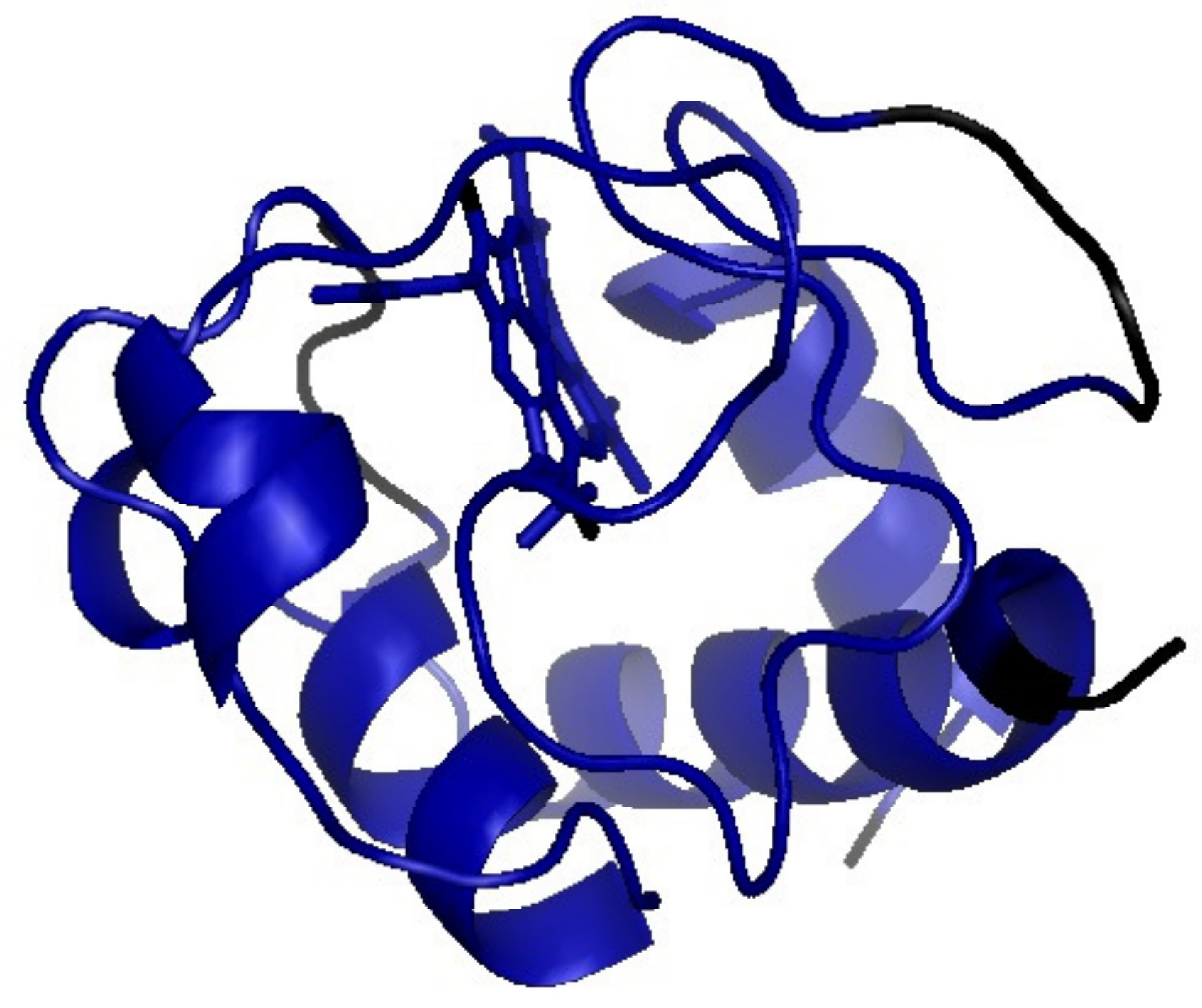}
(b)\includegraphics[width=0.47\columnwidth]{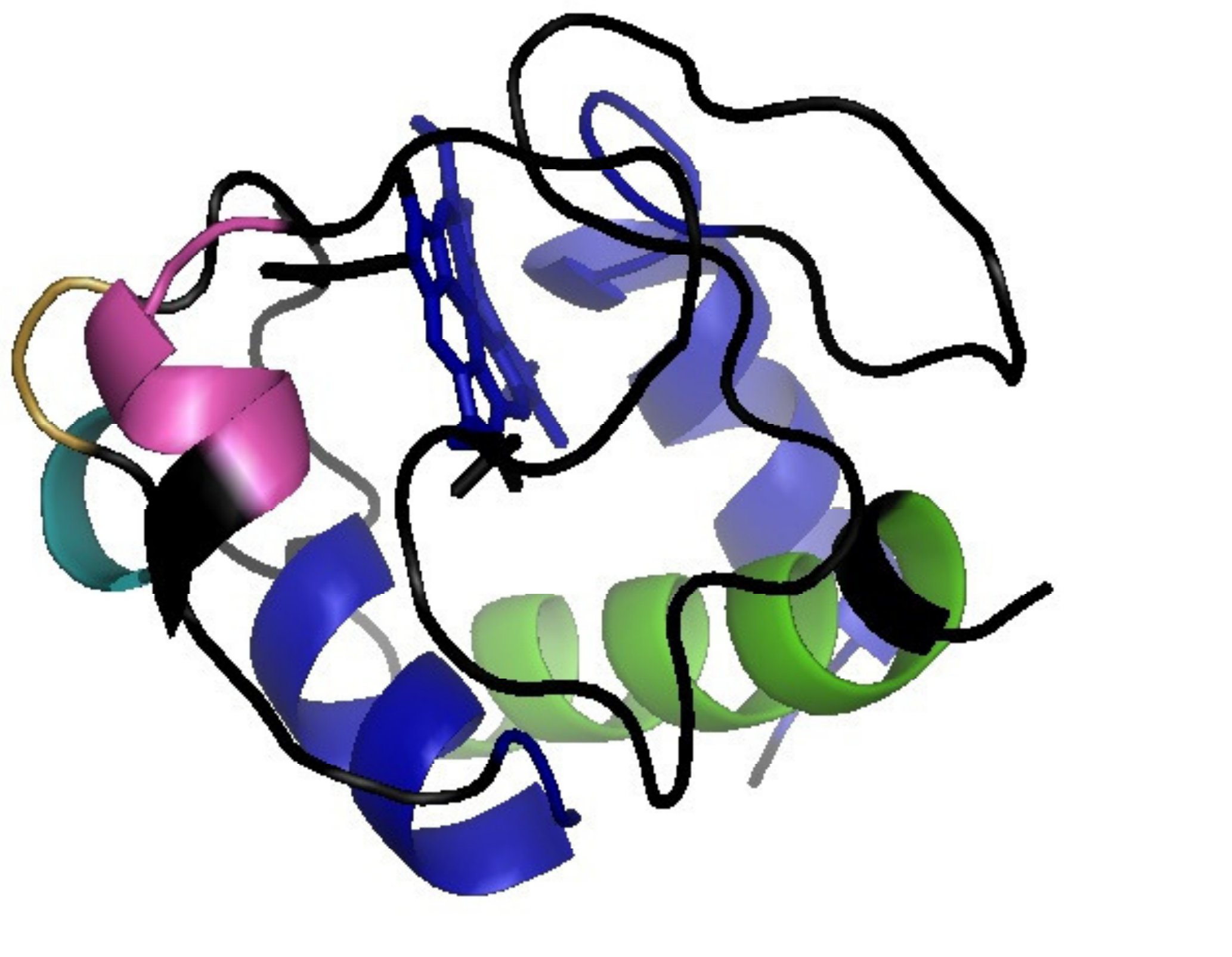}

}
\caption{Tertiary structure of cytochrome-c (1HRC). Colouring of the tertiary
structure is defined as in Figure \ref{3D_1BPI_Ec} but with cutoff energies (a) $E_{\rm cut}=-0.7$
kcal/mol and (b) $E_{\rm cut}=-1.2$ kcal/mol.}
\label{3D_1HRC_Ec}
\end{figure}

\begin{figure}
\hspace{1pc}
 \centering
\subfigure{
(a)\includegraphics[width=0.3\columnwidth]{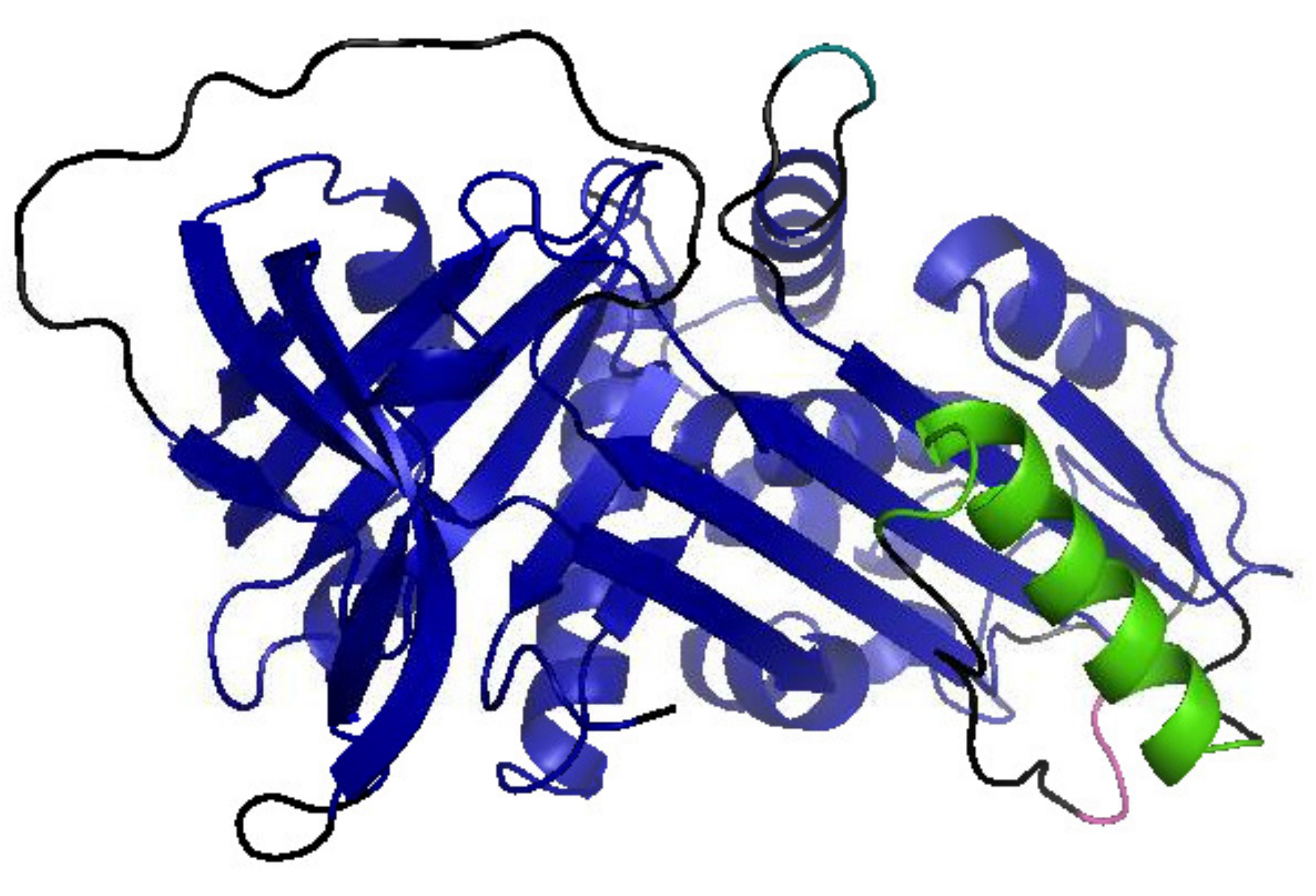}
(b)\includegraphics[width=0.3\columnwidth]{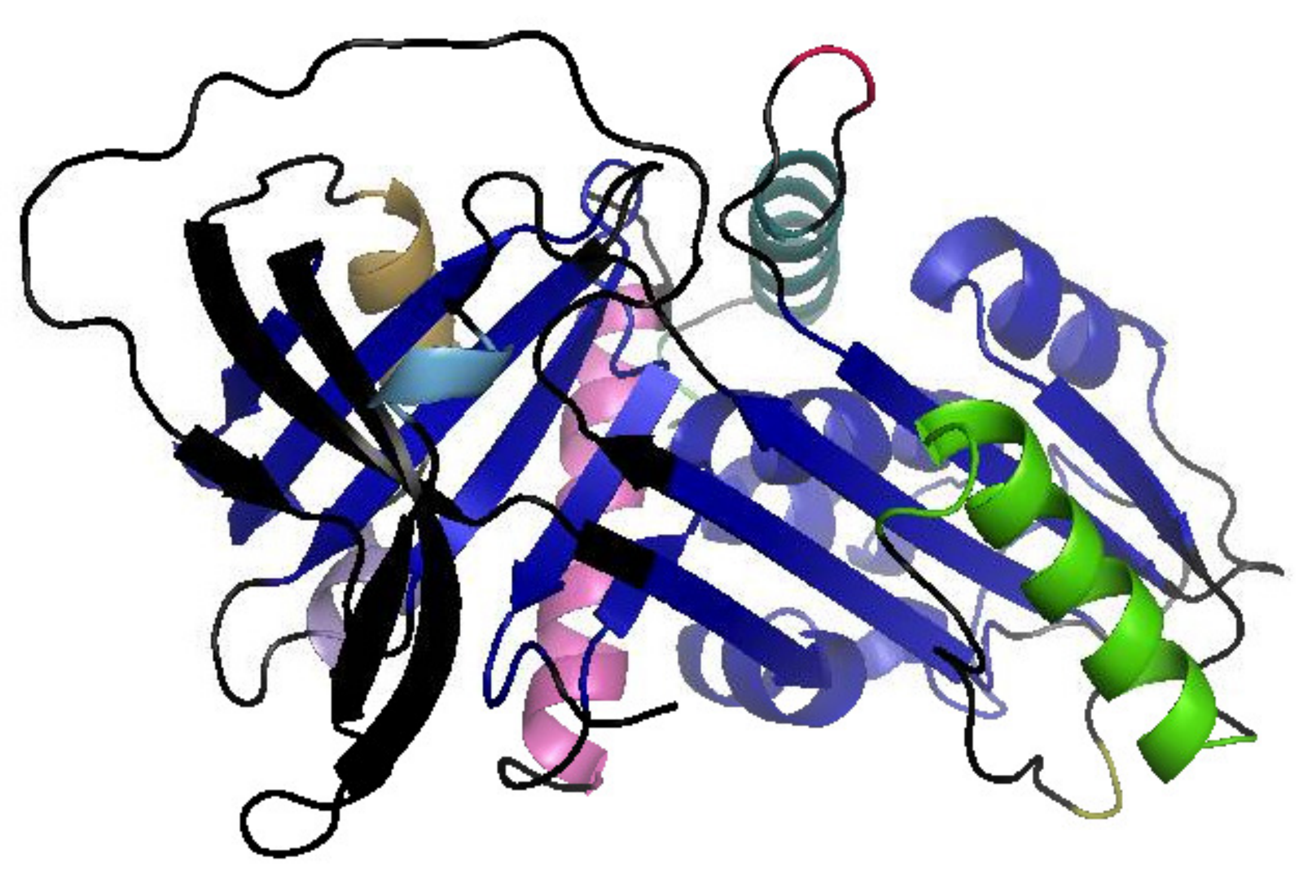}
(c)\includegraphics[width=0.3\columnwidth]{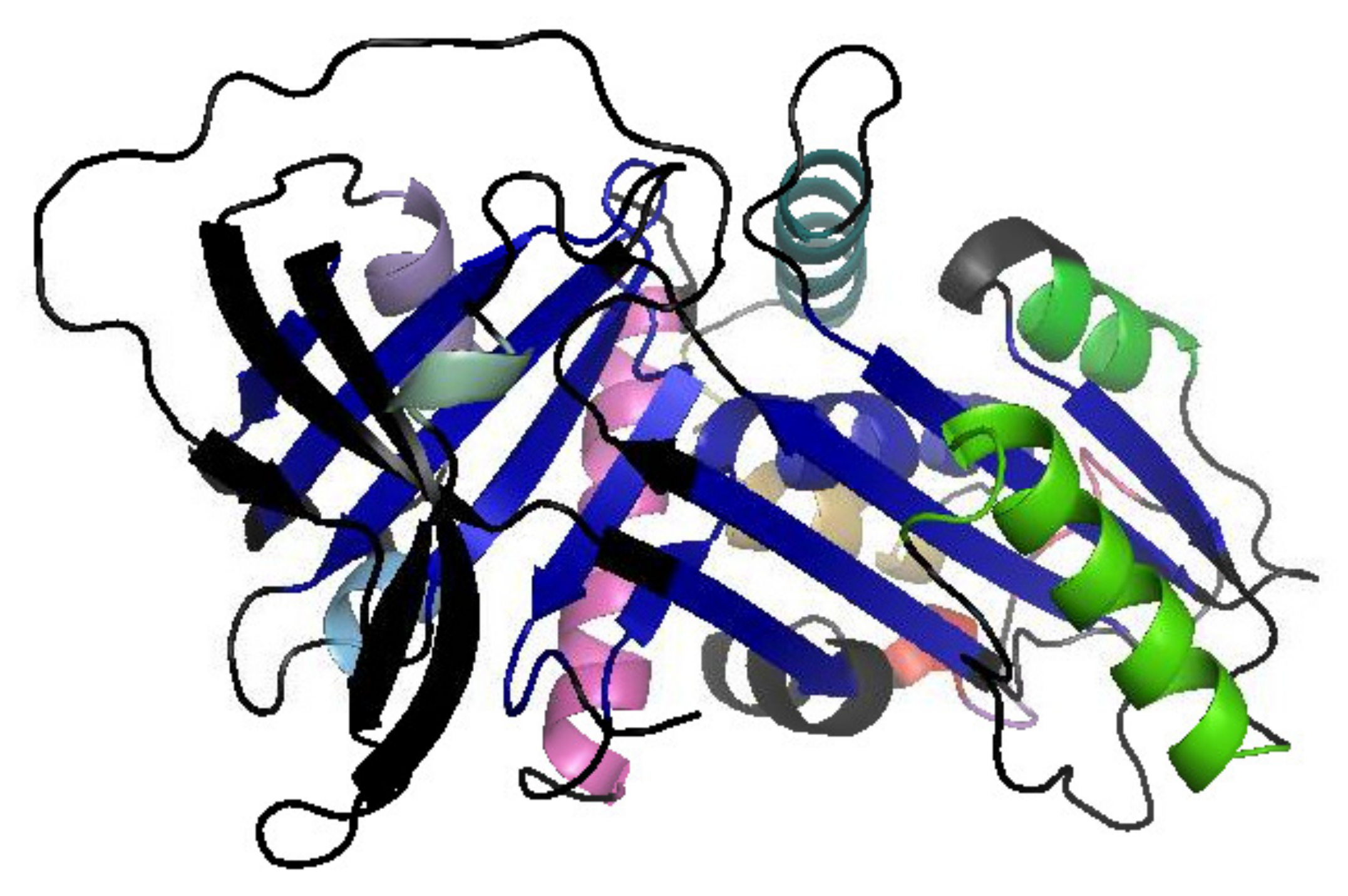}
}
\caption{Tertiary structure of $\alpha$1-antitrypsin (1QLP). Colouring of the tertiary
structure is defined as per Figure \ref{3D_1BPI_Ec} but at cutoff energies of (a) $E_{\rm
cut}=-0.1$ kcal/mol, (b) $E_{\rm cut}=-0.5$ kcal/mol and (c) $E_{\rm cut}=-1.1$ kcal/mol.}

\label{3D_1QLP_Ec}
\end{figure}

\begin{figure}
\hspace{1pc}
 \centering
\subfigure{
(a)\includegraphics[width=0.3\columnwidth]{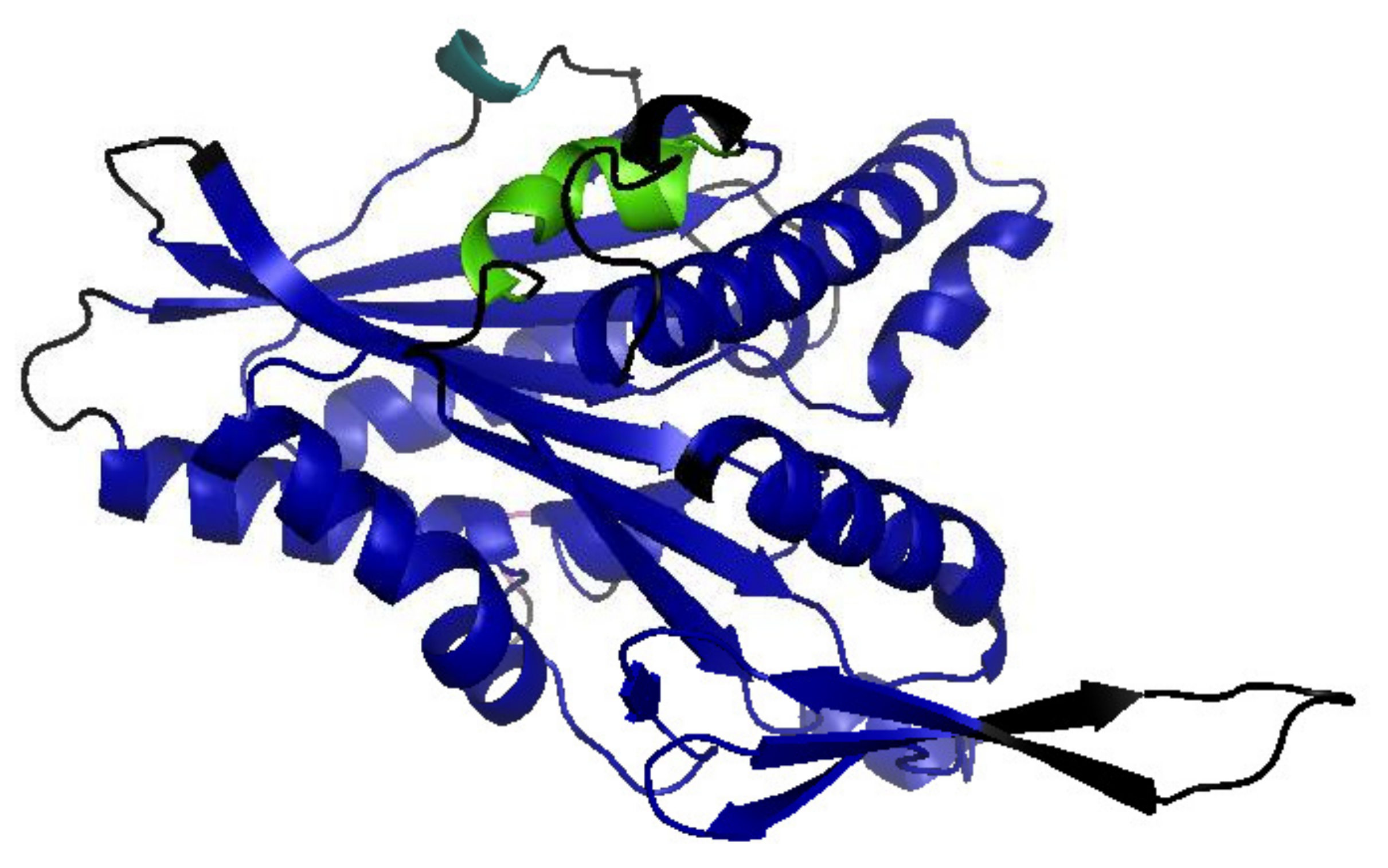}
(b)\includegraphics[width=0.3\columnwidth]{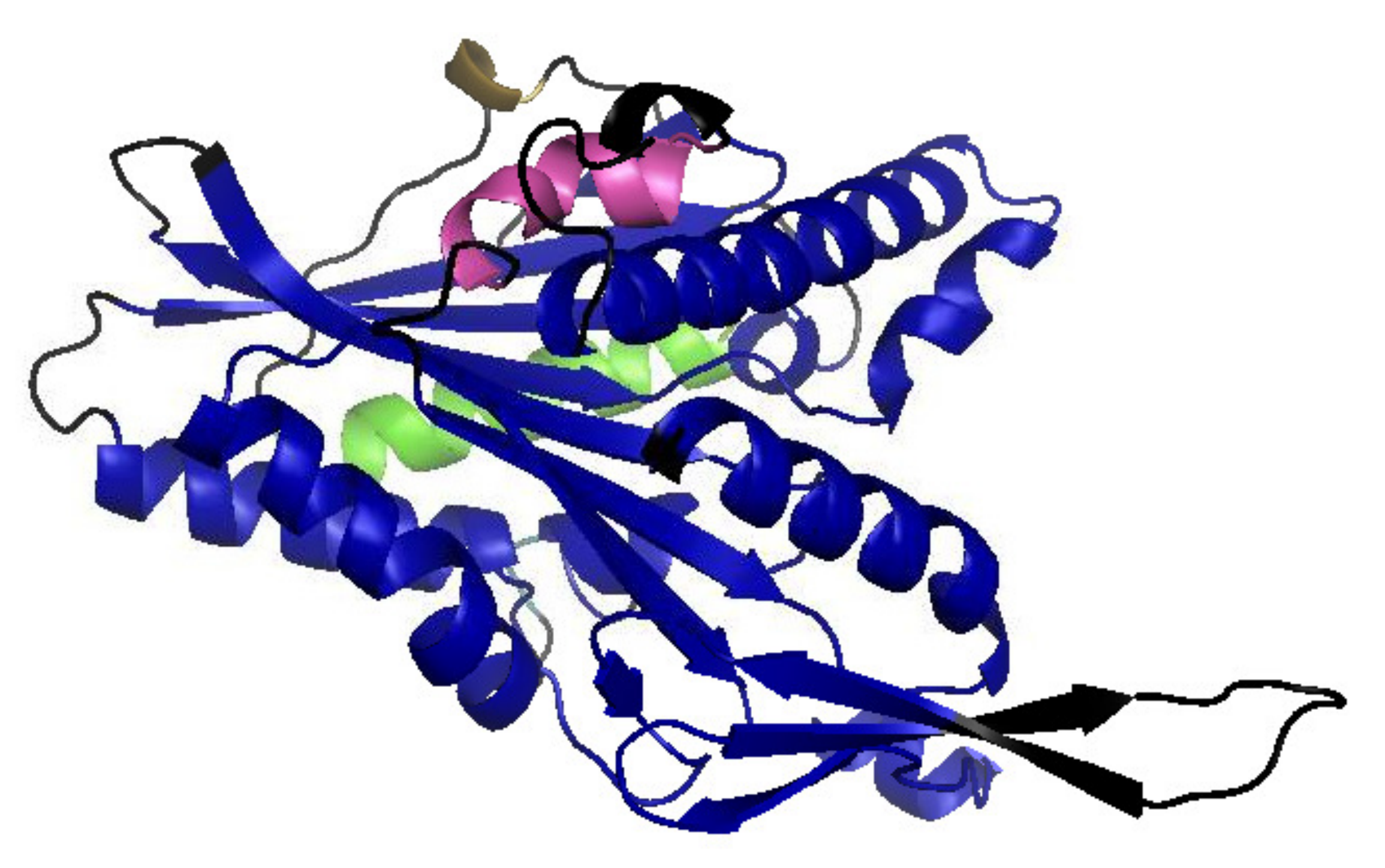}
(c)\includegraphics[width=0.3\columnwidth]{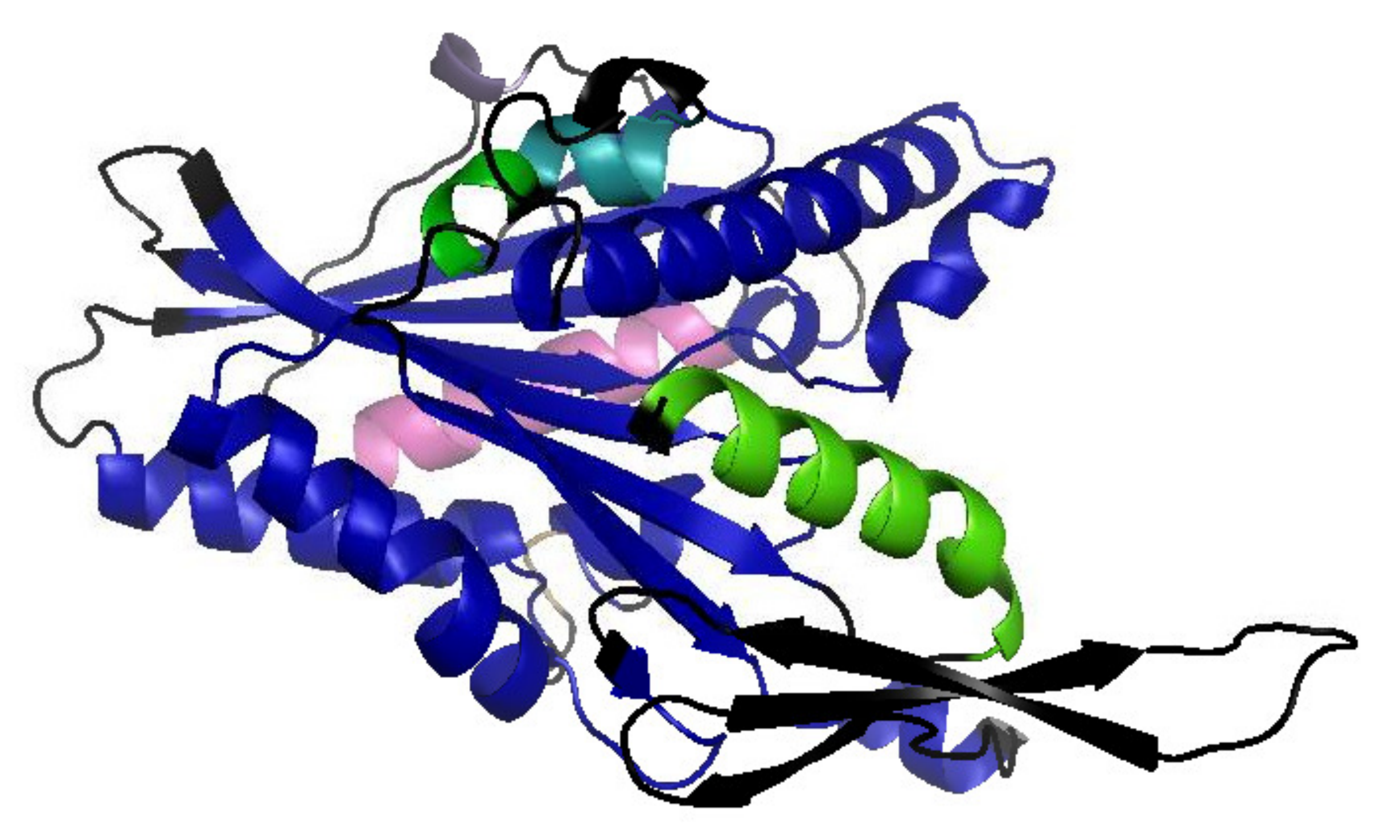}
}
\caption{Tertiary structure of internal kinesin motor domain (1RY6). Colouring of the
tertiary structure is defined as in Figure \ref{3D_1BPI_Ec} but with cutoff energies (a) $E_{\rm
cut}=-0.4$ kcal/mol, (b) $E_{\rm cut}=-0.6$ kcal/mol and (c) $E_{\rm cut}=-1.1$ kcal/mol.}
\label{3D_1RY6_Ec}
\end{figure}


\begin{figure}[tb]
\centering
(a)\includegraphics[width=0.45\textwidth]{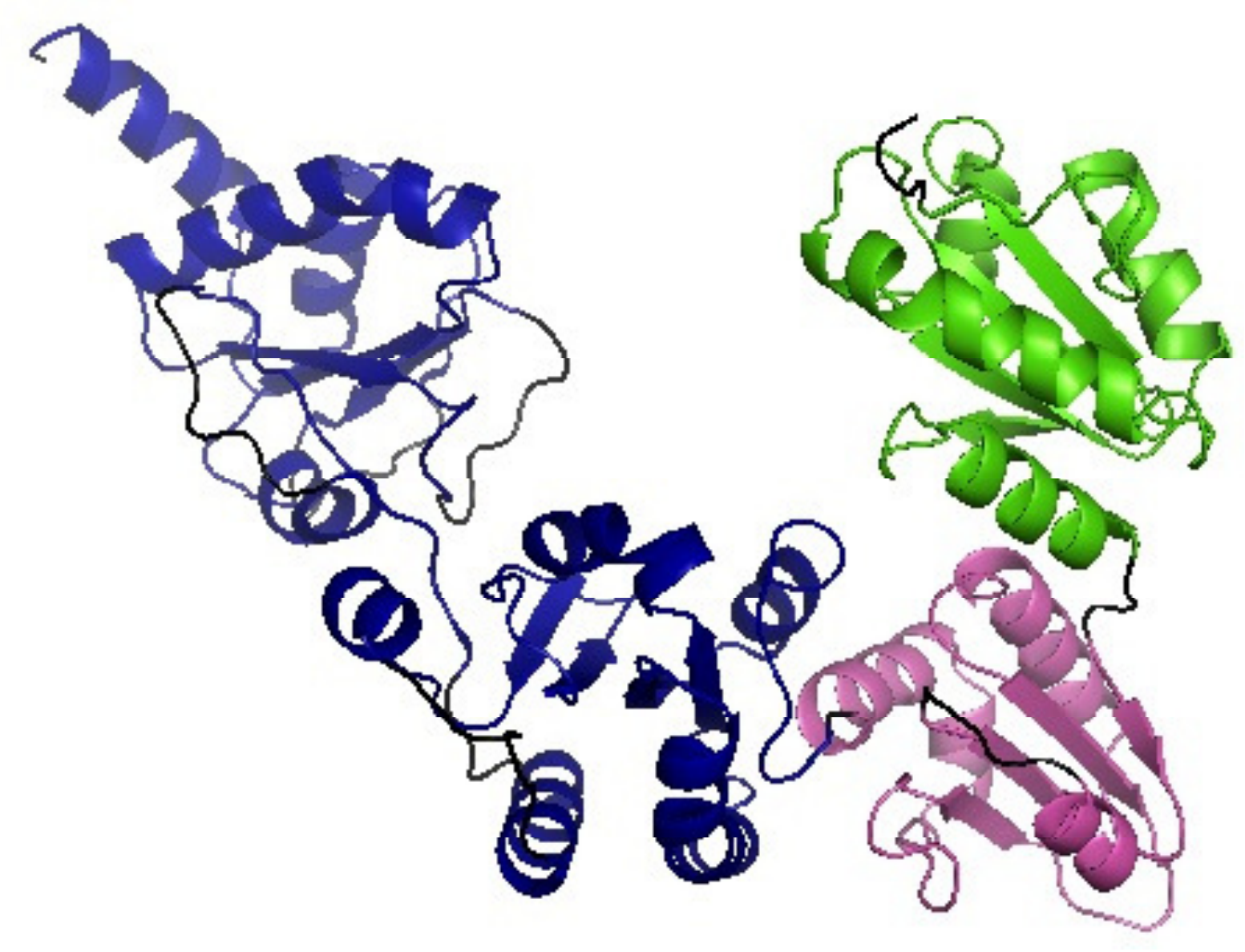}
(b)\includegraphics[width=0.45\textwidth]{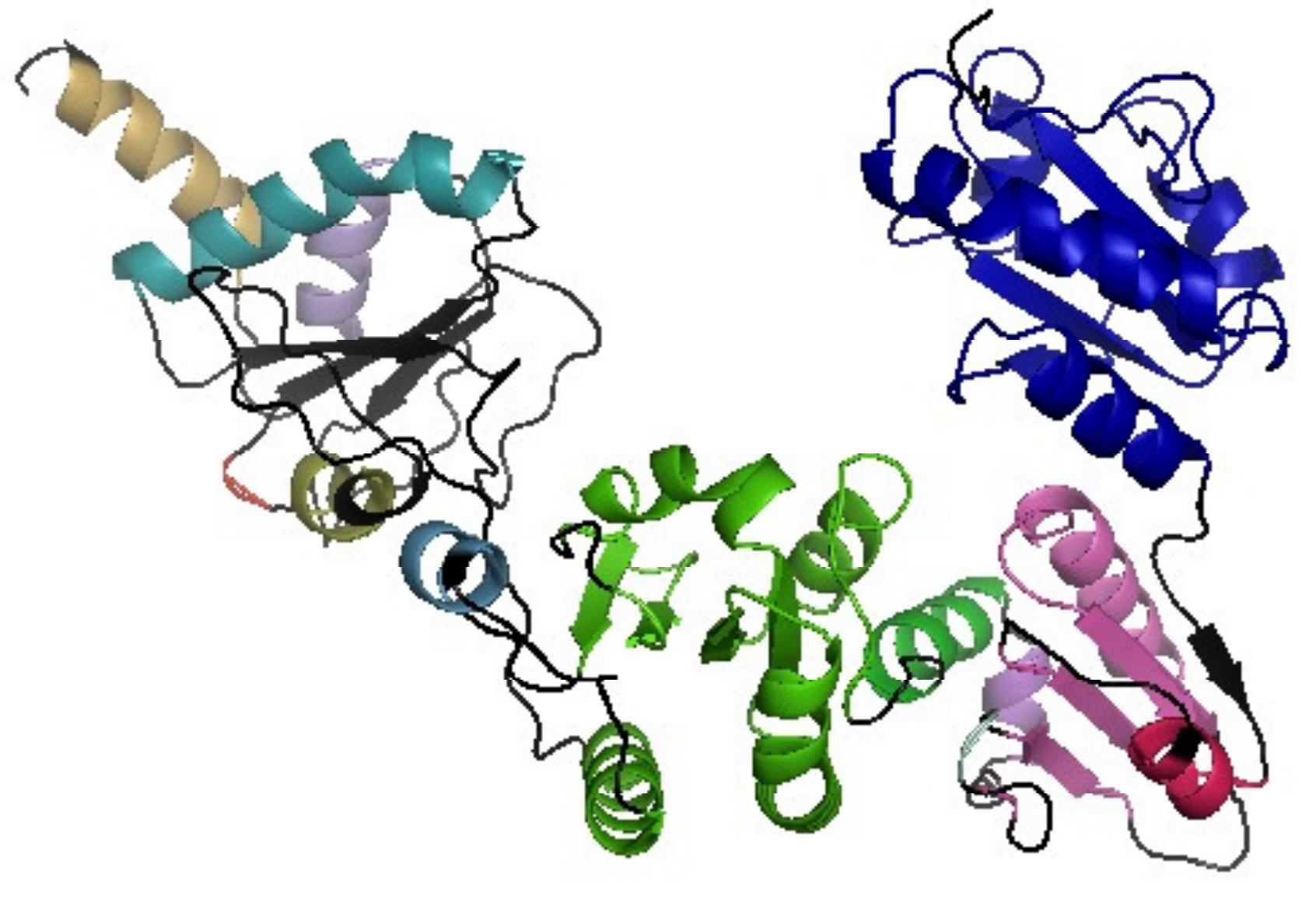}
(c)\includegraphics[width=0.45\textwidth]{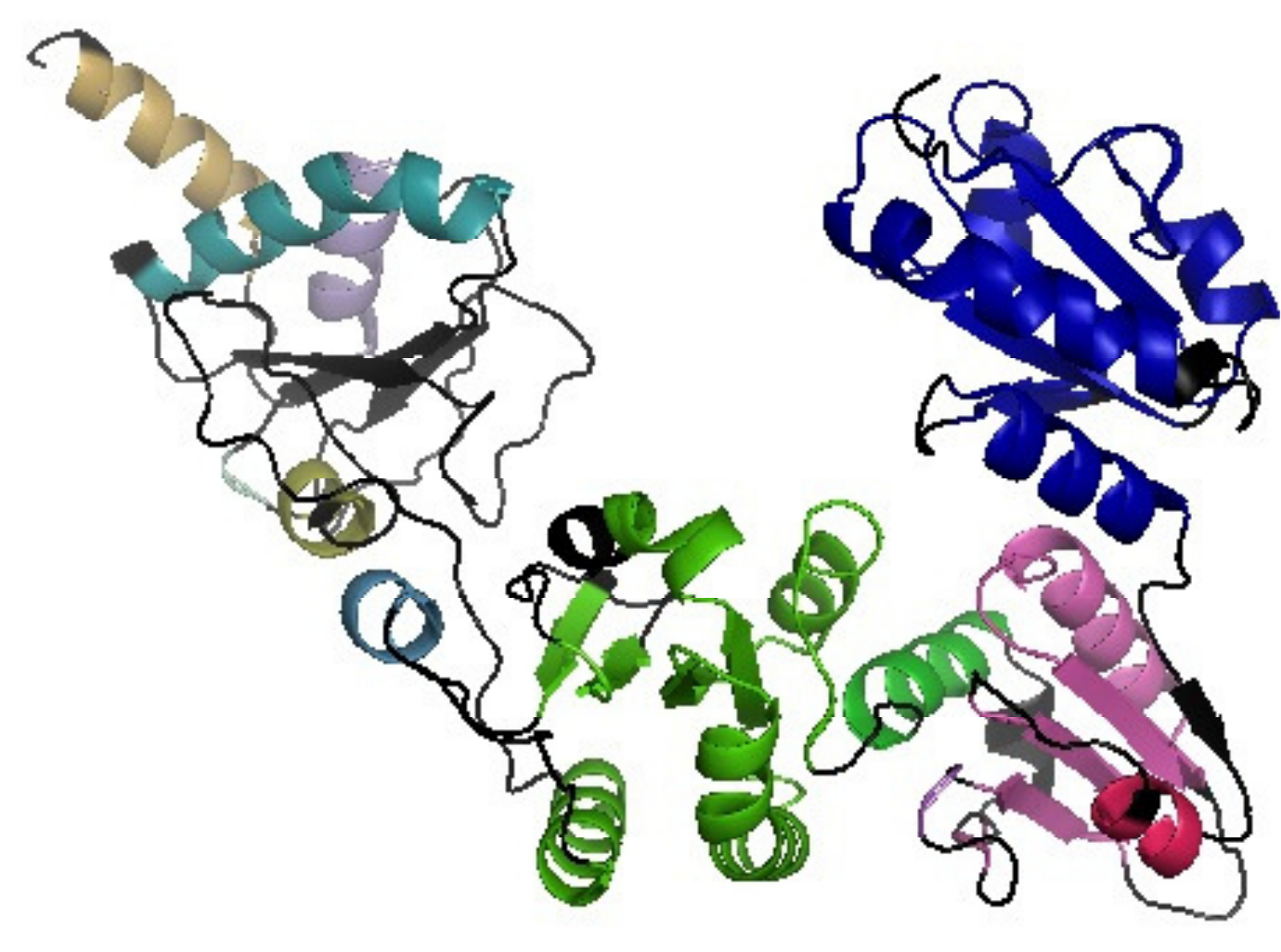}
(d)\includegraphics[width=0.45\textwidth]{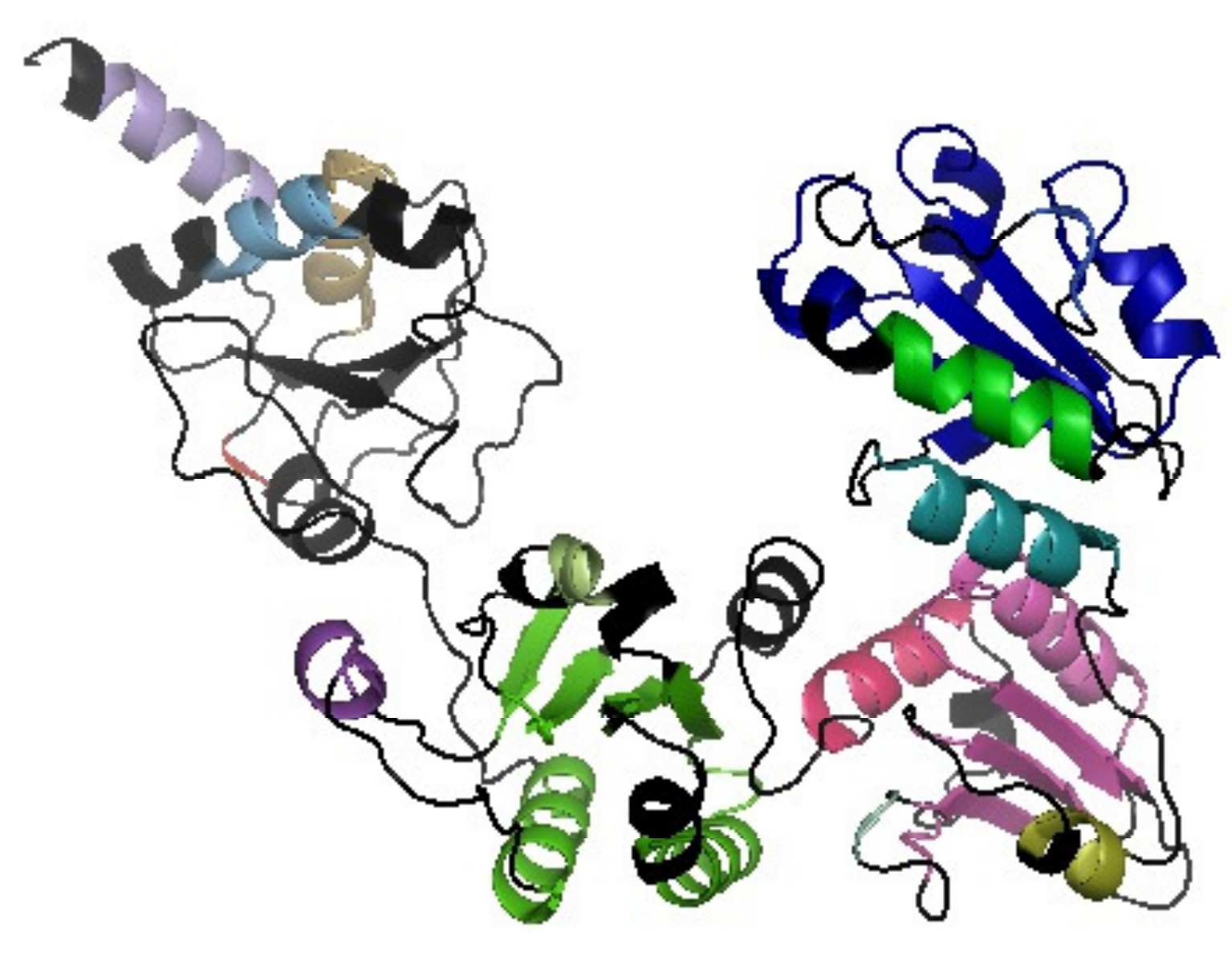}
\caption{Tertiary structure of yeast PDI (2B5E). Colouring of the tertiary
structure is defined as in Figure \ref{3D_1BPI_Ec} but with cutoff energies (a) $E_{\rm cut}$ =
$-0.015$ kcal/mol, (b) $E_{\rm cut}$ = $-0.522$ kcal/mol, (c) $E_{\rm cut}$ = $-0.885$ kcal/mol and
(d) $E_{\rm cut}$ = $-1.412$ kcal/mol. Note that colors are assigned according to cluster size which
changes depending on the cutoff energy.}
\label{3D_2B5E_Ec}
\end{figure}

\begin{figure}
\centering
(a)\includegraphics[width=0.45\textwidth]{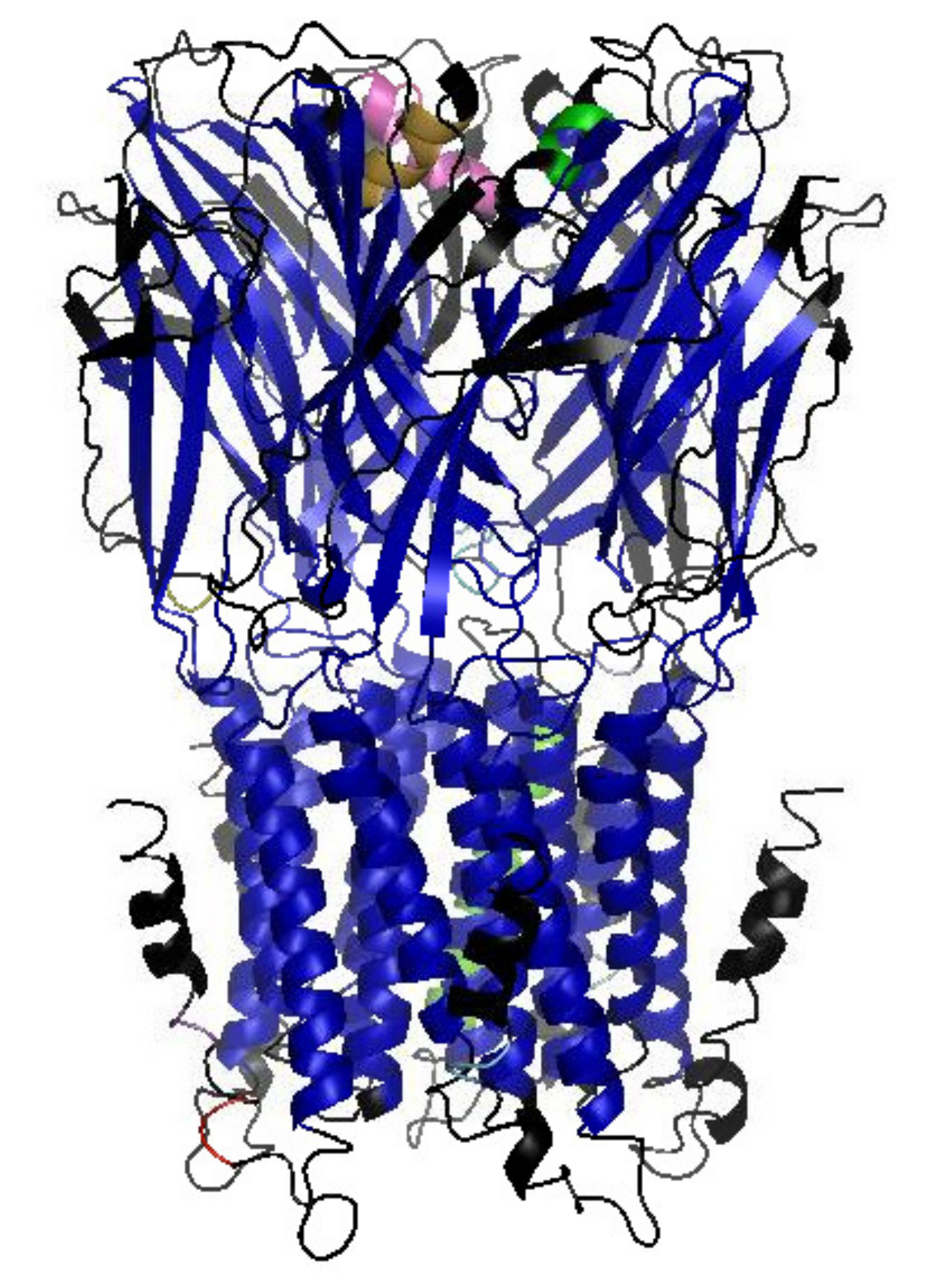}
(b)\includegraphics[width=0.45\textwidth]{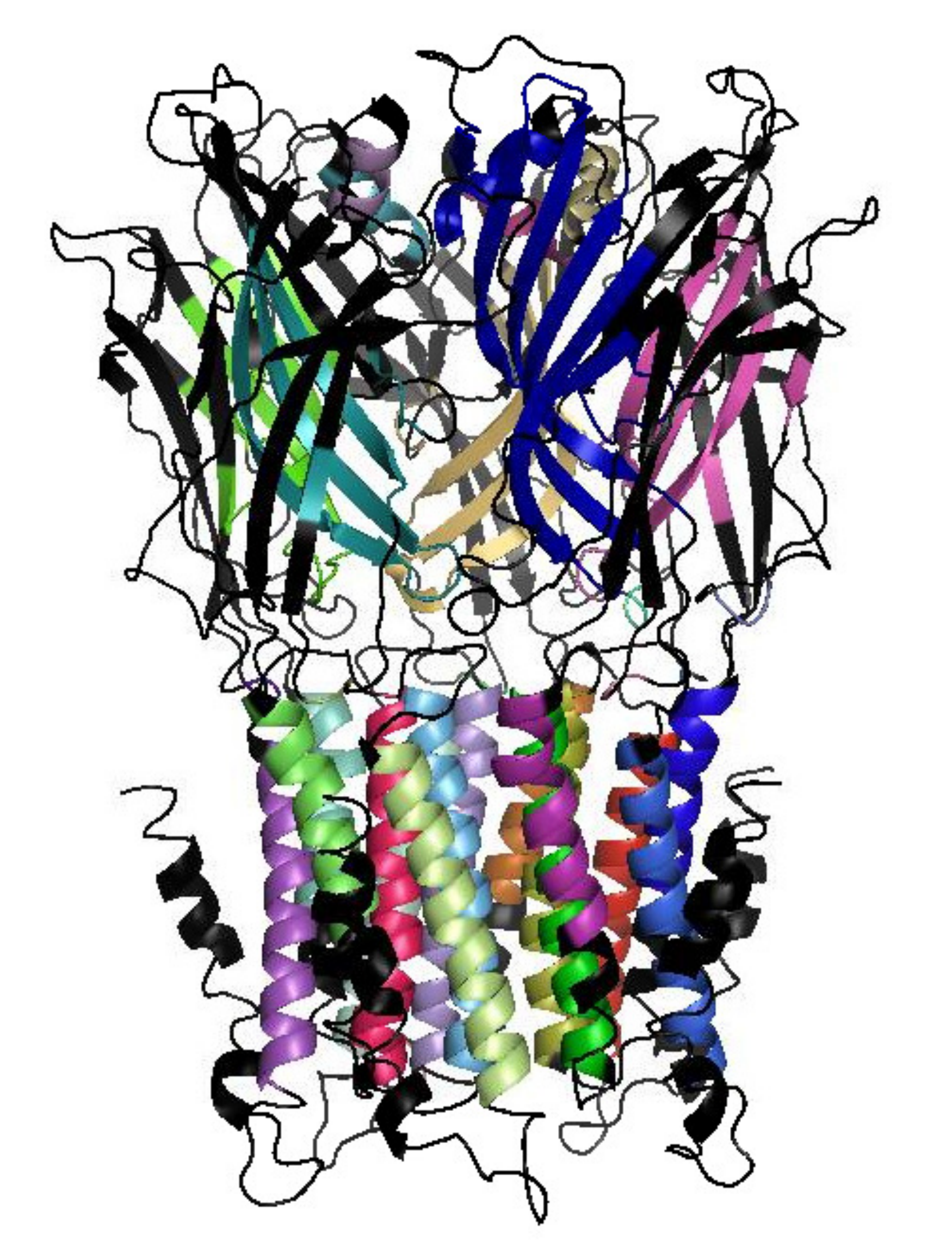}
\caption{Tertiary structure of pLGIC (2VL0). Colouring of the tertiary structure is defined as in
Figure \ref{3D_1BPI_Ec} but with cutoff energies (a) $E_{\rm cut}$ = $-0.4$ kcal/mol and (b) $E_{\rm
cut}$ = $-0.5$ kcal/mol. Note that the protein appears to be rigid for $E_{\rm cut}$ = $-0.4$
kcal/mol and that there is a switch-like first order rigidity transition at $E_{\rm cut}$ = $-0.5$
kcal/mol which reveals the most flexible parts of the secondary
structure which allow mobility.}
\label{3D_2VL0_Ec}
\end{figure}

\begin{figure}[tb]
\centering
(a)\includegraphics[width=0.45\columnwidth]{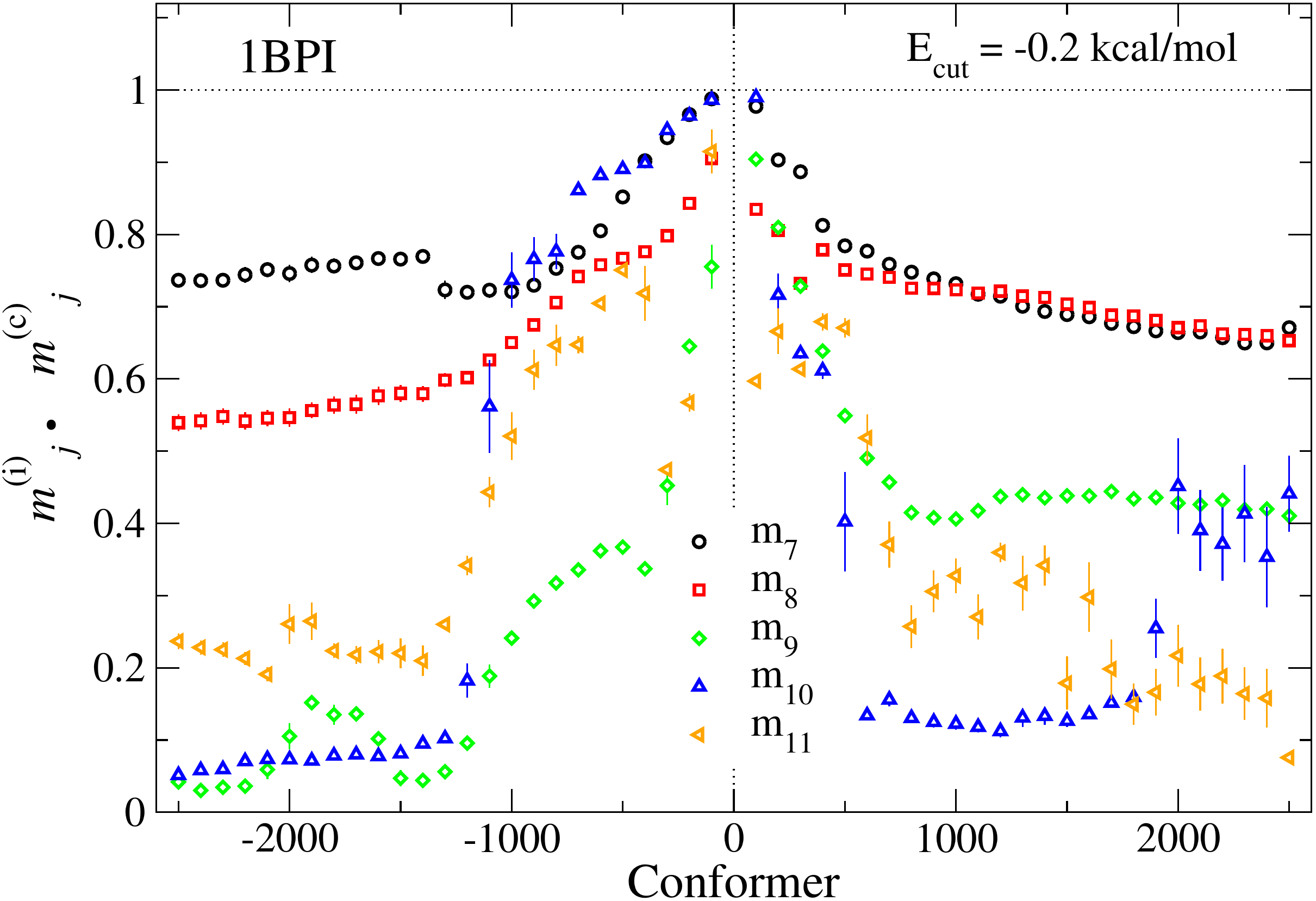}
(b)\includegraphics[width=0.45\columnwidth]{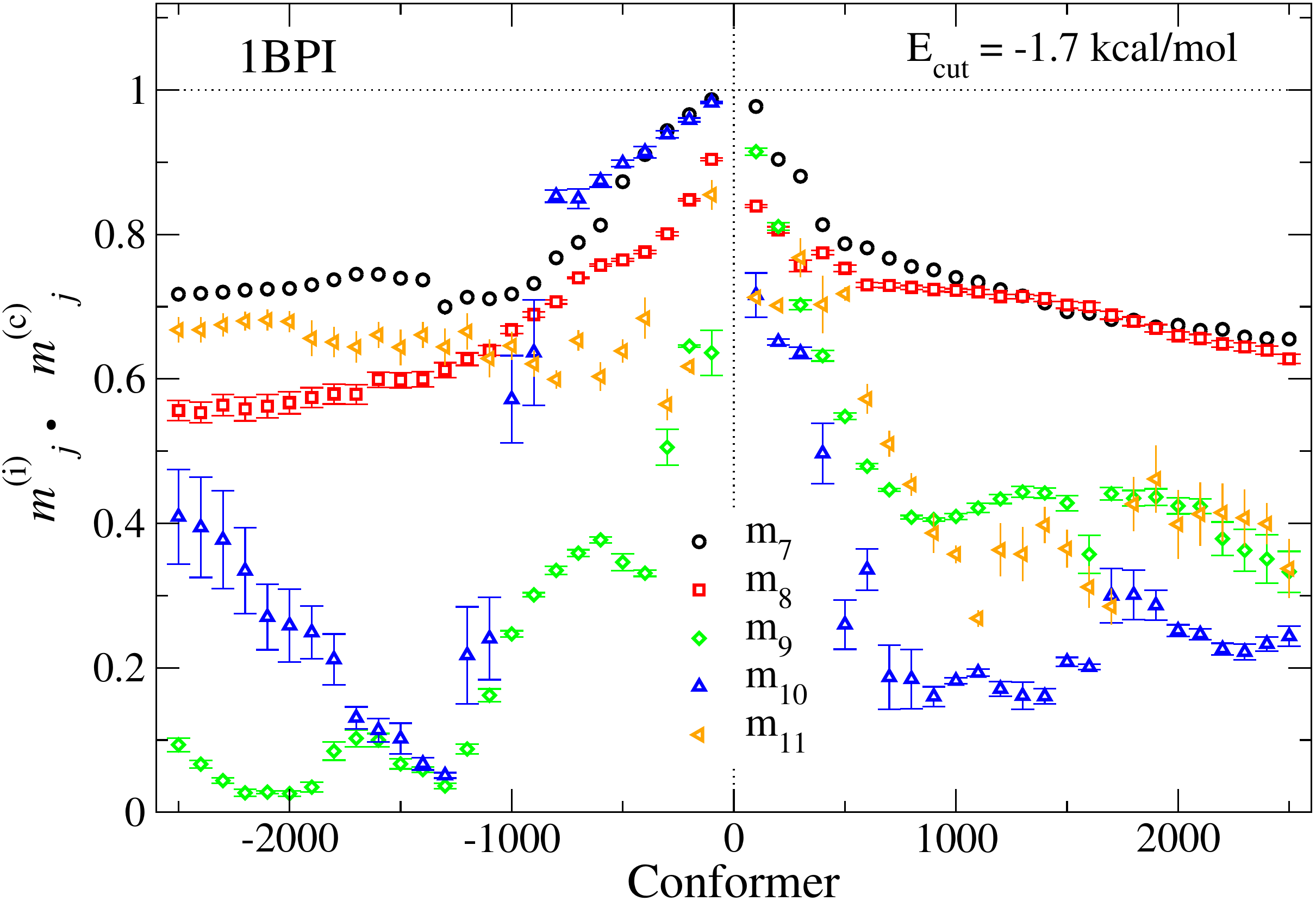}
(c)\includegraphics[width=0.45\columnwidth]{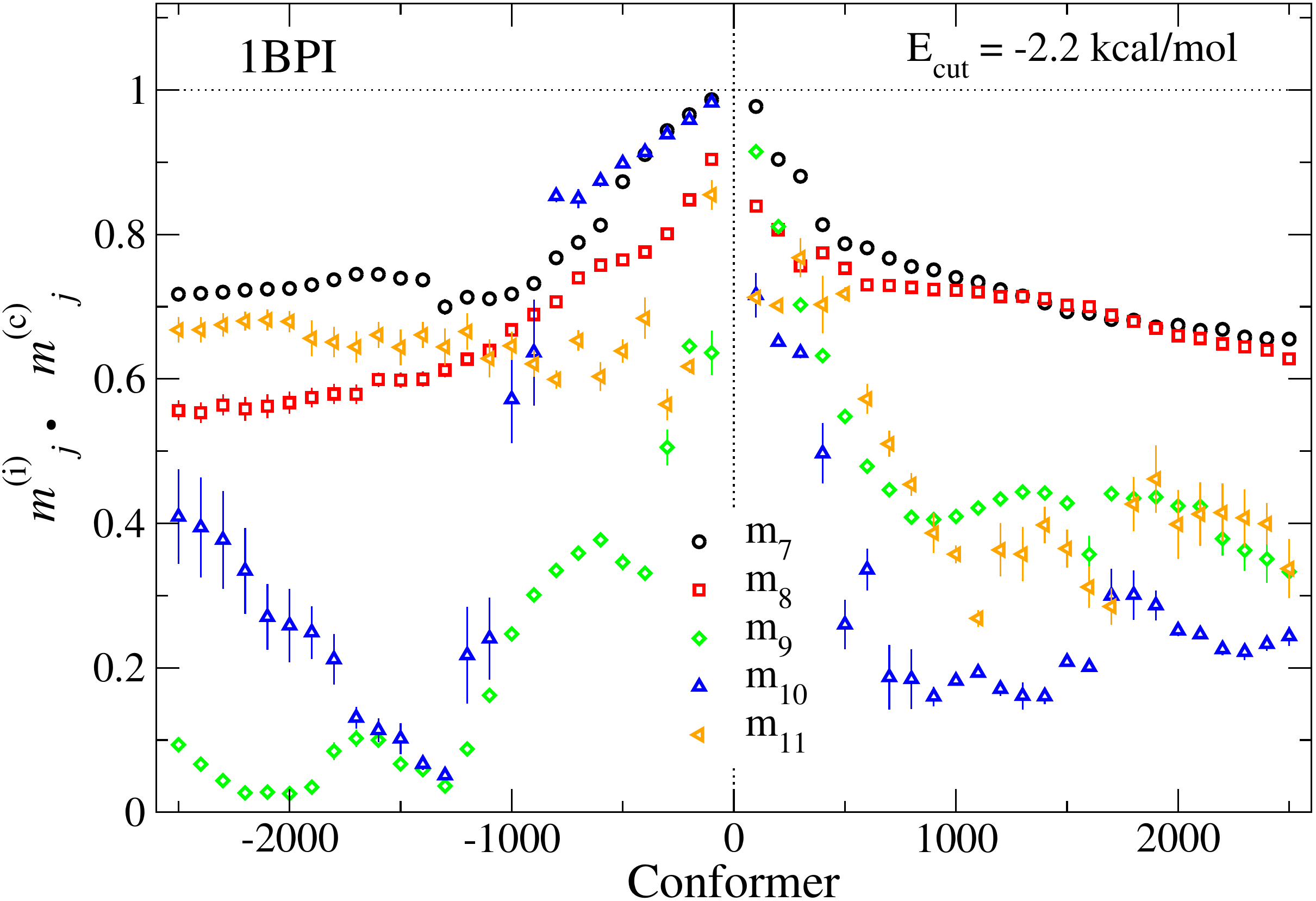}
\caption{\label{Fig:dot1BPI} Dot product graph for BPTI (1BPI). The dot product $m_{j}^{(i)} \cdot m_{j}^{(c)}$ between an
initial starting mode $m_{j}^{(i)} $ and its current mode $m_{j}^{(c)}$, $j=7, \ldots, 11$ as the
initial structure is projected along the
initial mode. The current modes, $m_{j}^{(c)}$, are obtained from performing normal mode analysis on
the current conformations as the initial structure is projected along an initial mode $m_{j}^{(i)}
$. For clarity, dot products for only $25$ conformations of
each direction of motion are shown.
The evolution of the dot product along the conformations is reported for
different cutoff energies, which for BPTI are (a) $E_{\rm cut} = -0.2$ kcal/mol, (b) $E_{\rm cut}= -1.7$ kcal/mol and (c) $E_{\rm cut} = -2.2$ kcal/mol. The horizontal and vertical dotted lines denote the largest possible value of $m_{j}^{(i)} \cdot m_{j}^{(c)}$ and the zero on the conformer axis, respectively. }
\end{figure}
\begin{figure}[tb]
\centering
(a)\includegraphics[width=0.45\columnwidth]{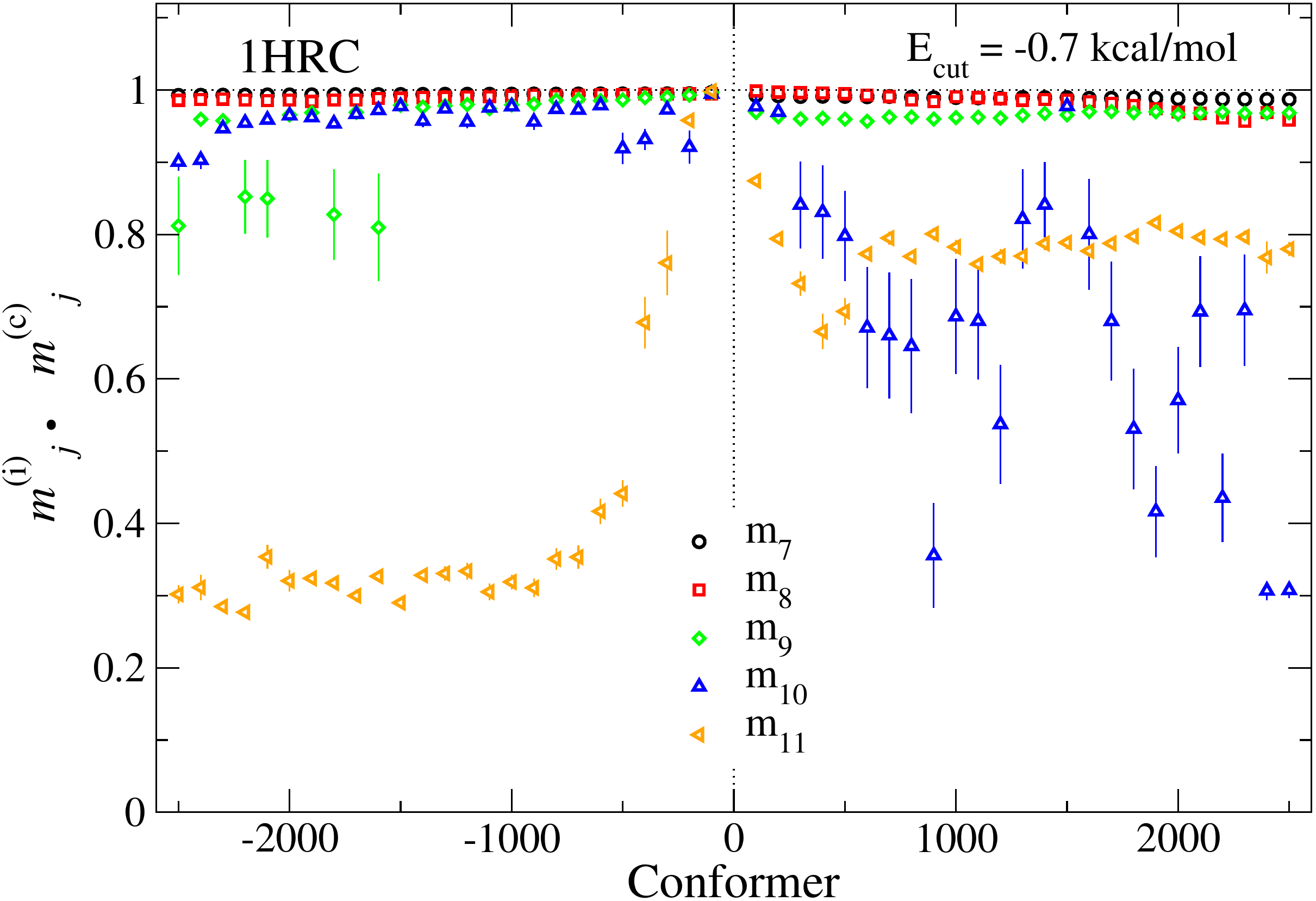}
(b)\includegraphics[width=0.45\columnwidth]{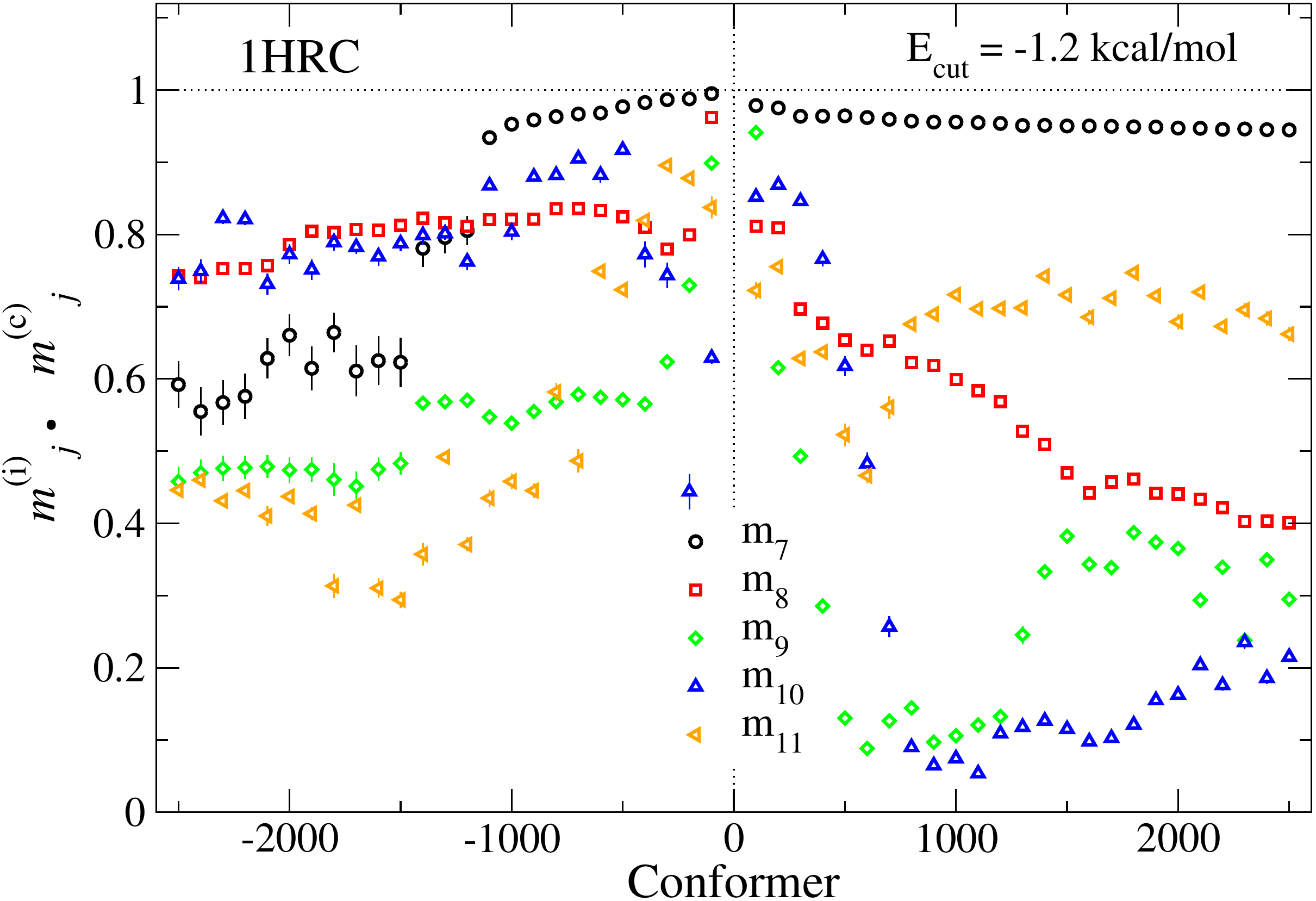}

\caption{\label{Fig:dot1HRC}Dot product graph for cytochrome-c (1HRC) as described in
Figure \ref{Fig:dot1BPI} but with $E_{\rm cut}$ values of (a)
$-0.7$ kcal/mol and (b) $-1.2$ kcal/mol.}
\end{figure}
\begin{figure}[tb]
\centering
(a)\includegraphics[width=0.45\columnwidth]{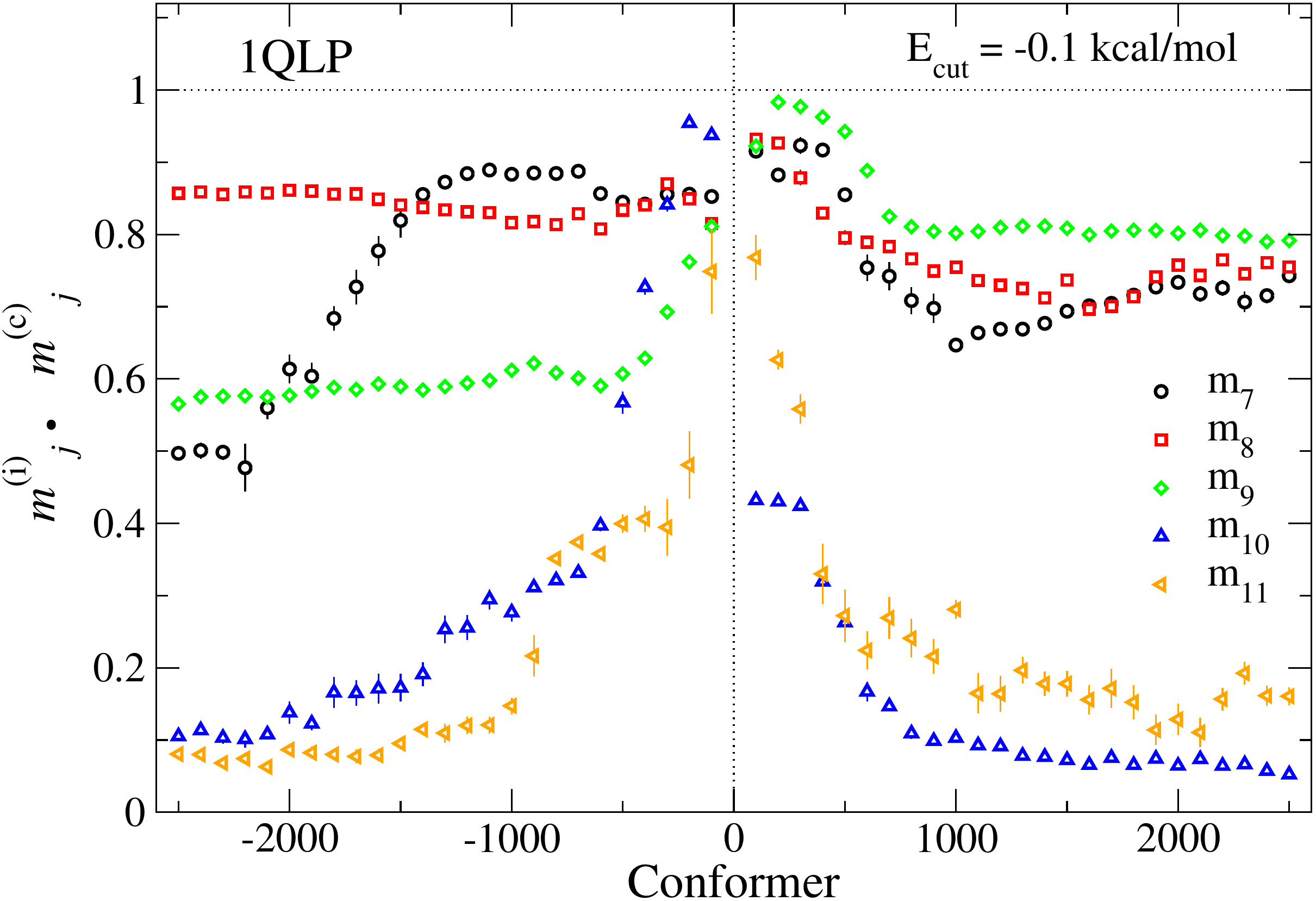}
(b)\includegraphics[width=0.45\columnwidth]{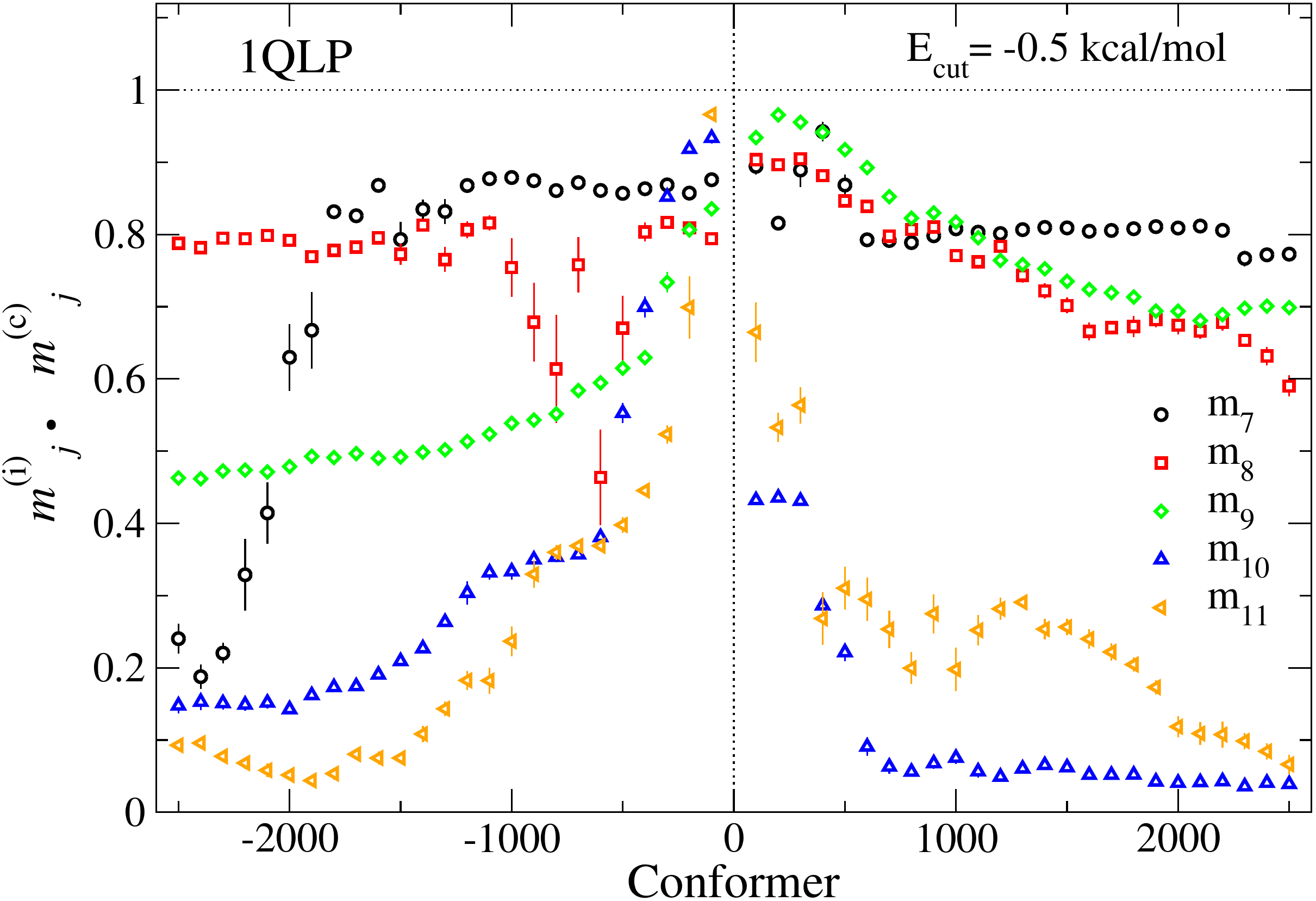}
(c)\includegraphics[width=0.45\columnwidth]{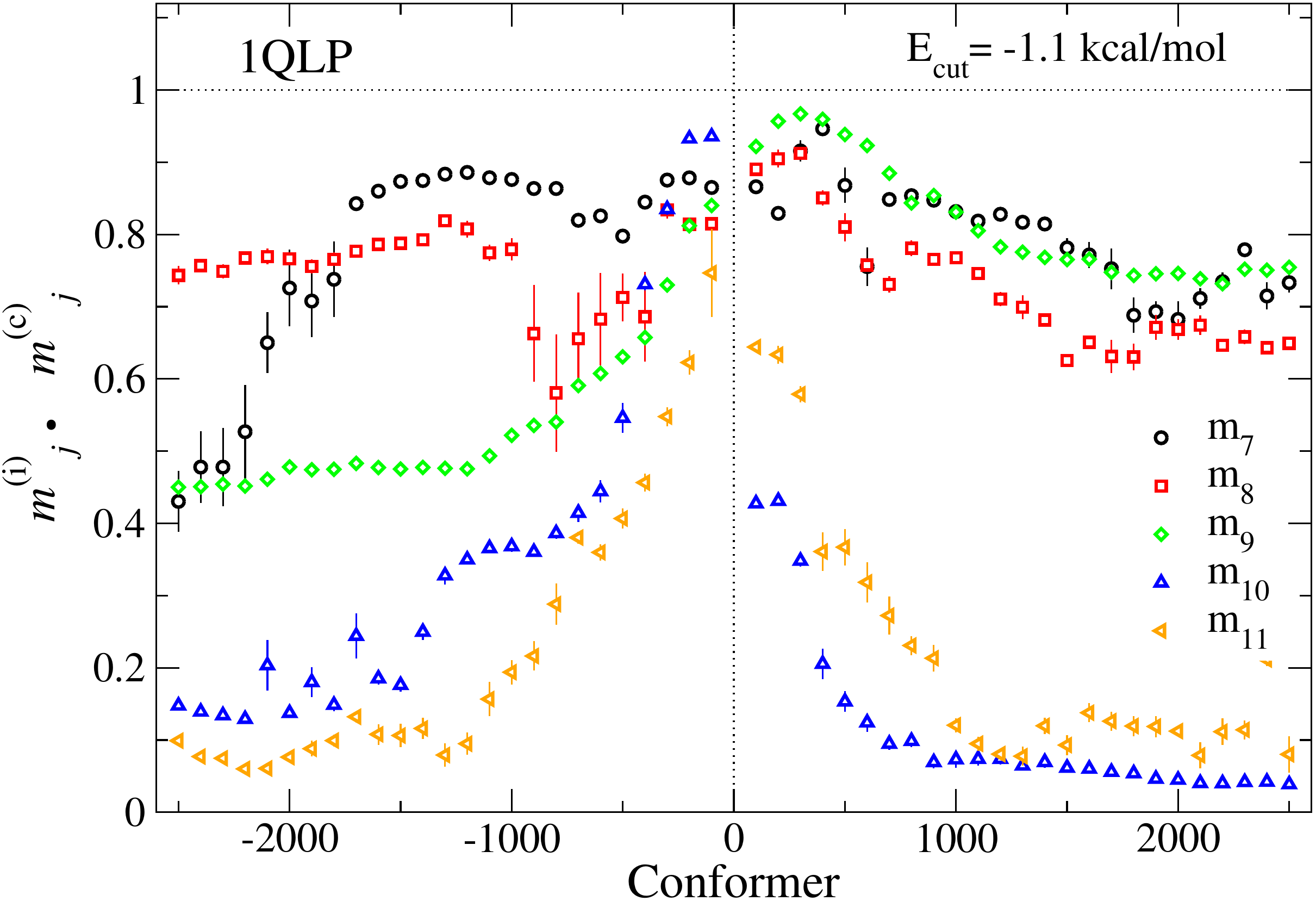}
\caption{\label{Fig:dot1QLP} Dot product graph for $\alpha$1-antitrypsin (1QLP) as
described in Figure \ref{Fig:dot1BPI} but with $E_{\rm cut}$ values of (a) $E_{\rm
cut} = -0.1$
kcal/mol, (b) $E_{\rm cut} = -0.5$ kcal/mol and (c) $E_{\rm cut} = -1.1$ kcal/mol.}
\end{figure}

\begin{figure}[tb]
\centering
(a)\includegraphics[width=0.45\columnwidth]{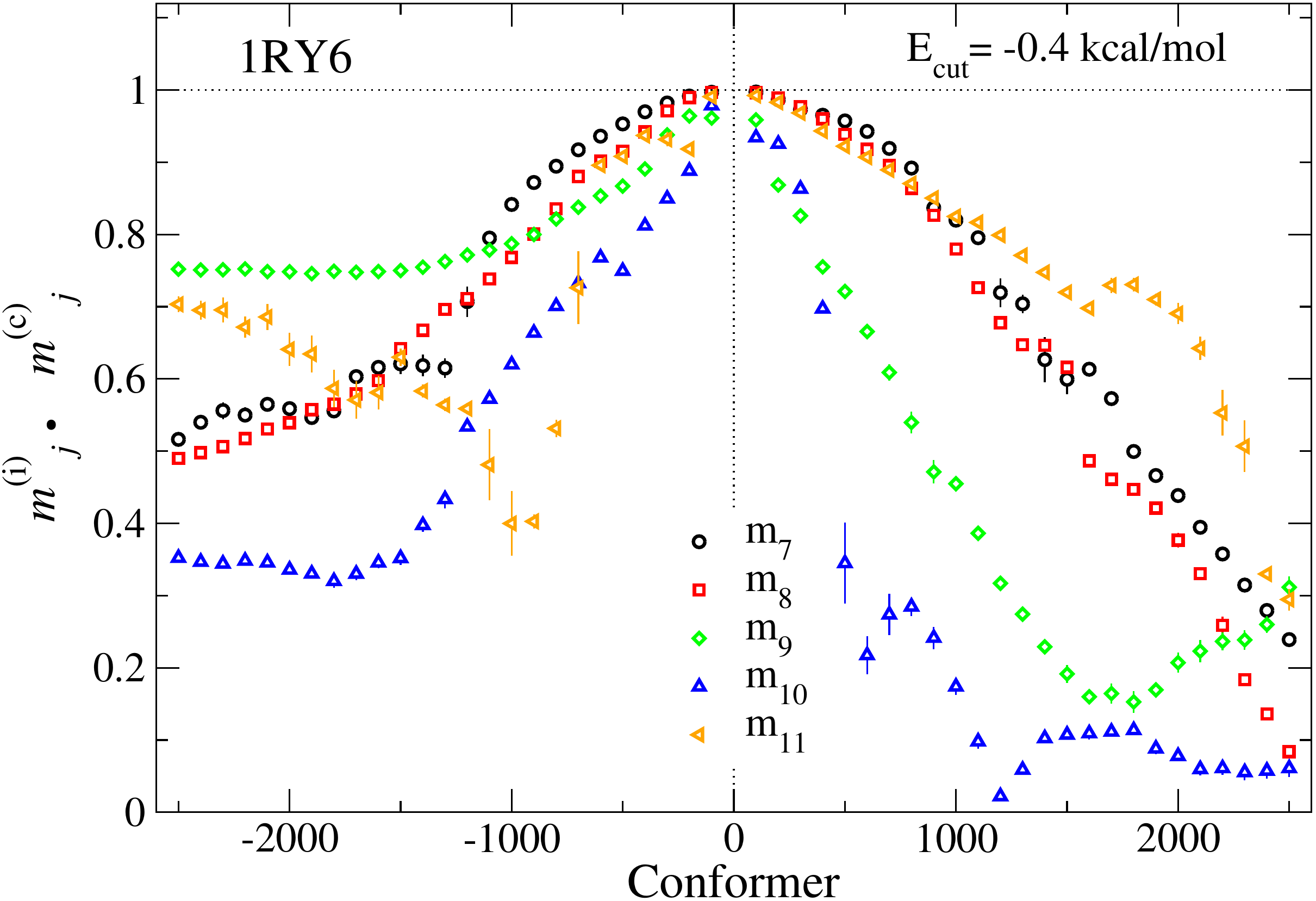}
(b)\includegraphics[width=0.45\columnwidth]{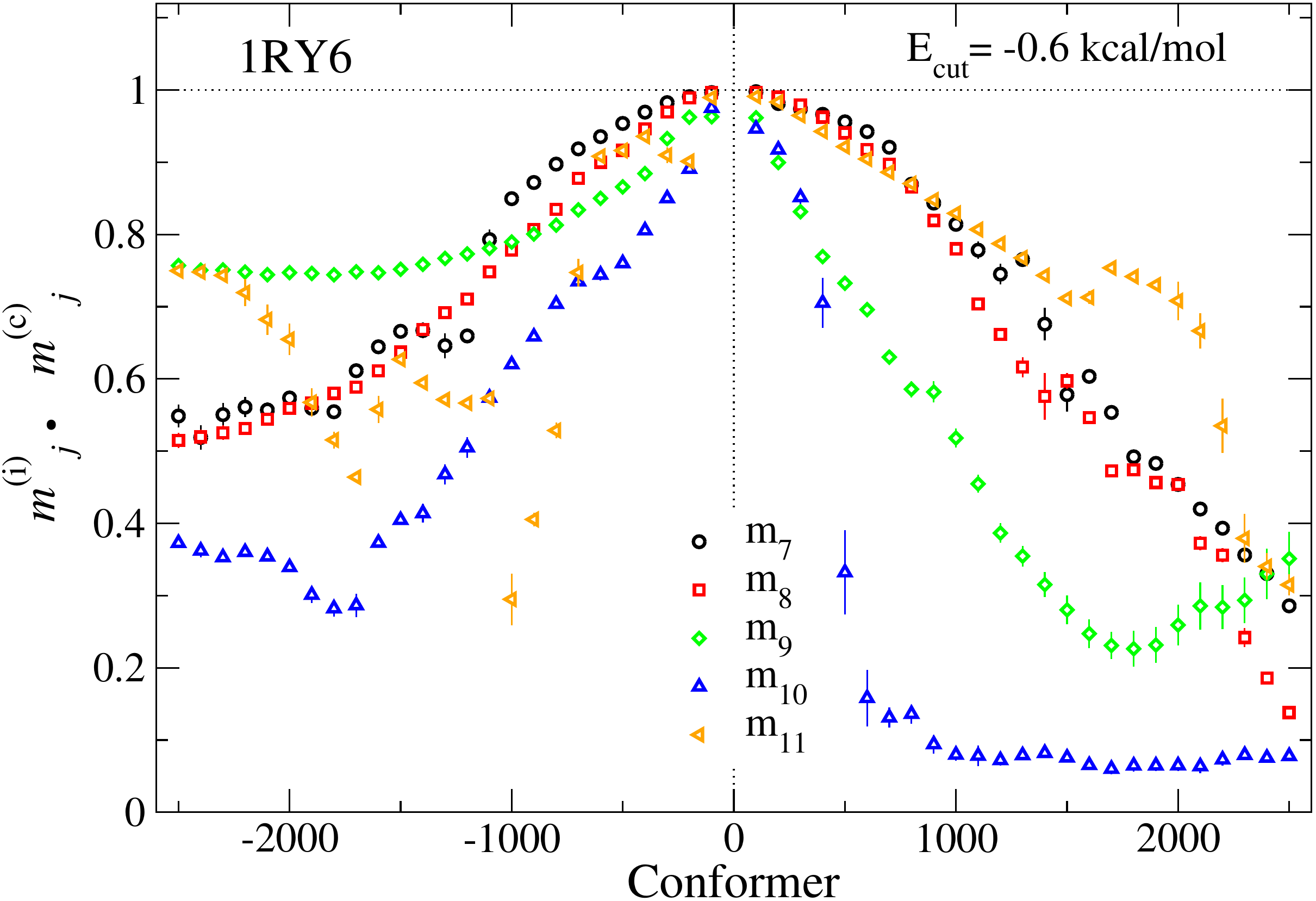}\\
(c)\includegraphics[width=0.45\columnwidth]{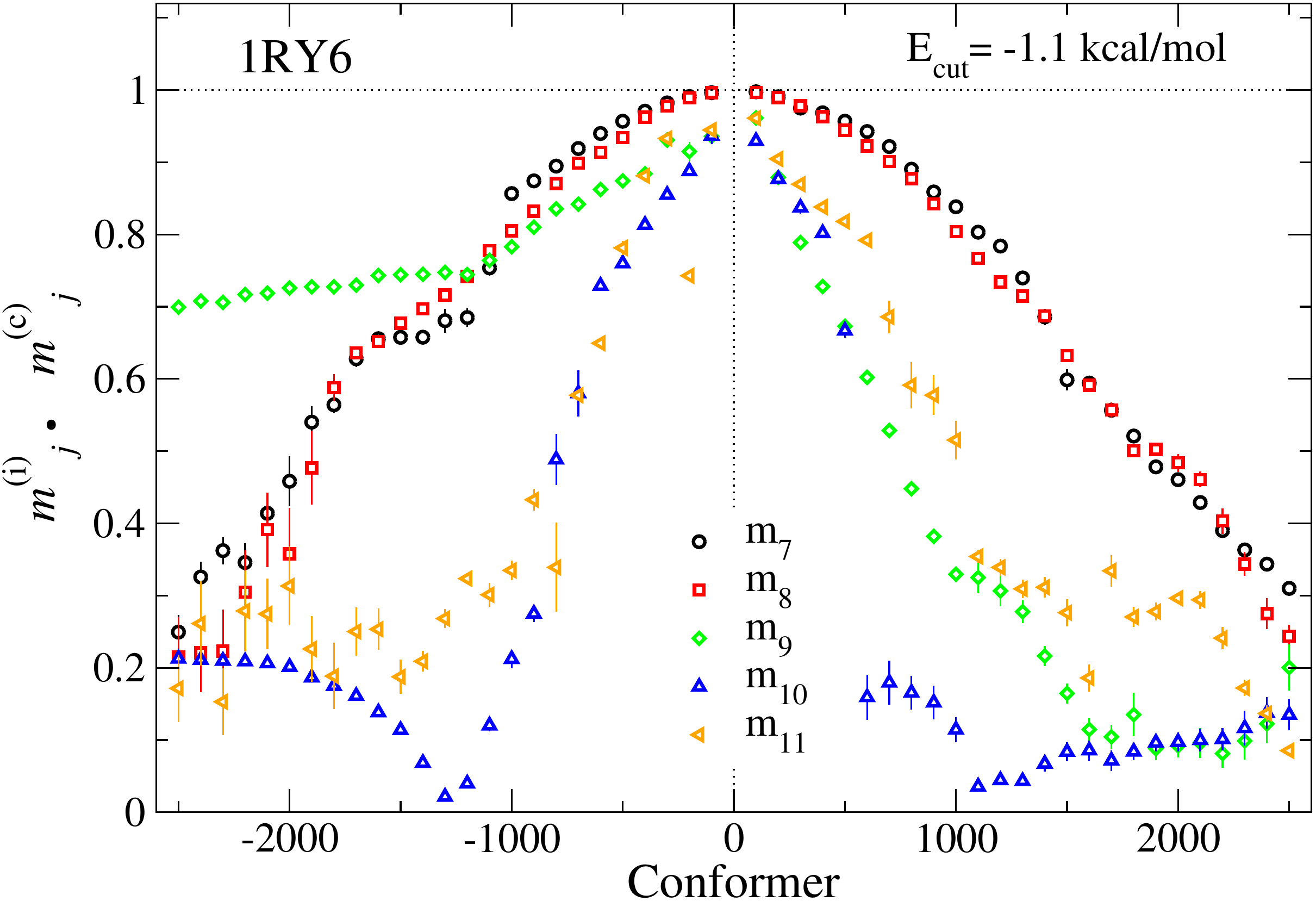}
\caption{\label{Fig:dot1RY6}Dot product graph for internal kinesin motor domain (1RY6) as
described in Figure \ref{Fig:dot1BPI} but with $E_{\rm cut}$ values of (a) $-0.4$ kcal/mol (b)
$-0.6$ kcal/mol and (c) $-1.1$ kcal/mol.}
\end{figure}

\begin{figure}[tb]
\centering

(a)\includegraphics[width=0.45\columnwidth]{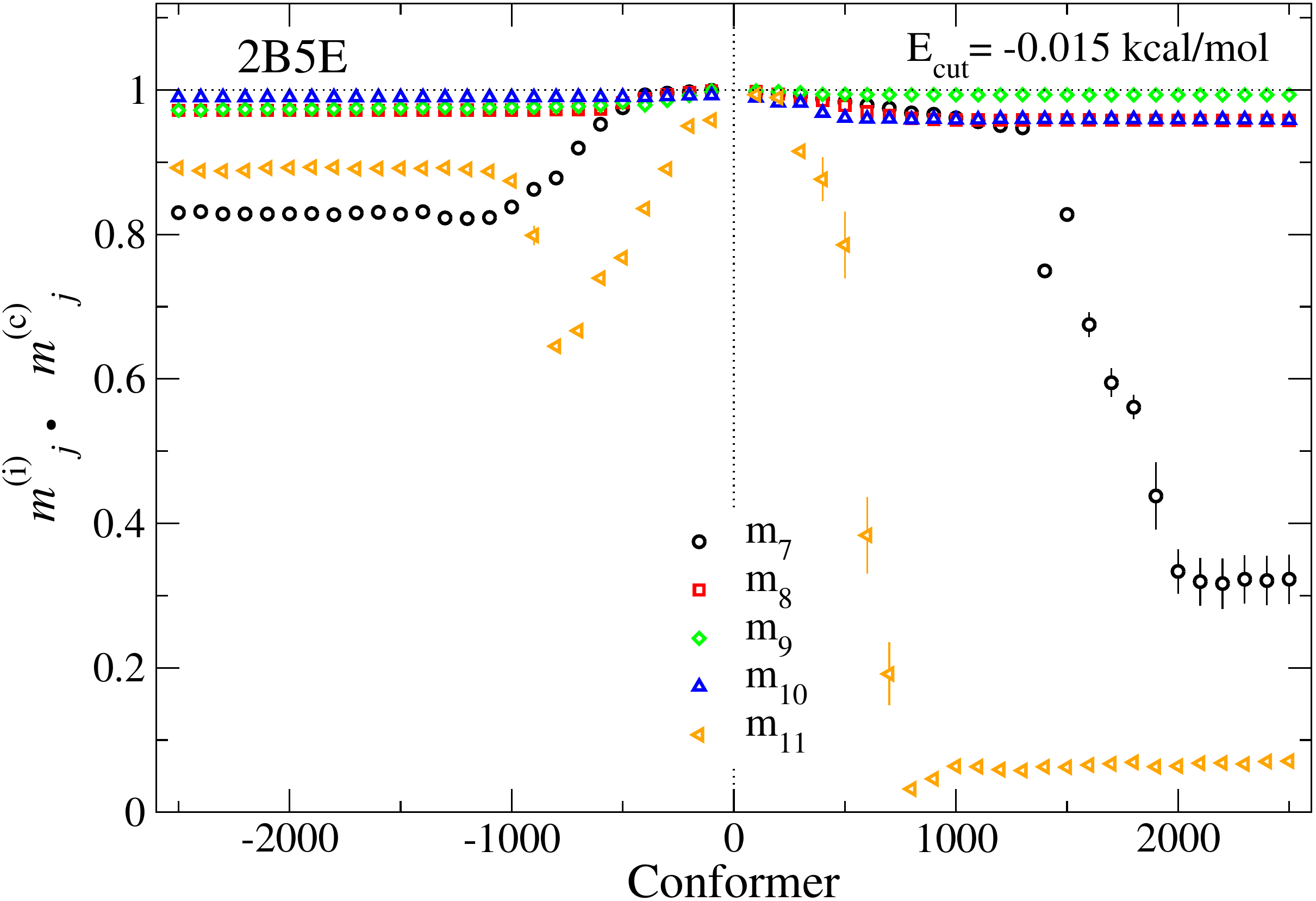}
(b)\includegraphics[width=0.45\columnwidth]{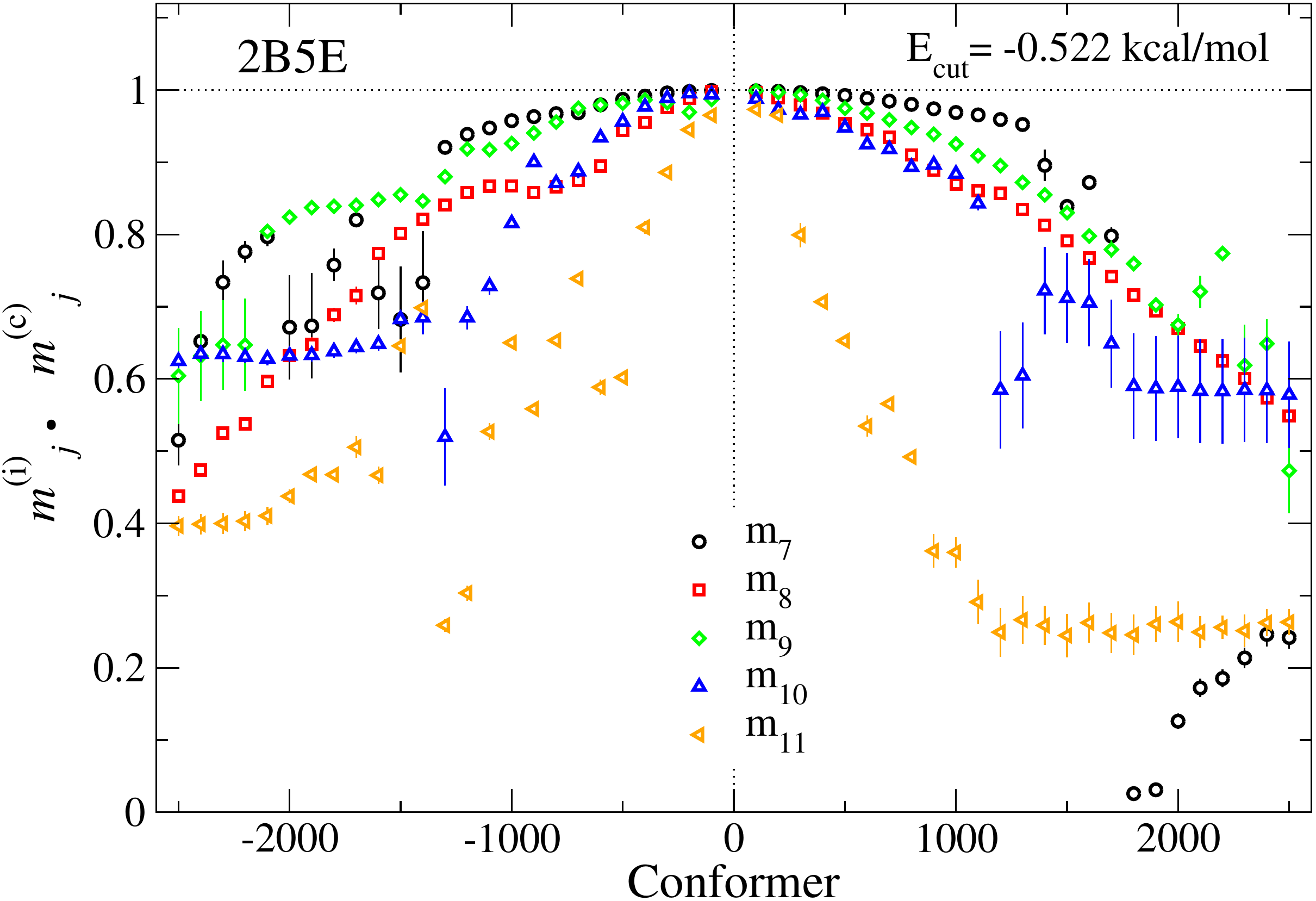}\\
(c)\includegraphics[width=0.45\columnwidth]{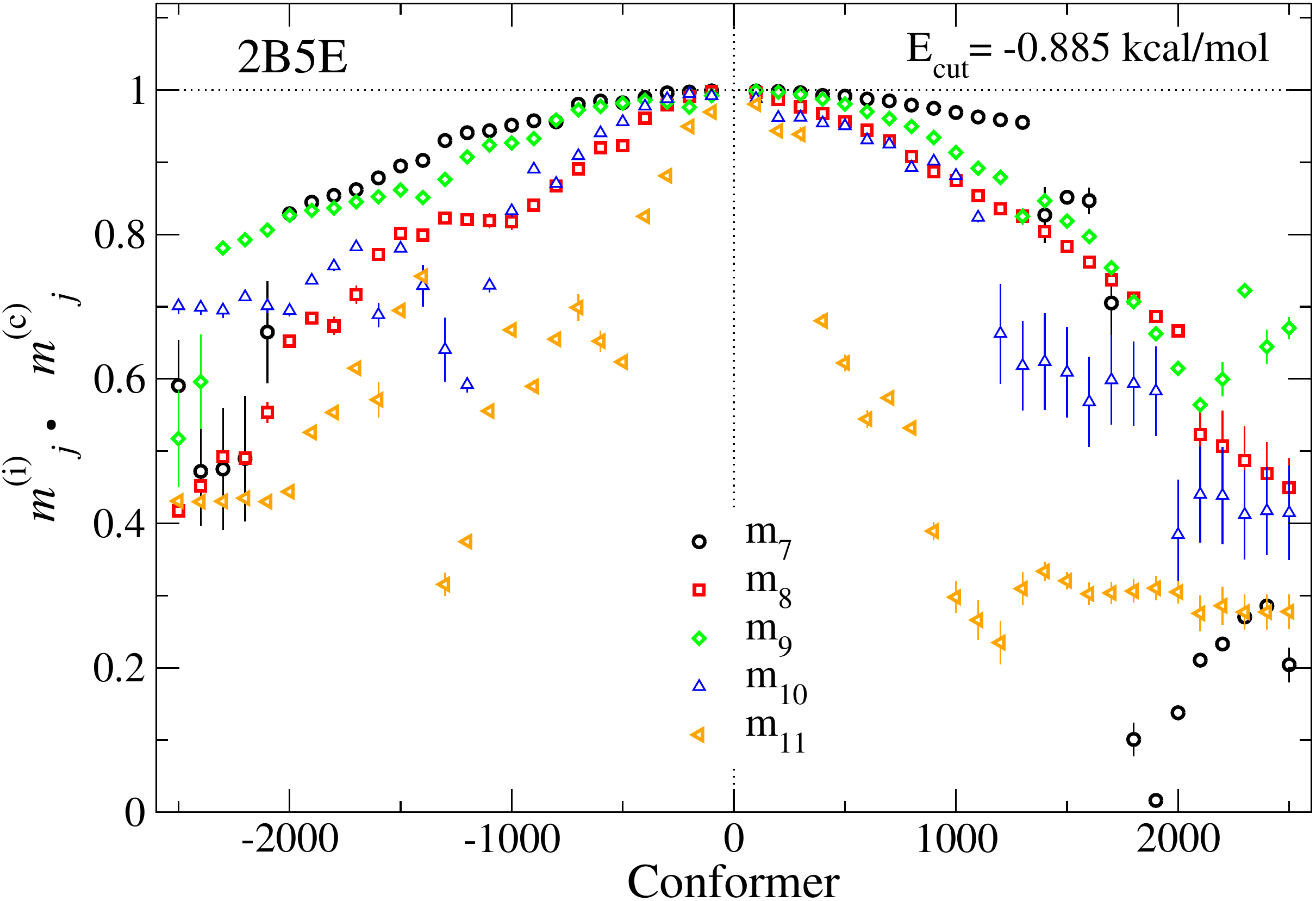}
(d)\includegraphics[width=0.45\columnwidth]{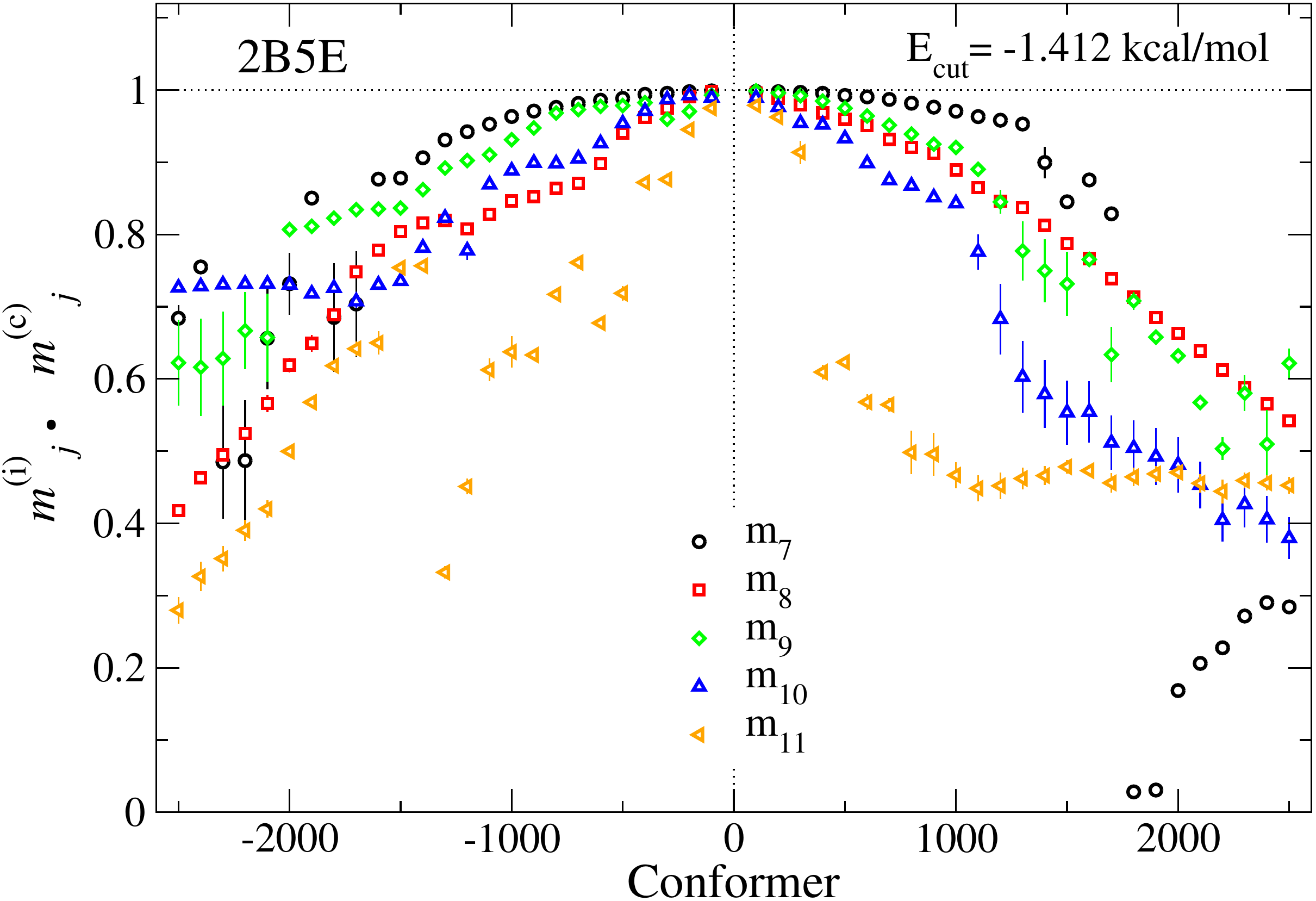}

\caption{\label{Fig:dot2B5E}Dot product graph for yeast PDI (2B5E) as described in
Figure \ref{Fig:dot1BPI} but with $E_{\rm cut}$ values of (a) $E_{\rm cut} = -0.015$ kcal/mol,
(b) $E_{\rm cut} = -0.522$ kcal/mol, (c) $E_{\rm cut} = -0.885$ kcal/mol and (d) $E_{\rm cut} =
-1.412$ kcal/mol.}
\end{figure}

\begin{figure}[tbh]
\centering
(a)\includegraphics[width=0.45\columnwidth]{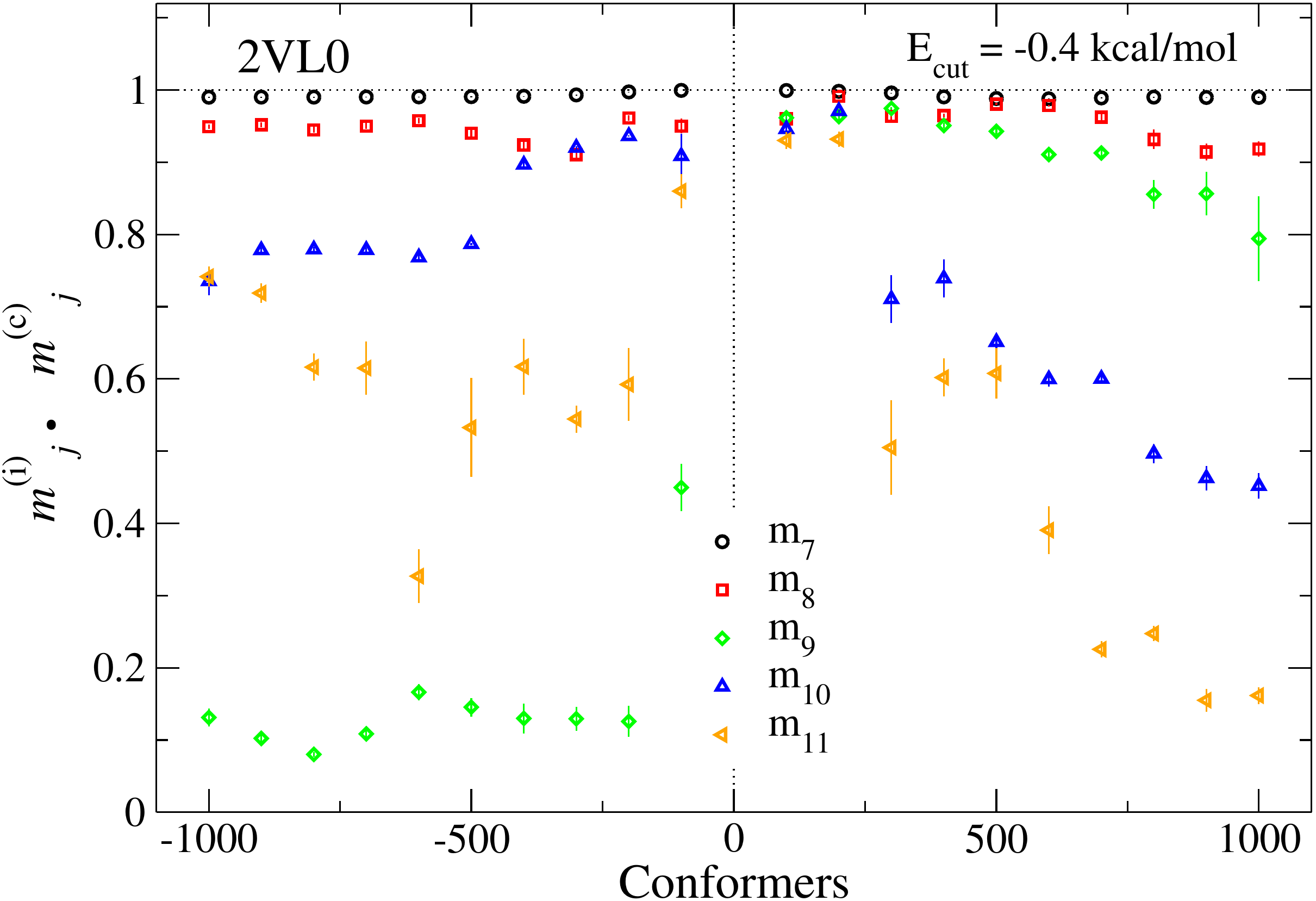}
(b)\includegraphics[width=0.45\columnwidth]{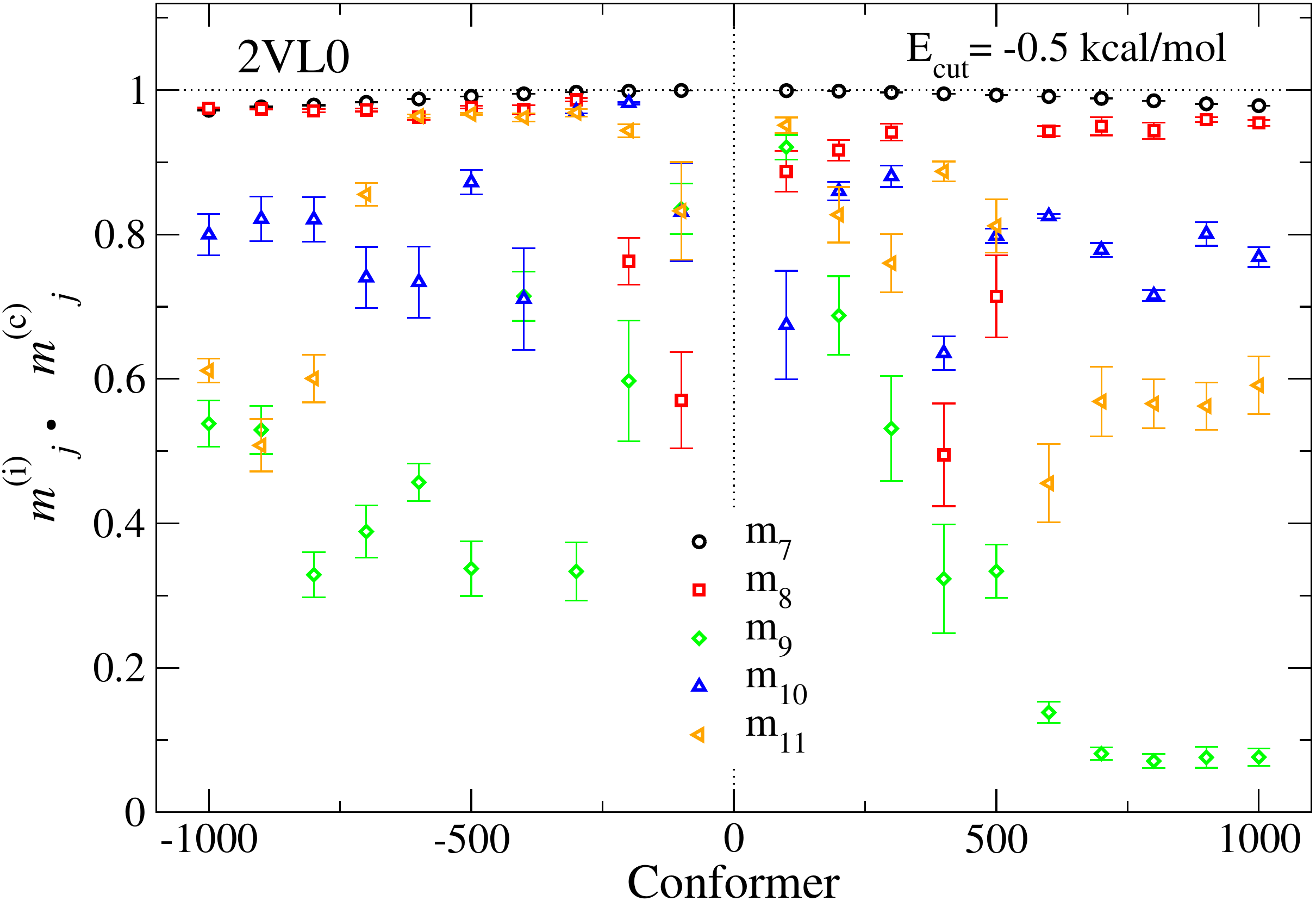}

\caption{Dot product graph for a ligand gated ion channel protein (2VL0) as described in
Figure \ref{Fig:dot1BPI}but with $E_{\rm cut}$ values of (a) $E_{\rm cut} = -0.4$ kcal/mol and
(b) at $E_{\rm cut} = -0.5$ kcal/mol.}
\label{Fig:dot2VL0}
\end{figure}


\begin{thebibliography}{10}

\bibitem{HenK07}
Henzler-Wildman K and Kern D.
\newblock Dynamic personalities of proteins.
\newblock {\em Nature}, 450(13):964--972, 2007.

\bibitem{ShaMLS10}
Shaw DE, Maragakis P, Lindorff-Larsen K, Piana S, Dror RO, Eastwood MP, Banks
  JA, Jumper JM, Salmon JK, Shan Y, and Wriggers W.
\newblock Atomic-level characterization of the structural dynamics of proteins.
\newblock {\em Science}, 330:341--346, 2010.

\bibitem{GieAET06}
Giepmans BNG, Adams SR, Ellison MH, and Tsien RY.
\newblock Review- the fluorescent toolbox for assessing protein location and
  function.
\newblock {\em Science}, 312:217--224, 2006.

\bibitem{RusLPS09}
Russell D, Lasker K, Phillips J, Schneidman-Duhovny D, Velazquez-Muriel JA, and
  Sali A.
\newblock The structural dynamics of macromolecular processes.
\newblock {\em Curr. Opin. Cell Biol.}, 21:97--108, 2009.

\bibitem{Cle08}
Clementi C.
\newblock Coarse-grained models of protein folding: toy models or predictive
  tools?
\newblock {\em Curr Opin Struct Biol}, 18:10--15, 2008.

\bibitem{YalB07}
Yaliraki SN and Barahona M.
\newblock Chemistry across scales: from molecules to cells.
\newblock {\em Phil Trans R Soc A Math Phys Eng Sci}, 365:2921--2934, 2007.

\bibitem{SheBS08}
P~Sherwood, B~R Brooks, and M~S~P Sansom.
\newblock Multiscale methods for macromolecular simulations.
\newblock {\em Curr Opin Struct Biol}, 18:630--640, 2008.

\bibitem{JacRKT01}
Jacobs DJ, Rader AJ, Kuhn LA, and Thorpe MF.
\newblock Protein flexibility predictions using graph theory.
\newblock {\em Prot: Struct. Func. Gen.}, 44:150--165, 2001.

\bibitem{WelJR09}
Wells SA, Jimenez-Roldan JE, and R{\"o}mer RA.
\newblock Comparative analysis of rigidity across protein families.
\newblock {\em Phys. Biol.}, 6(4):046005--11, 2009.

\bibitem{WelMHT05}
Wells SA, Menor S, Hespenheide BM, and Thorpe MF.
\newblock Constrained geometric simulation of diffusive motion in proteins.
\newblock {\em Phys. Biol.}, 2:S127--S136, 2005.

\bibitem{JolWHT06}
Jolley CC, Wells SA, Hespenheide BM, Thorpe MF, and Fromme P.
\newblock Docking of photosystem {I} subunit c using a constrained geometric
  simulation.
\newblock {\em J. Am. Chem. Soc.}, 128(27):8803--8812, 2006.

\bibitem{JolWFT08}
Jolley CC, Wells SA, Fromme P, and Thorpe MF.
\newblock Fitting low-resolution cryo-{EM} maps of proteins using constrained
  geometric simulations.
\newblock {\em Biophys. J.}, 94:1613--1621, 2008.

\bibitem{SuhS04}
Suhre K and Sanejouand Y-H.
\newblock {ElN{\'e}mo}: a normal mode web server for protein movement analysis
  and the generation of templates for molecular replacement.
\newblock {\em Nucl. Acids Res. (Web Issue)}, 32:610--614, 2004.

\bibitem{SuhSB04}
Suhre K and Sanejouand Y-H.
\newblock On the potential of normal mode analysis for solving difficult
  molecular replacement problems.
\newblock {\em Acta Cryst D}, 60:796--799, 2004.

\bibitem{BahLBS10}
Bahar I, Lezon TR, Bakan A, and Shrivastava IH.
\newblock Normal mode analysis of biomolecular structures: functional
  mechanisms of membrane proteins.
\newblock {\em Chem Rev}, 110:1463--1497, 2010.

\bibitem{Ma05}
Ma~J.
\newblock Usefulness and limitations of normal mode analysis in modelling
  dynamics of biomolecular complexes.
\newblock {\em Structure}, 13:373--380, 2005.

\bibitem{RueCO07}
Rueda M, Chac\'on P, and Orozco M.
\newblock Thorough validation of protein normal mode analysis: A comparative
  study with essential dynamics.
\newblock {\em Structure}, 15:565--575, 2007.

\bibitem{NakMKH10}
Nakasako M, Maeno A, Kurimoto E, Harada T, Yamaguchi Y, Oka T, Takayama Y,
  Iwata, and Kato K.
\newblock Redox-dependent domain rearrangement of protein disulfide isomerase
  from a thermophilic fungus.
\newblock {\em Biochemistry}, 49:6953--6962, 2010.

\bibitem{DykS10}
Dykeman EC and Sankey OF.
\newblock Normal mode analysis and applications in biological physics.
\newblock {\em J. Phys. Condens. Matter}, 22:423202, 2010.

\bibitem{BerWFG00}
Berman HM, Westbrook J, Feng Z, Gilliland G, Bhat TN, Weissig H, Shindyalov IN,
  and Bourne PE.
\newblock The protein data bank.
\newblock {\em Nucl. Acids Res.}, 28:235--242, 2000.
\newblock http://www.rcsb.org.

\bibitem{WloWHS84}
Wlodawer A, Walter J, Huber R, and Sjolin L.
\newblock Structure of bovine pancreatic trypsin inhibitor: Results of joint
  neutron and {X}-ray refinement of crystal form {II}.
\newblock {\em J. Mol. Biol.}, 180(2):301--329, 1984.

\bibitem{AmiH88}
Amir D and Haas E.
\newblock Reduced bovine pancreatic trypsin inhibitor has a compact structure.
\newblock {\em Biochemistry}, 27(25):8889--8893, 1988.

\bibitem{ShiHTM04}
K~Shipley, M~Hekmat-Nejad, J~Turner, C~Moores, R~Anderson, R~Milligan,
  R~Sakowicz, and R~Fletterick.
\newblock Structure of a kinesin microtubule depolymerization machine.
\newblock {\em EMBO {J}ournal}, 23(7):1422--1432, 2004.

\bibitem{EllPDL00}
Elliott PR, Pei XY, Dafforn TR, and Lomas DA.
\newblock Topography of a 2.0 {$\AA$} structure of $\alpha$-1-antitrypsin
  reveals targets for rational drug design to prevent conformational disease.
\newblock {\em Prot. Sci.}, 9(7):1274--1281, 2000.

\bibitem{FreKR02}
Freedman RB, Klappa P, and Ruddock LW.
\newblock Protein disulfide isomerases exploit synergy between catalytic and
  specific binding domains.
\newblock {\em {EMBO} {J}ournal}, 15:136--140, 2002.

\bibitem{TiaKLS08}
G~Tian, F~Kober, U~Lewandrowski, A~Sickmann, W~J Lennarz, and H~Schindelin.
\newblock The catalytic activity of protein-disulfide isomerase requires a
  conformationally flexible molecule.
\newblock {\em J. Biol. Phys.}, 283:33630--33640, 2008.

\bibitem{TiaXNLS06}
G~Tian, S~Xiang, R~Noiva, WJ~Lennarz, and H~Schindelin.
\newblock The crystal structure of yeast protein disulfide isomerase suggests
  cooperativity between its active sites.
\newblock {\em Cell}, 124:61--73, 2006.

\bibitem{HilD08}
Hilf RJC and Dutzler R.
\newblock X-ray structure of a prokaryotic pentameric ligand-gated ion channel.
\newblock {\em Nature}, 452(7185):375--379, 2008.

\bibitem{WorLRR99}
Word JM, Lovell SC, Richardson JS, and Richardsonzhed DC.
\newblock Asparagine and glutamine: Using hydrogen atoms contacts in the choice
  of side-chain amide orientation.
\newblock {\em J. Mol. Biol.}, 285:1735--1747, 1999.

\bibitem{Del02}
DeLano WL.
\newblock The {PyMOL} molecular graphics system.
\newblock {\em http://www.pymol.org}, 2002.

\bibitem{Tir96}
Tirion MM.
\newblock Large amplitude elastic motions in proteins from single-parameter
  atomic analysis.
\newblock {\em Phys. Rev. Lett.}, 77:1905--1908, 1996.

\bibitem{ZheD03}
Zheng W and Doniach S.
\newblock A comparative study of motor-protein motions using a simple elastic
  network model.
\newblock {\em Proc. Nat. Acad. Sci.}, 100:13253--58, 2003.

\bibitem{CheLGL06}
Cheng X, Lu~B, Grant B, Law RJ, and McCammon JA.
\newblock Channel opening motion of $\alpha$7 nicotinic acetylcholine receptor
  as suggested by normal mode analysis.
\newblock {\em J. Mol. Biol.}, 355:310--324, 2006.

\bibitem{XuTB03}
Xy~C, Tobi D, and Bahar I.
\newblock Allosteric changes in protein structure computed by a simple
  mechanical model: hemoglobin {T}---{R2} transition.
\newblock {\em J. Mol. Biol.}, 333:153--168, 2003.

\bibitem{MiyOW03}
O~Miyashita, JN~Onuchic, and PG~Wolynes.
\newblock Nonlinear elasticity, proteinquakes, and the energy landscapes of
  functional transitions in proteins.
\newblock {\em Proc. Nat. Acad. Sci.}, 100:12570--12575, 2003.

\bibitem{FarST10}
Farrell DW, Speranskiy K, and Thorpe MF.
\newblock Generating stereochemically-acceptable protein pathways.
\newblock {\em Proteins}, 78:2908--2921, 2010.

\bibitem{JimWFR11}
Jimenez-Roldan JE, Wells SA, Freedman RB, and RoemerRA.
\newblock Integration of {FIRST}, {FRODA} and {NMM} in a coarse grained method
  to study protein disulphide isomerase conformational change.
\newblock In {\em Journal of Physics: Conference Series}, volume 286, page
  012002. IOP Publishing, 2011.

\bibitem{KreAWE02}
Krebs WG, Alexandrov V, Wilson CA, Echols E, Yu~H, and Gerstein M.
\newblock Normal mode analysis of macromolecular motions in a database
  framework: developing mode concentration as a useful classifying statistic.
\newblock {\em Proteins}, 48:682--695, 2002.

\bibitem{JimBVF11}
Jimenez-Roldan JE, Bhattacharya M, Vishveshwara S, Freedman RB, and Roemer RA.
\newblock in preperation.

\bibitem{BelCM11}
Belfield W, Cole D, and Payne M.
\newblock in preparation.

\bibitem{BurVWW11}

Burkoff NS, Varnai C, Wells SA, and Wild DL.
\newblock Exploring the energy landscapes of protein folding simulations with
  bayesian computation.
\newblock {\em Proc. Nat. Acad. Sci.}, 2011.
\newblock submitted.

\end{thebibliography}
\end{document}